\documentclass[11pt]{article}

\usepackage[a4paper,
            left=2.4cm,
            right=2.4cm,
            top=3cm,
            bottom=3cm]{geometry}

\usepackage{amsmath,amssymb}
\usepackage[T1]{fontenc}
\usepackage{lmodern}
\usepackage{microtype}
\usepackage{graphicx}

\title{Finding the right path: statistical mechanics of connected solutions in constraint satisfaction problems}
\author{Barbier Damien\\
\small Department of Computing Sciences, BIDSA, Bocconi University, Milan, MI 20100, Italy\\
\small \texttt{damien.barbier@unibocconi.it}}
\date{}

\begin{document}

% IOP
%%%%%%%%%%%%%%%%%%%%%%%%
%%%%%%%%%%%%%%%%%%%%%%%%
%\articletype{Paper} 

%\title{Finding the right path: statistical mechanics of connected solutions in constraint satisfaction problems}

%\author{Barbier Damien}

%\affil{ Department of Computing Sciences, BIDSA, Bocconi University, Milan, MI 20100, Italy}

%\email{damien.barbier@unibocconi.it}

%\keywords{neural network, statistical mechanics, disordered system}

%%%%%%%%%%%%%%%%%%%%%%%%
%%%%%%%%%%%%%%%%%%%%%%%%

%arxiv
\maketitle

\begin{abstract}
We define and study a statistical mechanics ensemble that characterizes connected solutions in constraint satisfaction problems (CSPs). Built around a well-known local entropy bias, it allows us to better identify hardness transitions in problems where the energy landscape is dominated by isolated solutions. We apply this new device to the symmetric binary perceptron model (SBP), and study how its manifold of connected solutions behaves.
We choose this particular problem because, while its typical solutions are isolated, it can be solved using local algorithms for a certain range of constraint density $\alpha$ and threshold $\kappa$. 
With this new ensemble, we unveil the presence of a cluster composed of delocalized connected solutions. In particular, we demonstrate its stability until a critical threshold $\kappa^{\rm no-mem}_{\rm loc.\, stab.}$ (dependent on $\alpha$). This transition appears as paths of solutions become unstable, a phenomenon that more conventional statistical mechanics approaches fail to grasp. Finally, we compared our predictions to simulations. For this, we used a modified Monte-Carlo algorithm, designed specifically to target these delocalized solutions. We obtained, as predicted, that the algorithm finds solutions until $\kappa\approx\kappa^{\rm no-mem}_{\rm loc.\, stab.}$.
\end{abstract}

\section{Introduction}
Characterizing rugged and complex energy landscapes is a challenging task that many fields are grappling with. It spans disordered physics \cite{mezard1987spin,PhysRevLett.71.173,Ros2021,krzakala2007gibbs}, ecology \cite{Fear1998,Hatton2024}, chemistry \cite{Stadler2002,Mauri2023,Wu2016}, computer science \cite{Gamarnik2022,mezard2002analytic,DMM09,addario2018algorithmic} and even biology \cite{Papkou2023,Greenbury2022,Macadangdang2022} to name a few. While the initial focus was on describing the minima of these landscapes \cite{parisi1979infinite,thouless1977solution,wr32,Hopfield1982,gardner1988optimal}, recent years have seen a growing interest in a deeper understanding of their dynamical explorability. With different tools and terminologies, each field has unveiled important breakthroughs: the glassy transition and aging dynamics in disordered physics \cite{mezard1987spin,parisi1979infinite,PhysRevLett.71.173}, the overlap-gap property (OGP) and locally stable algorithms in computer science \cite{gamarnik2021survey,Achlioptas2011,gamarnik2014limits} or even the inference of phylogenetic trees in biology \cite{Rannala1996,Huelsenbeck2001}. In all these cases, the overall setup remains the same: a system with a large number of degrees of freedom evolves in a landscape with dynamics that only allows for incremental reconfigurations. Many questions then arise: Where does a given dynamic lead? Is there an optimal dynamic given an energy landscape? Can the geometry of the landscape provide us with guidance for designing new dynamics/algorithms?

A case of particular interest is the perceptron model.
It is a constrained satisfactory problem in which a $N$-dimensional vector ${\bf x}$ has to correctly classify an extensive number of random data points $\{\xi^\mu\}_{\mu \in [\![1,M]\!]}$ (for which extensive means $M\propto N$) given an activation function. Since its first appearance in the 80s \cite{gardner1988optimal}, many variants of this model have been scrutinized with different phase spaces for ${\bf x}$, data point distributions or activation functions \cite{Annesi2023,krauth89storage,baldassi2020clustering,aubin2019storage}. However, in most cases, one problem remains. The statistical mechanics approach used to characterize the minima of its landscape fails to predict why dynamics manages (or not) to access these low-energy configurations \cite{krauth89storage,braunstein2006learning,bansal2020line}. This failure does not come from the introduction of wrong $Ansatz$ or hypotheses in some computation steps (like the level of replica symmetry breaking when performing a standard disordered physics computation), as certain cases have been proven to be rigorously correct \cite{aubin2019storage,barbier2023atypical,Huang2024}. In fact, the problem is more basic. Standard statistical mechanics approaches probe low-energy configuration manifolds without considering their dynamical accessibility. This results in measurements dominated by minima that are irrelevant to any realistic dynamics (because they are inaccessible). Several attempts to correct for this shortcoming have been introduced over the last thirty years \cite{franz1995recipes,Franz_2013,Franz_2015,Barbier2025} and have given promising results \cite{KrzakalaEtAl,Barrat1997,Barbier2023,SaraoMannelli2020}. Yet, none of these newer methods have succeeded in properly describing the perceptron model and the arrangements of its accessible solutions.

In recent years, a series of works on this model have begun to unveil the properties of its particular low-energy manifold \cite{baldassi2016local,baldassi2015subdominant,baldassi2020clustering,baldassi2021unveiling,Baldassi2020,Baldassi2020_,Annesi2023}. With different specificities due to model variations, a general picture has emerged. While the dominant solutions (exponentially more numerous) lie in isolated regions of the landscape, subsets of minima form well-connected clusters that extend over a non-negligible part of the phase space. In particular, authors have claimed to be able to target these {\it dense clusters} with several new dynamics, all based on local-entropy biases \cite{baldassi2016local,Baldassi2020___,baldassi2016unreasonable}. Instead of simply searching for a minimum via a loss-minimization routine (for example Monte-Carlo or gradient descent), these newer dynamics have an additional mechanism that directs the search around low-energy configurations that have other low-energy configurations in their immediate surrounding. The two most common dynamics with this bias are the local-entropy probing \cite{baldassi2016local,Baldassi2020___} and the replicated Monte-Carlo \cite{baldassi2016unreasonable,Baldassi2018}. 

For now, the role of local-entropy biases has not been fully understood. Their implementations in algorithms still rely on heuristics and it remains difficult to estimate how well they target the {\it dense cluster} observed in the perceptron model. Additionally, if so many points remain unclear about these new dynamical schemes, it is also because these ``giant clusters'' lack a solid theoretical characterization. 

In this paper, we will focus our case study on the symmetric binary perceptron (SBP). On the one hand, this model is of particular interest for computer science as it is a constraint satisfaction problem that can be tackled with rigorous mathematical techniques. On the other hand, it is also a setup that shares a lot of characteristics with sequence-based evolution models -as we will see in the next section-. The aim of this paper will be to show that local-entropy biases are in fact a first conceptual step for targeting minima that form a web of connected configurations. We will detail how this technique can be generalized and why it gives a first theoretical characterization of the ``giant cluster'' that were only observed so far. This paper is organized as follows, in Sec.~\ref{sec: I} we introduce the perceptron model and define -in a generic form- the statistical ensemble for connected solutions. In Sec.~\ref{sec: II}, we detail the computation and observables that enable us to characterize delocalized clusters of solutions. Finally, in Sec.~\ref{sec: numerics} we provide numerical results in order to compare our theoretical predictions with large-size simulations.

\section{Definitions and first physics intuitions}
\label{sec: I}
\subsection{The model}
The symmetric binary perceptron (SBP) is a theoretical model that was first introduced in \cite{aubin2019storage}. It is a constraint satisfaction problem in which a $N$-dimensional binary vector ${\bf x} \in \{-1,+1\}^N$ (or, written in a compact form, ${\bf x}\in \Sigma^N$) must verify an extensive number of random inequalities. These random inequalities are constructed as follows. Given an ensemble of $M$ i.i.d. random patterns $\{\xi^\mu\}_{\mu\in[\![1,M]\!]}$, with $\xi^\mu\in {\rm I\!R}^N$, we want to find ${\bf x}$ on the hypercube  such that
\begin{equation} \begin{split}
\label{eq: constraints}
    \vert\xi^\mu\cdot{\bf x}\vert\leq\kappa\sqrt{N} ~~~~~ \mbox{for all}~ 1 \le \mu \le M\, .
\end{split}\end{equation} 
For simplicity, we choose normal distributed patterns, i.e. $\xi^\mu\sim\mathcal{N}(0,{\bf I})$. Considering the thermodynamic limit ($N\rightarrow+\infty$), this model is controlled by two positive parameters: the constraint density $\alpha=M/N$ and the threshold $\kappa$. Intuitively, one can understand that the SBP becomes more difficult to solve as $\alpha$ increases and vice versa as $\kappa$ decreases. More precisely, it was proven in~\cite{aubin2019storage} that this problem admits solutions with high probability if and only if the margin threshold $\kappa$ satisfies
\begin{equation} \begin{split}
\kappa>\kappa^\alpha_{\rm SAT}
\end{split}\end{equation}
with
\begin{equation} \begin{split}
\label{eq:SAT}
\log(2)+\alpha\log\left(\int \mathcal{D}u,\Theta(\kappa^\alpha_{\rm SAT}-\vert u\vert)\right)=0,.
\end{split}\end{equation}
Throughout this paper, we will use the notation $\mathcal{D}.$ to represent integration over a scalar normally distributed variable, and $\Theta(.)$ for the Heaviside function. In short, the authors of \cite{aubin2019storage} demonstrated this result by showing that the entropy of solutions for the SBP can be evaluated at the \emph{annealed} level:
\begin{align} 
s&=\lim_{N\rightarrow+\infty}\frac{1}{N}{\rm I\!E}_\xi\left[\log\left(\sum_{{\bf x}\in\Sigma^N}\prod_{\mu =1}^M\Theta\left(\kappa-\left\vert\frac{\xi^\mu\cdot {\bf x}}{\sqrt{N}}\right\vert\right)\right)\right]\\ &=\lim_{N\rightarrow+\infty}\frac{1}{N}\log\left(\sum_{{\bf x}\in\Sigma^N}\prod_{\mu =1}^M{\rm I\!E}_\xi\left[\Theta\left(\kappa-\left\vert\frac{\xi^\mu\cdot {\bf x}}{\sqrt{N}}\right\vert\right)\right]\right)\\ 
&=\log(2)+\alpha\log\left(\int \mathcal{D}u\,\Theta(\kappa^\alpha_{\rm SAT}-\vert u\vert)\right)\nonumber \end{align}
When the entropy becomes negative, the SBP admits no solutions. Usually, the solution space that follows from an \emph{annealed} computation is trivially convex. Taking a typical solution ${\bf x}^a$, one finds an ever-increasing number of surrounding solutions ${\bf x}$ as the distance $d=1-{\bf x}\cdot{\bf x}^a/N$ increases. One simple consequence of this is that two solutions ${\bf x}^a$, ${\bf x}^b$ are typically at distance $d=1-{\bf x}^a\cdot{\bf x}^b/N=0$ from each other. However, as first described in the seminal paper by M\'ezard and Krauth \cite{krauth89storage}, the solution space of the SBP does not behave exactly like this. Rather, it follows a more complex geometrical structure known as {\it frozen replica symmetry breaking} \cite{martin2004frozen,zdeborova2008locked,huang2013entropy,huang2014origin,Semerjian}. As in the \emph{annealed} case, two solutions ${\bf x}^a$, ${\bf x}^b$ are typically at distance $d=1-{\bf x}^a\cdot{\bf x}^b/N=0$; it is for this reason that the entropy evaluated at the \emph{annealed} level is correct. But in their surroundings, the number of solutions does not increase with the distance. In fact, typical solutions have no other solutions in their immediate surroundings; they are \emph{isolated}. To reach any other solution from a typical solution, one must perform an extensive number of bit flips.

\begin{figure}
\centering
   \includegraphics[width=1\linewidth]{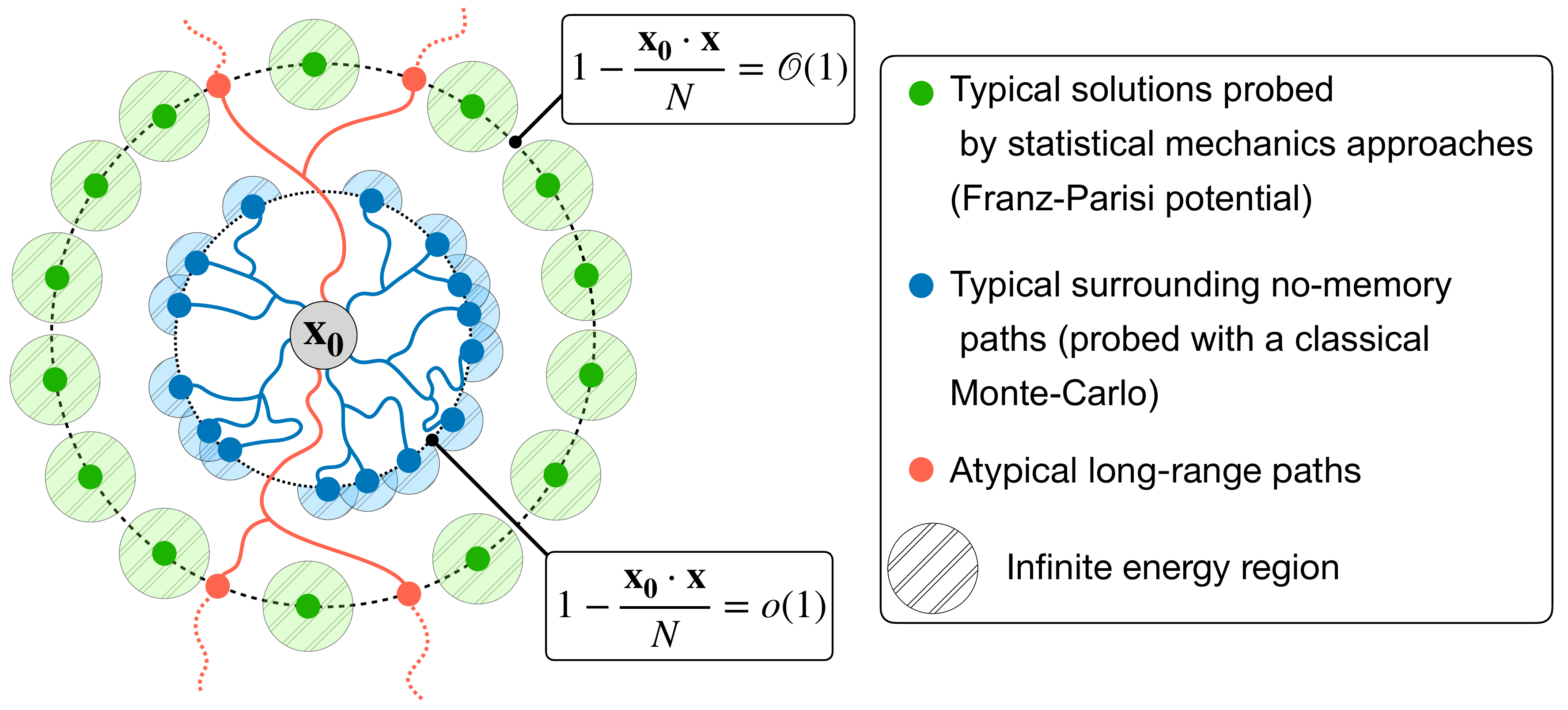}
    \caption{Drawing representing the local arrangement of solutions $\bf x$ around a robust reference minima $\bf x_0$. At any extensive distance from ${\bf x}_0$ -$1-{\bf x}\cdot{\bf x}_0/N=\mathcal{O}(1)$-, typical solutions are isolated as infinite energy barriers surround them \cite{krauth89storage}. This known as an overlap-gap property (OGP) \cite{gamarnik2021survey}. Due to these barriers, these solutions ${\bf x}$ cannot be accessed with local algorithms. Extremely close to ${\bf x}_0$  -for $1-{\bf x}\!\cdot\!{\bf x}_0/\!N\!=\!o(1)$-, paths of connected solutions are dominated in number by Markov chains \cite{Barbier2025}. Such paths can be explored by a classical Monte-Carlo algorithm, but their length is not extensive. This work proposes to find the atypical paths (pictured in orange) that would allow us to decorrelate from ${\bf x}_0$.}
    \label{fig: paths in SBP}
\end{figure}

Further investigations on the SBP showed that, although certain atypical/robust solutions are not isolated -meaning they are surrounded by other low-energy configurations- \cite{barbier2023atypical}, simple dynamics like the standard Monte-Carlo algorithm fail to navigate in short times inside these well-chosen regions \cite{Barbier2025}.  In Fig.~\ref{fig: paths in SBP}, we sum up in a schematic way how low-energy configurations are arranged around an atypical/robust solution of the SBP (that we label ${\bf x}_0$).

Despite these challenges, more sophisticated algorithms can find (and sometimes provably) atypical solutions of the SBP within a finite range of parameters $\alpha$ and $\kappa$ \cite{bansal2020line,baldassi2016local,braunstein2006learning}. Several works have attributed this success to the presence of so-called atypical \emph{dense} regions of solutions \cite{baldassi2016unreasonable,baldassi2016local,baldassi2015subdominant,baldassi2020clustering,baldassi2021unveiling}. To characterize these minima, the authors rely on replica techniques and use them to probe low-energy configurations tightly confined within small regions of the phase space. In short, they studied how $n$-tuples of configurations $\{{\bf x}^a\}_{a\in[\![1,\dots,n]\!]}$ -constrained to verify ${\bf x}^a\cdot {\bf x}^{b(\neq a)}/N=q$ (with $a,b\in[\![1,\dots,n]\!]$ and $a\neq b$)- populate the SBP landscape. Although this method shows promising agreement with numerical experiments, a recent study \cite{Barbier2025} demonstrated that such a statistical-mechanics approach cannot capture the algorithmic hardness of the SBP. 
This work showed that standard statistical mechanics fails to characterize how a Monte Carlo algorithm explores the vicinity of a robust solution ${\bf x}_0$. Instead, a computation based on a chain of equilibrium, where minima $\{{{\bf x}_t\}}_{t\in{\rm I\!N}}$ are sampled according to a transition probability distribution $P[{\bf x}_t\vert {\bf x}_{t-1}]$, is required to capture the correct phenomenology. This shows that standard statistical mechanics approaches, including those mentioned above, cannot describe how algorithms navigate the SBP landscape. Finally, while the work in \cite{Barbier2025} better characterizes this landscape, it only explains why algorithms succeed to solve the SBP in a high-threshold regime, $\kappa\sim\sqrt{\log{N}}$.

Interestingly, this architecture for the SBP and its solution manifold appears to reproduce many recently discovered features in sequence-based models (of proteins or bacteria) and their associated landscapes. For the interested reader, we will describe in a few lines the main parallels between these models. Firstly, both models are discrete-variable systems ($\pm 1$ for the SBP and amino acids for proteins). This choice of phase-space is known to affect the geometry of low-energy manifolds for many problems \cite{Kosterlitz1976,parisi1979infinite,gardner1988optimal}. In particular, compared to continuous-variable systems, this makes energy barriers more difficult to bypass dynamically. Secondly, the landscape in both setups is rugged: it is dominated by a large number of isolated minima \cite{aubin2019storage,Barbier2025,Mauri2023,Greenbury2022} (surrounded by energetic barriers). These prevalent minima can prevent standard statistical mechanics approaches from predicting the behavior of local-move dynamics. Indeed, if the low-energy manifold is dominated by isolated configurations, statistical mechanics tools will inevitably characterize such inaccessible minima. Finally, the paths linking accessible solutions appear to be complex. In fact, in both cases, simple direct paths fail to join two solutions without crossing high-energy barriers \cite{Barbier2025,Mauri2023,Greenbury2022}.

After these phenomenological considerations, we will introduce the three central quantities for the study of the SBP. The first one is the margins (given a configuration $\bf x$)
\begin{equation} 
\begin{split}
    w^\mu=\frac{\xi^\mu\cdot{\bf x}}{\sqrt{N}}\quad \mbox{for}\quad \mu\in[\![1,M]\!]
\end{split}
\end{equation} 
and more particularly their statistical distribution
\begin{equation} \begin{split}
    P(w)={\rm I\!E}_{\xi,{\bf x}}\left[\frac{1}{M}\sum_{\mu=1}^M\delta\left(w-\frac{\xi^\mu\cdot{\bf x}}{\sqrt{N}}\right)\right]\,.
\end{split}\end{equation} 
where the average over ${\bf x}$ remains to be detailed at this point.
Emphasized in several works \cite{Barbier2025,barbier2023atypical}, this quantity appears to dictate if a given configuration $\bf x$ is isolated or not. We will see in the following that it also plays a crucial role within our framework for connectivity. The second quantity are overlaps. Namely, if we take two configurations ${\bf x}_a$ and ${\bf x}_b$ their overlap is defined as
\begin{equation} \begin{split}
    m=\frac{{\bf x}_a \cdot {\bf x}_b}{N}\, .
\end{split}\end{equation} 
In high dimensions, it turns out (as we will see) that a configuration path $\{x_{t}\}_{t\in[\![0,t_f]\!]}$ can be fully described by the overlap matrix \cite{PhysRevLett.71.173,Barbier2025,Franz_2013}
\begin{equation} \begin{split}
    {\bf Q}_{t,t'}=\frac{{\bf x}_{t}\cdot {\bf x}_{t'}}{N}\, .
\end{split}\end{equation} 
This arises naturally in any statistical mechanics approach, it is the result of measure-concentration properties. Finally, the last quantity we have to introduce is the partition function $\mathcal{Z}$ (with its associated free energy $\phi$) as
\begin{equation} \begin{split}
\label{eq: partition func def}
    \mathcal{Z}=\sum_{{\bf x}\in \Sigma^N}e^{-\mathcal{L}_{\rm SBP}({\bf x})}
    \quad \mbox{with}\quad\phi=\log{\mathcal{Z}}
\end{split}\end{equation} 
and
\begin{equation} \begin{split}
\label{eq: SBP loss}
    e^{-\mathcal{L}_{\rm SBP}({\bf x})}= \prod_{\mu =1}^M\Theta\left(\kappa-\left\vert\frac{\xi^\mu\cdot {\bf x}}{\sqrt{N}}\right\vert\right)\, .
\end{split}\end{equation} 
Thus, if a configuration $\bf x$ does not verify all the constraints from Eq.~(\ref{eq: constraints}), it will not contribute in the SBP partition function ($e^{-\mathcal{L}_{\rm SBP}({\bf x})}=0$) . Conversely, any configurations solving the problem are given the same weight ($e^{-\mathcal{L}_{\rm SBP}({\bf x})}=1$). As mentioned earlier, this partition function is in fact dominated by isolated solutions. A standard method to show this is to evaluate with the replica method its disordered-averaged free energy \cite{gardner1988optimal,krauth89storage,aubin2019storage}
\begin{equation} \begin{split}
    \phi={\rm I\!E}_{\xi}[\log(\mathcal{Z})]\underset{n\rightarrow 0}{=}\frac{{\rm I\!E}_{\xi}[\mathcal{Z}^n]-1}{n}
\end{split}\end{equation} 
where ${\rm I\!E}_{\xi}$ indicates the average over the patterns distribution -$\xi^\mu\sim\mathcal{N}(0,{\bf I})$ ($i\in[\![1,N]\!]$) in our case-. We will use this technique later to compute other partition functions.

\subsection{Local entropy probing and generalization}
We will now introduce the first tool to move towards sampling connected configurations. Given an energy landscape -with associated loss function $\mathcal{L}(\cdot)$-, the idea is to modify the setting by adding a local-entropy cost to the initial loss \cite{baldassi2016local}. This means that the partition function $\mathcal{Z}$ that we will now try to evaluate is
\begin{equation} \begin{split}
    \mathcal{Z}=\sum_{{\bf x}_0\in \Sigma^N}e^{-\mathcal{L}({\bf x}_0)+y_1\phi_1({\bf x}_0,m)}
\end{split}\end{equation} 
with the local-entropy cost
\begin{equation} \begin{split}
    \phi_1({\bf x}_0,m)=\log\left[\sum_{\underset{{\rm s.t.}\, \frac{{\bf x}_1\cdot {\bf x}_0}{N}=m}{{\bf x}_1\in \Sigma^N}}e^{-\mathcal{L}({\bf x}_1)}\right]\,.
\end{split}\end{equation} 
More explicitly, $\phi_1({\bf x}_0,m)$ counts the low-energy configurations (labeled ${\bf x}_1$) lying in the vicinity of a reference vector ${\bf x}_0$. For binary vectors, this constraint imposes that ${\bf x}_1$ and ${\bf x}_0$ share $N(1+m)/2$ bits in common, or written differently $\frac{{\bf x_1}\cdot {\bf x}_0}{N}=m$. Therefore, a configuration ${\bf x}_0$ with a great local-entropy cost -i.e. with a large number of low-energy states surrounding it- will start to count more in the new partition function with its shifted weight: $e^{-\mathcal{L}({\bf x}_0)}\rightarrow e^{-\mathcal{L}({\bf x}_0)+y_1\phi_1({\bf x}_0,m)}$. In the thermodynamic limit, $\mathcal{Z}$ is dominated by a manifold of configurations with fixed loss $\mathcal{L}({\bf x}_0)+y_1\phi({\bf x}_0,m)$, which shows a competition between optimizing the original loss $\mathcal{L}(\cdot)$ and finding non-isolated configurations. The Lagrange multiplier $y_1$ tunes this effect, setting $y_1=0$ gives the original setup while $y_1=+\infty$ gives only the local-entropy bias. In general, this bias has often been seen as a mean to target ``dense'' regions of the energy landscape \cite{baldassi2015subdominant}, in the sense that a large number of minima are close to each other. We will see that it is also the main ingredient for targeting clusters containing connected minima. 

Interestingly, this local-entropy potential seems to draw another parallel between the SBP and sequence-based models. For the interested reader, this bias appears to be very close to a mechanism known as diversity-generating retroelements (DGRs) in biology \cite{Medhekar2007,Laurenceau2025,Macadangdang2022}. First observed in the Bordetella phage (BPP-1), it is a mechanism which allows a piece of DNA sequence -called variable repeat (VR)- to be highly diversified in a bacteria population. In fact, when a BPP-1 bacterium multiplies, this particular region of DNA is not passed on to descendants via a usual copy-and-paste mechanism. Actually, there exists an almost identical DNA sequence located elsewhere in the genome -called template repeat (TR)- that acts like a parent sequence for generating VR sequences during multiplications. This copy-and-replace mechanism induces errors, which means that the VR sequence is highly variable in a population (while the TR remains relatively constant). In the end, only the VR sequence is expressed and influences the organism's behavior. In terms of landscape fitness, this means that the TR sequence is surrounded by low-energy sequences (the VR mutants).
The parallel with the local-entropy cost is therefore direct. The TR sequence plays the same role as the reference configuration ${\bf x}_0$ in the local-entropy setting. In this special case, it follows the limit $y_1=+\infty$ because the TR is not directly expressed, i.e. it does not feel the energy landscape directly (also called the fitness landscape in biology). Nevertheless, it must be close to viable DNA sequences (the VRs) so that the overall population can survive. With this parallel, the VRs intuitively correspond to the low-energy states labeled ${\bf x}_1$ in our setting.

Although promising, the local-entropy bias is not sufficient to obtain algorithmically accessible solutions in the SBP \cite{barbier2023atypical}. The problem is as follows: ${\bf x}_0$ now has minima in its vicinity, but these minima ${\bf x}_1$ are actually isolated. This means that the configurations that bias the sampling of ${\bf x}_0$ lie in a rugged region of the landscape and, consequently, are  not good dynamical attractors. If we want a solution ${\bf x}_0$ that is dynamically accessible, we must ensure that its direct vicinity is composed of non-isolated minima. Only such well-connected regions can be exploited by dynamics. To obtain these configurations, we propose to study a chain of nested local-entropy biases. The idea behind this is to bias not only the weights for the reference configuration ${\bf x}_0$ (with a local-entropy cost), but also those of ${\bf x}_1$ with another local-entropy cost. In other words, we will also constrain the states for ${\bf x}_1$ to be surrounded by yet other nearby low-energy configurations. This intuitively adds a sum over binary vectors ${\bf x}_2$ with an associated set of weights. Continuing this construction, we can also bias the ${\bf x}_2$ configurations with a local-entropy cost to ensure that they are not isolated. Intuitively, this scheme can be iterated $k_f$ times, with each states for ${\bf x}_k$ drawn in the vicinity of a configuration ${\bf x}_{k-1}$ and biased to have low-energy states ${\bf x}_{k+1}$ around it. Ultimately, this construction guaranties that the configurations ${\bf x}_0$ are included in a connected-minima cluster: in the vicinity of ${\bf x}_0$ we can now find non-isolated low-energy configurations.
The partition function associated with such a construction is
\begin{equation} \begin{split}
\label{eq: connected measure}
    \mathcal{Z}=\sum_{{\bf x}_0\in \Sigma^N}e^{-\mathcal{L}({\bf x}_0)+y_1\phi_1({\bf x}_0,m)}
\end{split}\end{equation} 
with generalized local-entropy costs ($k\in[\![1,k_f-1]\!]$)
\begin{equation} \begin{split}
    \hspace{-0.115cm}\phi_k({\bf x}_{k-1},m)\!=\!
    \log\!\left[\!\sum_{{\bf x}_k\in \Sigma^{N,m}_{{\bf x}_{k-1}}}\hspace{-0.25cm}e^{-\mathcal{L}({\bf x}_k)+y_{k+1}\phi_{k+1}({\bf x}_{k},m)}\!\right]
\end{split}\end{equation} 
and
\begin{equation} \begin{split}
    \phi_{k_f}({\bf x}_{k_f-1},m)=
    \log\left[\sum_{{\bf x}_{k_f}\in \Sigma^{N,m}_{{\bf x}_{k_f-1}}}e^{-\mathcal{L}\left({\bf x}_{k_f}\right)}\right].
\end{split}\end{equation} 
For compact notation, we use $\Sigma^{N,m}_{{\bf x}^*}$ to represent the space of binary configurations $\bf x$ that verify ${\bf x}\cdot {\bf x}^*/N=m$. 

As a reminder, the SBP loss involves random patterns -see Eq.~(\ref{eq: SBP loss})-. This means that we will have to add an average over the patterns' distribution. In this case, the standard quantity to evaluate is the associated $quenched$ free energy
\begin{equation} \begin{split}
    \phi={\rm I\!E}_\xi\left[\log(\mathcal{Z})\right]\, .
\end{split}\end{equation} 
Its associated measure can be written as
\begin{align}
    {\rm I\!E}_{\xi,{\bf x}_k}\left[f(\{\xi^\mu\},{\bf x}_k)\right]={\rm I\!E}_{\xi}\left[\prod_{j=0}^{k_f}\left(\frac{\underset{{{\bf x}_{j}\in\Sigma^{N,m}_{{\bf x}_{j-1}}}}{\sum}\!\!e^{-\mathcal{L}({\bf x}_{j})+y_{j+1}\phi_{j+1}({{\bf x}_{j},m})}}{\underset{{\underline{\bf x}_{j}\in\Sigma^{N,m}_{{\bf x}_{j-1}}}}{\sum}\!\!e^{-\mathcal{L}(\underline{\bf x}_{j})+y_{j+1}\phi_{j+1}({\underline{\bf x}_{j},m})}}\right)f(\{\xi^\mu\},{\bf x}_k)\!\right] \, .
\end{align}

Before trying to evaluate this free energy, we must consider which value to assign to $m$. Although $m$ is usually seen as an adjustable parameter, we will take  the limit $m\rightarrow1$ in our setting. This will ensure a strong connectivity constraint among minima. Indeed, it will ensure that two connected minima ${\bf x}_{k}$ and ${\bf x}_{k+1}$ become increasingly close to one another, while still remaining distinct. Why do we take this limit? We recall that our focal point is to obtain a vicinity around ${\bf x}_0$ that is composed of non-isolated minima. Only such regions can be dynamical attractors. Taking $m\rightarrow1$, we ensure that this requirement is met: every configuration ${{\bf x}_k}$ will have nearby neighbors ${{\bf x}_{k+1}}$. Quite intuitively, this limit will require taking the total number of iteration $k_f$ to infinity, so as to have a well extended cluster of connected minima, i.e.
\begin{equation} \begin{split}
    \lim_{m\rightarrow1}\frac{{\bf x}_{k_f}\cdot {\bf x}_0}{N}\neq 1 \,.
\end{split}\end{equation} 

In the following, we will detail what $Ansatz$ and approximations can be used to evaluate the free energy $\phi$.

\section{The no-memory cluster(s)}
\label{sec: II}
\subsection{From the general setting to the no-memory $Ansatz$}
To evaluate the $quenched$ free energy for the SBP
\begin{align}
\label{eq: quenched free energ connected cluster}
     \phi={\rm I\!E}_\xi\left[\log(\mathcal{Z})\right]={\rm I\!E}_\xi\left[\log\left(\sum_{{\bf x}_0\in \Sigma^N}e^{-\mathcal{L}_{\rm SBP}({\bf x}_0)+y_1\phi_1({\bf x}_0,m)}\right)\right]\, ,
\end{align}
we use a standard technique known as the replica method \cite{mezard1987spin}.  It begins by computing -for $y_0\in{\rm I\!N}$-
\begin{align}
{\rm I\!E}_\xi[\mathcal{Z}^{y_0}]={\rm I\!E}_\xi\left[\prod_{j_0=1}^{y_0}\left(\sum_{{\bf x}_0^{j_0}\in\Sigma^N}e^{-\mathcal{L}_{\rm SBP}\left({\bf x}_0^{j_0}\right)+y_1\phi_1\left({\bf x}_0^{j_0},m\right)}\right)\right]\, ,
\end{align}
then extending this quantity to $y_0\in{\rm I\!R}$ and finally performing the limit
\begin{align}
    \phi={\rm I\!E}_{\xi}[\log(\mathcal{Z})]\underset{y_0\rightarrow 0}{=}\frac{{\rm I\!E}_{\xi}[\mathcal{Z}^{y_0}]-1}{y_0}\,.
\end{align}
We can see that the expression for ${\rm I\!E}_\xi[\mathcal{Z}^{y_0}]$ can be further developed as
\begin{align}
{\rm I\!E}_\xi[\mathcal{Z}^{y_0}]&={\rm I\!E}_\xi\left\{\prod_{j_0=1}^{y_0}\left[\sum_{{\bf x}_0^{j_0}\in\Sigma^N}e^{-\mathcal{L}_{\rm SBP}\left({\bf x}_0^{j_0}\right)}\prod_{j_1=1}^{y_1}\left(\sum_{{\bf x}^{j_1}_{1}\in \Sigma^{N,m}_{{\bf x}^{j_0}_{0}}}e^{-\mathcal{L}_{\rm SBP}\left({\bf x}_1^{j_1}\right)+y_2\phi_2\left({\bf x}_1^{j_1},m\right)}\right)\right]\right\}\, .
\end{align}
This expression can be made more compact by introducing a path notation,
\begin{align}
    {\bf P}_0=\{j_0\}_{j_0\in[\![1,y_0]\!]}~~{\rm and }~~ {\bf P}_1= \left\{j_0,j_1\right\}_{\substack{j_0\in[\![1,y_0]\!] \\j_1\in[\![1,y_1]\!]}},
\end{align}
which yields
\begin{align}
{\rm I\!E}_\xi[\mathcal{Z}^{y_0}]&={\rm I\!E}_\xi\left[\prod_{{\bf P}_0}\sum_{{\bf x}^{{\bf P}_0}\in\Sigma^N}e^{-\mathcal{L}_{\rm SBP}\left({\bf x}^{{\bf P}_0}\right)}\prod_{{\bf P}_1}\sum_{{\bf x}^{{\bf P}_1}\in \Sigma^{N,m}_{{\bf x}^{{\bf P}^*_1}}}e^{-\mathcal{L}_{\rm SBP}\left({\bf x}^{{\bf P}_1}\right)+y_2\phi_2\left({\bf x}^{{\bf P}_1},m\right)}\right]\, .
\end{align}
with the direct ancestor vector ${\bf P}^*_1={\bf P}_1[1]$. As an example, if we take ${\bf P}_1=\{2,3\}$ then ${\bf P}^*_1=\{2\}$.

\begin{figure}
    \centering
    \includegraphics[width=1\linewidth]{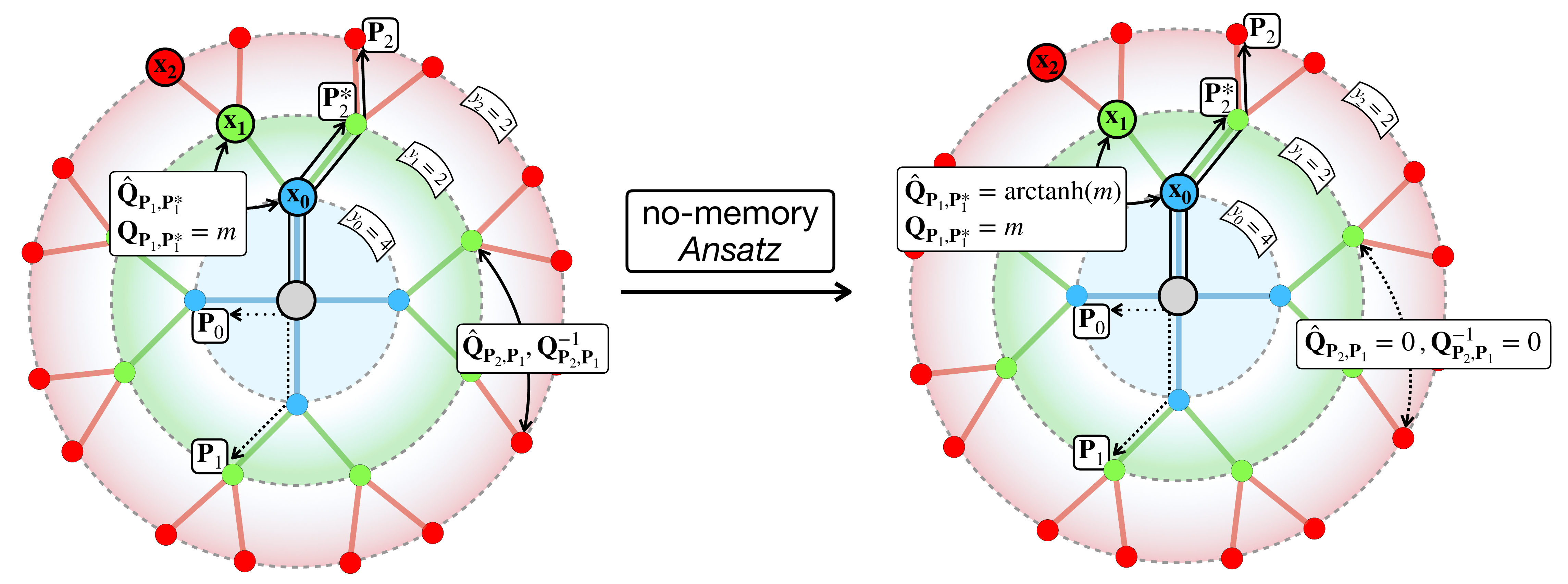}
    \caption{Drawing representing the connected structure introduced to evaluate $\phi$ -see Eq.~(\ref{eq: quenched free energ connected cluster})-. On the left, we have the generic structure where all configurations can couple with each other. The only constraint is that each configuration ${\bf x}^{{\bf P}_k}$ overlaps with its direct ancestor: ${\bf x}^{{\bf P}^*_k}\cdot{\bf x}^{{\bf P}_k}/N=m$. On the right, we detailed how almost all correlating fields are dropped once the no-memory $Ansatz$ is taken. }
    \label{fig: no-mem sketch}
\end{figure}

If we now develop all the $k_f$ local-entropy costs, we will have to consider $\{y_k\}_{k\in[\![0,k_f]\!]}\in {\rm I\!N}^{k_f+1}$ in order to perform the average over the patterns' distribution. A computation that we will then extend for $\{y_k\}_{k\in[\![0,k_f]\!]}\in {\rm I\!R}^{k_f+1}$. In other words, we consider the quantity
\begin{align}
\hspace{-0.1cm}{\rm I\!E}_\xi[\mathcal{Z}^{y_0}]={\rm I\!E}_{\xi}\!\left\{\prod_{k=0}^{k_f}\prod_{{\bf P}_{ k}} \sum_{{\bf x}^{{\bf P}_{ k}}\in\Sigma^{N,m}_{{\bf x}^{{\bf P}^*_{k}}}}\hspace{-0.46cm}e^{-\mathcal{L}_{\rm SBP}\left({\bf x}^{{\bf P}_{ k}}\right)}\!\right\}
\end{align}
where ${\bf P}_{ k} (=\{j_0,\dots j_{k_f}\}_{j_k\in[\![1,k]\!]})$ indexes each connected configuration. It corresponds to the unique path from level $k'=0$ to level $k'=k$ that reaches its associated binary configuration ${\bf x}^{{\bf P}_k}$ (${\bf x}^{{\bf P}_k}\in \Sigma^N$). For clarity, we illustrate this indexing in Fig.~\ref{fig: no-mem sketch}. 
Also, for better readability in the following, we must introduce the direct ancestor of a configuration ${\bf x}^{{\bf P}_k}$. This configuration, labeled ${\bf x}^{{\bf P}^*_k}$, is the unique configuration at level $k'=k-1$ whose overlap with ${\bf x}^{{\bf P}_k}$ is fixed: ${\bf x}^{{\bf P}_k}\cdot {\bf x}^{{\bf P}^*_k}/N=m$. In terms of path notation, this means that ${\bf P}_k^*={\bf P}_k[1:k-1]$. As an example, if we have ${\bf P}_4=\{2,5,3,7\}$ then ${\bf P}_4^*=\{2,5,3\}$. Consequently, the space $\Sigma^{N,m}_{{\bf x}^{{\bf P}^*_k}}$ corresponds to all the configurations on the hypercube that have ${{\bf x}^{{\bf P}^*_k}}$  as a direct ancestor. Again, we illustrate this definition in Fig.~\ref{fig: no-mem sketch}. Finally, with this construction, we have at the root (level $k=0$) a number $\Omega({\bf P}_0) =y_0$ of paths ${\bf P}_0$. Each configuration ${\bf x}^{{\bf P}_k}$ is the direct ancestor of $y_{k+1}$ configurations, which means that the number of paths ${\bf P}_{k+1}$ is
\begin{align}
    \Omega({\bf P}_{k+1}) =y_{k+1}\Omega({\bf P}_{k})=\prod_{j=0}^{k+1}y_j\, .
\end{align}
We emphasize these details of the computation, as we believe they will make the following results easier to understand.

As usual with replica techniques, averaging over the patterns' distribution results in a saddle point optimization problem with the overlap matrix ${\bf Q}_{{\bf P}_k,{{\bf P}'_{k'}}}={\rm I\!E}_\xi[{\bf x}^{{\bf P}_k}\cdot {\bf x}^{{\bf P}'_{k'}}/N]$ \cite{mezard1987spin}. After standard computational steps, we obtain in the thermodynamic limit (see App.~\ref{app: General comp} for the detailed computation)
\begin{align}
\label{eq: opt partition func}
  \lim_{N\rightarrow+\infty} \frac{\phi}{N}&=\underset{\underset{\left({\rm s.t.}\,{\bf Q}_{{\bf P}_{k},{\bf P}^*_{k}}=m\,,\, {\bf Q}_{{\bf P}_{k},{\bf P}_{k}}=1\right)}{\hat{\bf Q},{\bf Q}}}{\rm opt}\hspace{-0.5cm}\left\{\frac{\log\tilde{\mathcal{Z}}}{N}\right\}
\end{align}
with
\begin{align}
\label{eq: free energ true}
    \frac{\log\tilde{\mathcal{Z}}}{N}=&-\sum_{k,k'}\sum_{{\bf P}_k,\,{{\bf P}'_{k'}}}\hat{\bf Q}_{{\bf P}_k,{{\bf P}'_{k'}}}{\bf Q}_{{\bf P}_k,{{\bf P}'_{k'}}}+\!\log\!\left\{\prod_{k=0}^{k_f}\prod_{{\bf P}_k} \sum_{{ x}^{{\bf P}_k}=\pm 1}e^{\underset{{k,\,{k'}}}{\sum}\,\underset{{{\bf P}_k,\,{\bf P}'_{k'}}}{\sum} \hat{\bf Q}_{{\bf P}_k,{{\bf P}'_{k'}}}x^{{\bf P}_k}x^{{\bf P}'_{k'}}}\right\}\\
    &+\alpha\log\left\{\frac{1}{\mathcal{N}}\prod_{k=0}^{k_f}\prod_{{\bf P}_k}\int_{-\kappa}^\kappa d{{ w}^{{\bf P}_k}} e^{\frac{-\underset{{k,\,{k'}}}{\sum}\,\underset{{{\bf P}_k,\,{\bf P}'_{k'}}}{\sum}{\bf Q}^{-1}_{{\bf P}_k,{{\bf P}'_{k'}}}w^{{\bf P}_k}w^{{\bf P}'_{k'}}}{2}}\right\} \nonumber
\end{align}
and the normalization
\begin{align}
\label{eq: norm true}
    \mathcal{N}\!\!=\!\prod_{k=0}^{k_f}\prod_{{\bf P}_k} \int\! d{{ w}^{{\bf P}_k}}e^{\frac{-\underset{{k,\,{k'}}}{\sum}\,\underset{{{\bf P}_k,\,{\bf P}'_{k'}}}{\sum}{\bf Q}^{-1}_{{\bf P}_k,{{\bf P}'_{k'}}}w^{{\bf P}_k}w^{{\bf P}'_{k'}}}{2}}\! .
\end{align}
With this notation, ${ w}^{{\bf P}_k}$ corresponds to the margins of ${\bf x}^{{\bf P}_k}$, in other words ${ w}^{{\bf P}_k,\, \mu}=\xi^\mu\cdot {\bf x}^{{\bf P}_k}/\sqrt{N}$. To be part of the SBP solution-manifold, these margins (for a given configuration ${\bf x}^{{\bf P}_k}$) must be smaller than $\kappa$ in absolute value. As is often the case with statistical mechanics approaches, our computation introduces a matrix $\hat{\bf Q}$ that corresponds to a set of magnetic fields imposing the overlaps ${\bf Q}$ between the different configurations.

Finding the optimum of $\tilde{\mathcal{Z}}$ is in general difficult. Thus, we will have to evaluate the free energy using an $Ansatz$ and check to what extent it gives relevant results. The simplification we will take comes from \cite{Barbier2025} and is referred to as the no-memory $Ansatz$. It ascribes each solution to correlate only with its direct ancestor:
\begin{align}
    {\bf Q}^{-1}_{{\bf P}_k,{{\bf P}'_{k'}}}&\neq 0 \quad \mbox{iff}\quad  {\bf P}'_{k'}\in\{{\bf P}_{k},{\bf P}^*_{k}\}\, \,(k'\leq k)\, ,\\
        \hat{\bf Q}_{{\bf P}_k,{{\bf P}'_{k'}}}&\neq 0 \quad \mbox{iff}\quad  {\bf P}'_{k'}\in\{{\bf P}_{k},{\bf P}^*_{k}\}\, \,(k'\leq k)\, ,
\end{align}
with the condition that
\begin{align}
\label{eq: cnstr 1}
    {\bf Q}_{{\bf P}_k,{{\bf P}_{k}}}&={\rm I\!E}_\xi\left[{\bf x}^{{\bf P}_k}\cdot {\bf x}^{{\bf P}_{k}}/N\right]=1\, ,\\
\label{eq: cnstr 2}
    {\bf Q}_{{\bf P}_k,{{\bf P}^*_{k}}}&={\rm I\!E}_\xi\left[{\bf x}^{{\bf P}_k}\cdot {\bf x}^{{\bf P}^*_{k}}/N\right]=m\,.
\end{align}
We recall that the inverse of the overlap matrix, i.e. ${\bf Q}^{-1}$, quantifies the correlations between the connected minima. We chose specifically this $Ansatz$ because it is the minimal form of ${\bf Q}$ and $\hat{\bf Q}$ that still ensures the connected construction with ${\bf x}^{{\bf P}_k}\cdot {\bf x}^{{\bf P}^*_k}/N=m$. It also has the advantage of being analytically tractable, as the computation of the free energy then becomes a summation over a Bethe tree. 

More practically, given the constraint equations (\ref{eq: cnstr 1}, \ref{eq: cnstr 2}) and our $Ansatz$, all entries of ${\bf Q}^{-1}$ become fixed. In other words, with the no-memory $Ansatz$, the optimization for estimating the connected free energy only involves $\hat{\bf Q}$. It also turns out that this form of correlation matrix corresponds to an ultrametric geometry. In short, if two configurations ${\bf x}_{k}^{{\bf P}_{k}}$ and ${\bf x}_{k'}^{{\bf P}'_{k'}}$ have their closest common ancestor at level $k''$ ($\leq k,k'$), their overlap is given by
\begin{align}
\label{eq: mag no-mem}
\hspace{-0.25cm}
     {\bf Q}_{{\bf P}_k,{{\bf P}'_{k'}}}=m^{\vert k''-k\vert+\vert k''-k'\vert}\, .
\end{align}
Similarly, along the same path of minima (${\bf P}_k$ contains ${\bf P}'_{k'}$ or vice-versa), we have
\begin{align}
\label{eq: mag no-mem path}
{\bf Q}_{{\bf P}_k,{{\bf P}'_{k'}}}=m^{\vert k-k'\vert}\,.
\end{align}
In Fig.\ref{fig: no-mem sketch} the interested reader can find an illustration of the no-memory $Ansatz$.

With the no-memory geometry, the nested sums and integrals in Eqs.~(\ref{eq: free energ true}, \ref{eq: norm true}) take the form of a Bethe tree partition function. With this, we can integrate all configurations ${\bf x}^{{\bf P}_{k(\neq 0)}}$, optimize over $\hat{\bf Q}$ and obtain an effective free energy for ${\bf x}^{{\bf P}_{0}}$ (see App.~\ref{app: no-mem computational steps} for more details):
\begin{align}
   \lim_{N\rightarrow+\infty}  \frac{\phi}{N}&=\log2+\tilde{\phi}^{\rm ent.}(m)\sum_{k=1}^{k_f}\left(\prod_{k'=1}^{k}y_{k'}\right)+\!\alpha\log\!\left\{\!\int_{-\kappa}^\kappa \!dw_0\,\frac{e^{\frac{-w_0^2}{2}}}{\sqrt{2\pi}}\,G_0^{\rm energ.}\left[w_0,m,\{y_l\}_{l\in[\![1,k_f]\!]}\right]\!\right\}
\end{align}
with the binary entropy
\begin{align}
\label{eq: ent binary}
    \tilde{\phi}^{\rm ent.}(m)=&-\sum_{X=\pm1}\frac{1+mX}{2}\log\left(\frac{1+mX}{2}\right)
\end{align}
and the iterative construction ($G_{k_f}^{\rm energ.}[\cdot]=1$)
\begin{align}
    \label{eq: iteration energ no-mem}
    G_{k}^{\rm energ.}\left[w_k,m,\{y_l\}_{l\in[\![k+1,k_f]\!]}\right]=\Bigg\{\int_{-\kappa}^\kappa dw_{k+1}\, \frac{e^{\frac{-(w_{k+1}-mw_{k})^2}{2(1-m^2)}}}{\sqrt{2\pi(1-m^2)}} G_{k+1}^{\rm energ.}\left[w_{k+1},m,\{y_l\}_{l\in[\![k+2,k_f]\!]}\right]\Bigg\}^{y_{k+1}} .
\end{align}
Let us take a moment to interpret the free energy of the no-memory connected minima. First, we can observe a shift in the SBP loss for ${\bf x}_0$. In the original problem, the loss simply ensures that ${\bf x}^{{\bf P}_0}$ satisfies the constraints. Now, for ${\bf x}^{{\bf P}_0}$ to be both a solution and connected, we should apply the loss
\begin{align}
\label{eq: loss eff}
        e^{-\mathcal{L}^{\rm eff.\, k}_{\rm SBP}\left({\bf x}^{{\bf P}_0}\right)}&= \prod_{\mu =1}^M\Theta\left(\kappa-\left\vert\frac{\xi^\mu\cdot {\bf x}^{{\bf P}_0}}{\sqrt{N}}\right\vert\right) G_{0}^{\rm energ.}\left[\frac{\xi^\mu\cdot {\bf x}^{{\bf P}_0}}{\sqrt{N}},m,\{y_{l}\}_{l\in [\![1,k_f]\!]}\right] .
\end{align}

As we shall see in Sec.~\ref{sec: numerics}, $G_{0}[\cdot,\cdot,\cdot]$ has the effect of compressing the margins closer to zero. Thus, to target non-isolated configurations, we must select minima that do not accumulate margin values close to the thresholds $\kappa$ and $-\kappa$. These accessible solutions are more robust than the typical ones. In the same vein, our computation allows us to derive the distribution of margins for these configurations:
\begin{align}
    P^{{\bf x}_0}(w)={\rm I\!E}_{\xi,{\bf x}_0}\left[\frac{1}{M}\sum_{\mu=1}^M\delta\left(w-\frac{\xi^\mu\cdot{\bf x}_0}{\sqrt{N}}\right)\right]= \!\frac{{e^{\frac{-w^2}{2}}}G_0^{\rm energ.}\left[w,m,\{y_l\}_{l\in[\![1,k_f]\!]}\right] }{\int_{-\kappa}^\kappa \!d\underline{w} {e^{\frac{-\underline{w}^2}{2}}}G_0^{\rm energ.}\!\left[\underline{w},m,\{y_l\}_{l\in[\![1,k_f]\!]}\right] }
\end{align}
with $w_0\in[-\kappa, \kappa]$. As a point of comparison, we recall that the margin distribution for the typical solutions of the SBP is 
\begin{align}
    \label{eq: typical distrib}
     P^{\rm typical}(w)= \!\frac{{e^{\frac{-w^2}{2}}}\Theta(\kappa-\vert w\vert) }{\int_{-\kappa}^\kappa \!d\underline{w} {e^{\frac{-\underline{w}^2}{2}}} }\,.
\end{align}

In order to keep a relatively concise analysis, we will focus in the next section on a special case of connected minima. Indeed, we will set $y_k=1$ (for all $k\in[\![1,k_f]\!]$)
and show that there exists at least one fully-delocalized cluster -i.e.
 ${\bf x}^{{\bf P}_0}\cdot {\bf x}^{{\bf P}_{k_f}}/N = 0$- formed by non-equivalent minima. 

\subsection{Star-shaped delocalized cluster(s), local entropy and edge entropy}
\label{sec: deloc cluster}
\subsubsection{Identification of a delocalized cluster}
Looking more closely at the iteration  from Eq.~(\ref{eq: iteration energ no-mem}), we see that it admits a fixed point if we set $y_k=1$ ($\forall k\in[\![1,k_f]\!]$). Indeed, in this special case the iteration becomes a linear problem with associated eigenvectors ${G}_\lambda^{\rm energ.}$ and eigenvalues $\lambda$:
\begin{align}
    \label{eq: iteration energ no-mem y=1}
    \lambda G_\lambda^{\rm energ.}\left[w_k,m\right]&=\int_{-\kappa}^\kappa dw_{k+1}\, \frac{e^{\frac{-(w_{k+1}-mw_{k})^2}{2(1-m^2)}}}{\sqrt{2\pi(1-m^2)}} G_{\lambda}^{\rm energ.}\left[w_{k+1},m\right]\\
    &\hspace{-1.5cm}=\int_{-\kappa}^\kappa dw_{k+1}\, \frac{e^{\frac{-(w_{k+1}-mw_{k})^2}{2(1-m^2)}}}{\sqrt{2\pi(1-m^2)}} \left[1+(w_{k+1}-w_k)\partial_w+\frac{(w_{k+1}-w_k)^2}{2}\partial^2_w\right]G_{\lambda}^{\rm energ.}\left[w=w_{k+1},m\right]\nonumber\\
    &\hspace{-1.cm}+o(1-m)\nonumber\\
    &\hspace{-2.4cm}\underset{\kappa-\vert w_k\vert\gg\sqrt{1-m}}{=}G_{\lambda}^{\rm energ.}\left[w_{k},m\right]+(1-m)\left\{-w_k\partial_wG_{\lambda}^{\rm energ.}\left[w_{k},m\right]+\partial^2_wG_{\lambda}^{\rm energ.}\left[w_{k},m\right]\right\}+o(1-m)\nonumber
\end{align}
The Perron-Frobenius theorem \cite{book_Potters} specifies that the equation above has a non-degenerate top eigenvector/eigenvalue (labeled $\lambda^{\rm top}$). For our iteration to converge to this top eigenvector, we need the total number of iterations $k_f$ to diverge. As we can see with the second line of Eq.~\ref{eq: iteration energ no-mem y=1}, the top eigenvalues typically scale as 
\begin{align}
    \lambda\sim 1-\tilde{\lambda}(1-m)\,.
\end{align}
Therefore, $G_{k}^{\rm energ.}[\cdot,\cdot,\cdot]$ is projected onto each eigenvector $G_{\lambda}^{\rm energ.}[\cdot,\cdot]$ at a rate $e^{-(k_f-k)(1-m)(\tilde{\lambda}^{\rm top}-\tilde{\lambda})}$. To keep only the largest one for $G_0^{\rm energ.}$, we need to set $k_f=\tilde{k}_f/(1-m)$ and take the limit $\tilde{k}_f\rightarrow+\infty$ as we send $m$ to one. When we consider this regime, the free energy further simplifies to
\begin{align}
    \label{eq: no-mem pot final ver.}
    \lim_{\underset{{\rm s.t.} \tilde{k}_f=k_f(1-m)}{m\rightarrow1,\, \tilde{k}_f\rightarrow+\infty}}\,\lim_{N\rightarrow+\infty}\frac{\phi}{N}=&\log(2)+k_f\left[\tilde{\phi}^{\rm ent.}(m)+\alpha\log(\lambda^{\rm top})\right]\\
    & +\alpha\log\left\{\int_{-\kappa}^\kappa dw_0\,\frac{e^{-\frac{w_0^2}{2}}}{\sqrt{2\pi}}\,G_{\lambda^{\rm top}}^{\rm energ.}\left[w_0,m\right]\right\}\,.\nonumber
\end{align}

With this limit, connected minima become delocalized. Indeed, the overlap between levels $k=0$ and $k_f$ along the same path goes to zero:
\begin{align}
\frac{{\bf x}^{{\bf P}_0}\cdot {\bf x}^{{\bf P}_{k_f}}}{N} = m^{k_f}=e^{\frac{\tilde{k}_f\ln(m)}{1-m}}\underset{m\rightarrow1}{=}e^{-\tilde{k}_f}\underset{\underset{\tilde{k}_f\rightarrow+\infty }{
     m\rightarrow1}  }{=}0\, .
\end{align}
Therefore, we are now characterizing an accessible region that spans the entire phase space.

\begin{figure}
    \centering
    \includegraphics[width=0.6\linewidth]{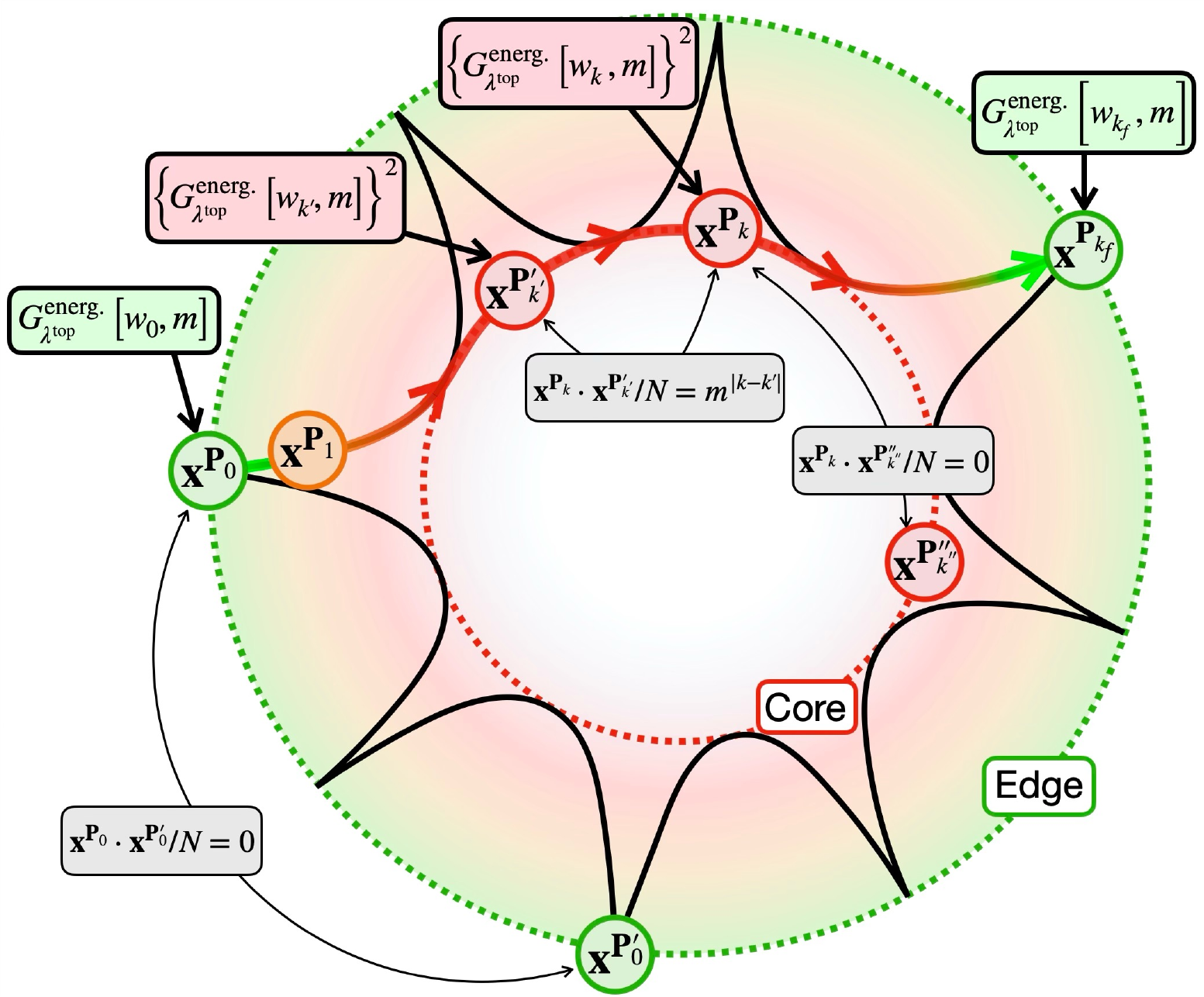}
    \caption{Drawing representing a star-shaped cluster of the connected minima. To obtain this structure within our formalism, we applied two successive simplifications: first, the no-memory $Ansatz$, and second, the chain arrangement ($y_k=1$ for all $k\in[\![1,k_f]\!]$). This cluster can be divided into two main components: an edge and a core. The typical minima are contained in the edge, they are the least robust ones in the cluster. To pass from one to the other, our statistical measure goes through the core of the cluster -where the minima are the most robust-.}
    \label{fig: star-shape}
\end{figure}

More interestingly, having set all $y$'s to one, the connected minima are arranged into a chain. Concretely,  ${\bf x}^{{\bf P}_k}$ has exactly one direct ancestor (at level $k-1$) and also one direct descendant (at level $k+1$) -for every level $k(\neq0,k_f)$-. At level $k=0$, ${\bf x}^{{\bf P}_0}$ has one direct descendant but no ancestor, whereas at level $k = k_f$, ${\bf x}^{{\bf P}_{k_f}}$ has one direct ancestor but no descendant. Because of this geometry and because the overlap matrix is translational invariant -see Eq.~(\ref{eq: mag no-mem path})-, the system becomes invariant by mirror symmetry: ${\bf x}^{{\bf P}_k}\rightarrow {\bf x}^{{\bf P}_{k_f-k}}$. This causes the two edge configurations, ${\bf x}^{{\bf P}_0}$ and ${\bf x}^{{\bf P}_{k_f}}$, to share the same effective loss
\begin{align}
\label{eq: eff loss deloc}
        e^{-\mathcal{L}^{\rm eff.}_{\rm SBP}\left({\bf x}\right)}&=\! \prod_{\mu =1}^M\!\Theta\!\left(\!\kappa-\left\vert\frac{\xi^\mu\!\cdot\! {\bf x}}{\sqrt{N}}\right\vert\right) G_{\lambda^{\rm top}}^{\rm energ.}\!\left[\frac{\xi^\mu\!\cdot\! {\bf x}}{\sqrt{N}},m\right]
\end{align}
and margins distribution
\begin{align}
        P^{{\rm edge}}(w)=P^{{\bf x}_0}(w)= \frac{e^{\frac{-w^2}{2}}\Theta\left(\kappa-\vert w\vert\right)\,G_{\lambda^{\rm top}}^{\rm energ.}\left[w,m\right]}{\int_{-\kappa}^\kappa d\underline{w}\, e^{\frac{-\underline{w}^2}{2}}\,G_{\lambda^{\rm top}}^{\rm energ.}\left[\underline{w},m\right]}\, .
\end{align}
Deep in the core of the chain (for $1\ll k\ll k_f$ ), the minima ${\bf x}^{{\bf P}_k}$ see the effect of connectivity twice: from its ancestors and its descendants. Reproducing the computation of the  no-memory free energy (detailed in App.\ref{app: no-mem computational steps}), this time integrating all configurations but ${\bf x}^{{\bf P}_k}$, we can derive the effective loss in the core
\begin{align}
\label{eq: eff loss deloc2}
        &e^{-\mathcal{L}^{\rm eff. 2}_{\rm SBP}\left({\bf x}\right)}=\prod_{\mu =1}^M\!\Theta\!\left(\!\kappa\!-\!\left\vert\frac{\xi^\mu\!\cdot\! {\bf x}}{\sqrt{N}}\right\vert\right) \!\left\{G_{\lambda^{\rm top}}^{\rm energ.}\!\left[\frac{\xi^\mu\!\cdot\! {\bf x}}{\sqrt{N}},m\right]\right\}^2
\end{align}
and margins distribution
\begin{align}
        P^{{\rm core}}(w)=\left.{\rm I\!E}_{\xi,{\bf x}_k}\left[\frac{1}{M}\sum_{\mu=1}^M\delta\left(w-\frac{\xi^\mu\cdot{\bf x}_k}{\sqrt{N}}\right)\right]\right\vert_{1\ll k\ll k_f}= \frac{e^{\frac{-w^2}{2}}\Theta\left(\kappa-\vert w\vert\right)\left\{G_{\lambda^{\rm top}}^{\rm energ.}\left[w,m\right]\right\}^2}{\int_{-\kappa}^\kappa d\underline{w}\, e^{\frac{-\underline{w}^2}{2}}\,\left\{G_{\lambda^{\rm top}}^{\rm energ.}\left[\underline{w},m\right]\right\}^2}\, .
\end{align}
Numerically, we can observe that $\left\{G_{\lambda^{\rm top}}^{\rm energ.}\left[\cdot,\cdot\right]\right\}^2$ compresses the margins closer to zero. This means that the configurations in the core of the chain are more robust than those at the edge. We recover here the star-shaped geometry observed for the first time in the negative margin perceptron \cite{Annesi2023}. Concretely, having fixed all $y$'s to one, the typical connected minima ${\bf x}^{{\bf P}_0}$ (and ${\bf x}^{{\bf P}_{k_f}}$ by symmetry) lie at the edge of a delocalized cluster. These configurations can be reached from one another by passing through a core of highly robust minima. We illustrated this star-shaped geometry in Fig.\ref{fig: star-shape}.

\subsubsection{Local entropy and edge entropy}
To characterize this delocalized cluster even more precisely, we can note that the free energy $\phi$ actually hides two crucial quantities. To unveil them, let us rewrite the potential as 
\begin{align}
    \phi=&\phi-y_1\partial_{y_1}\phi+\sum_{k=1}^{k_f}\left(y_k\partial_{y_k}-y_{k+1}\partial_{y_{k+1}}\right)\phi=s_{{\bf x}_0}+\sum_{k=1}^{k_f}\left[\left(\prod_{j=1}^{k}y_j\right)s_{\rm loc}(k)\right]\,.
\end{align}
These two quantities are the local entropy ($k\geq1$)
\begin{align}
    s_{\rm loc}(k)&={\rm I\!E}_{\xi,{\bf x}_{k-1}}\left[
    \log\Bigg[\sum_{{\bf x}_k\in\Sigma^{N,m}_{{\bf x}_{k-1}}}e^{\mathcal{L}_{\rm SBP}({\bf x}_k)}e^{y_{k+1}\{\phi_{k+1}({{\bf x}_k,m})-{\rm I\!E}_{\xi,{\bf x}_{k}}[\phi_{k+1}({{\bf x}_k,m})]\}}\Bigg]\right]\\
    &={\rm I\!E}_{\xi,{\bf x}_{k-1}}[\phi_{k}({{\bf x}_{k-1},m})]-y_{k+1}{\rm I\!E}_{\xi,{\bf x}_{k}}[\phi_{k+1}({{\bf x}_{k},m})]\nonumber\\
    &=\frac{\left(y_k\partial_{y_k}-y_{k+1}\partial_{y_{k+1}}\right)\phi}{\prod_{j=1}^k y_j}\nonumber
\end{align}
and the edge entropy
\begin{align}
    s_{{\bf x}_0}&\!=\!{\rm I\!E}_{\xi}\left\{
    \!\log\left[\sum_{{{\bf x}_0\in\Sigma^N}}\!e^{\mathcal{L}_{\rm SBP}({\bf x}_0)+y_{1}[\phi_{1}({{\bf x}_0,m})-\langle\phi_{1}({{\bf x}_0,m})\rangle]}\!\right]\!\right\}\\
    &=\phi-y_{1}{\rm I\!E}_{\xi,{\bf x}_0}[\phi_{1}({{\bf x}_{0},m})]\nonumber\\
    &=\phi-y_{1}\partial_{y_{1}}\phi\, \nonumber
\end{align}
where we recall that the expectation over the connected minima is defined as
\begin{align}
\label{eq: connected measure 2}
    {\rm I\!E}_{\xi,{\bf x}_k}\left[f(\{\xi^\mu\},{\bf x}_k)\right]={\rm I\!E}_{\xi}\left[\prod_{j=0}^{k_f}\left(\frac{\underset{{{\bf x}_{j}\in\Sigma^{N,m}_{{\bf x}_{j-1}}}}{\sum}\!\!e^{\mathcal{L}_{\rm SBP}({\bf x}_{j})+y_{j+1}\phi_{j+1}({{\bf x}_{j},m})}}{\underset{{\underline{\bf x}_{j}\in\Sigma^{N,m}_{{\bf x}_{j-1}}}}{\sum}\!\!e^{\mathcal{L}_{\rm SBP}(\underline{\bf x}_{j})+y_{j+1}\phi_{j+1}({\underline{\bf x}_{j},m})}}\right)f(\{\xi^\mu\},{\bf x}_k)\!\right] .
\end{align}

On the one hand, the local entropy $s_{\rm loc}(k)$ quantifies how many solutions are gained when the size of the cluster is increased from ${k}$ to ${k}+1$. Put differently, it counts how many configurations one has (when sitting on a minima in layer $k-1$) to move to layer $k$.  On the other hand, the edge entropy counts how many centers $\bf{x}^{{\bf P}_0}$ can be taken to start a connected chain. To see this more clearly, we can simply note that these quantities are derived in the same way as any entropy function. In a general setting, we recall that the entropy of an equilibrated system is given by $s=\phi-\beta\partial_{\beta}\phi$. In our case, we just generalized this approach by computing the potential on a layer $k$ and subtracting the effective potential imposed by the system on this layer, i.e. $s\propto y_{k}\partial_{y_{k}}\phi-y_{k+1}\partial_{y_{k+1}}\phi$. A priori, the local entropy $s_{\rm loc}(\cdot)$ is a function of $k$, i.e. it depends on which layer it is estimated. However, taking $y_k=1$ for all $k\in[\![1,k_f]\!]$ and supposing $G^{\rm energ.}_{k}=G^{\rm energ}_{\lambda^{\rm top}}$, the local entropy becomes constant in the core. This simply shows that in this region all minima are equivalent, they share the same effective loss and the same margin distribution. After a few computation steps (see App.~\ref{app: entropy loc}) we have 
\begin{align}
\label{eq: loc entro simple}
    s_{\rm loc}(k)= N\left[\tilde{\phi}^{\rm ent.}(m)+\alpha\log(\lambda)\right]\,\,(1\ll k\ll k_f ).
\end{align}
Importantly, this computation implicitly relies on the same chain formalism introduced in \cite{Barbier2025}. This is evident from the structure of the expectation $\langle\cdot\rangle_k$ -see Eq.~(\ref{eq: connected measure 2})-. Decomposing the product in this expression, we first generate a sequential probability distribution that constructs a chain of minima $\{{\bf x}_0,\dots,{\bf x}_{k-1}\}$ (with ${\bf x}_j\cdot {\bf x}_{j+1}/N=m$). We then evaluate an observable that depends on ${\bf x}_{k}$ in the neighborhood of ${\bf x}_{k-1}$  (${\bf x}_{k-1}\cdot {\bf x}_{k}/N=m$). The only difference from the setup in \cite{Barbier2025} is that, in our framework, the chain of minima is built with an additional entropy bias $\{\phi_j({\bf x}_j,m)\}_{j\in[\![0,\cdots,k_f]\!]}$, whereas in \cite{Barbier2025} the chain relies solely on the original SBP loss $\mathcal{L}_{\rm SBP}(\cdot)$. As a consequence, the chain without the entropy bias have been shown to exist only in a high-threshold regime $\kappa \sim \sqrt{\log N}$.

Finally, the edge entropy is in this exact case (see details in App.~\ref{app: entropy loc})
\begin{align}
    s_{{\bf x}_0}=&N\log(2)+M\log\left\{\int dw_0\,\frac{e^{\frac{-w_0^2}{2}}\Theta(\kappa-\vert w_0\vert)}{\sqrt{2\pi}}\,G_{\lambda^{\rm top}}^{\rm energ.}\left[w_0,m\right]\right\}\\
     & -M\int dw_{0}\,P^{{\rm edge}}(w_{0})\log \left\{ G_{\lambda^{\rm top}}^{\rm energ.}\left[w_{0},m\right]\right\}\nonumber
\end{align}
with the edge distribution
\begin{align}
\label{eq: distrib no-mem cluster}
    P^{{\rm edge}}(w)=\frac{e^{\frac{-w^2}{2}}\,G_{\lambda^{\rm top}}^{\rm energ.}\left[w,m\right]}{\int_{-\kappa}^\kappa d\underline{w}\, e^{\frac{-\underline{w}^2}{2}}\,G_{\lambda^{\rm top}}^{\rm energ.}\left[\underline{w},m\right]}\,.
\end{align}
Again, we can recognize a very familiar entropy. The function $s_{{\bf x}_0}$ is simply the $annealed$ entropy for a binary system with the effective loss $\mathcal{L}_{\rm SBP}^{\rm eff.}(\cdot)$, and with an effective temperature $\beta=1$. This entropy counts the number of minima located at the edge of the cluster. Since these minima dominate the measure defined in Eq.~(\ref{eq: connected measure}), implying that the core of the cluster contains exponentially fewer minima, $s_{{\bf x}_0}$ effectively counts the total number of minima in the entire cluster (including those in the core).

Before moving on to other physics quantities, we can highlight that the delocalized-cluster manifold disappears if either the edge or the local entropy becomes negative. In the first case, no edge ${\bf x}^{{\bf P}_0}$ can be taken to start a chain, regardless of its connectivity properties. In the second, the edge configuration ${\bf x}^{{\bf P}_0}$ may exist (it still requires $s_{{\bf x}_0}>0$), but the rest of the connected region cannot be built. For example, if $s_{\rm loc}(k)$ is negative, the move ${\bf x}^{{\bf P}^*_k}\rightarrow {\bf x}^{{\bf P}_k}$ does not exists.
Taking the limit $m\rightarrow1$, we obtain that $s_{\rm loc}(k)$ is always positive in the core of cluster. In fact, the entropic contribution exceeds the energetic one as we have $\tilde{\phi}^{\rm ent.}(m)\sim (1-m)\log(1-m)$ and $1-\lambda\sim 1-m$. The second scaling can be obtained easily with Eq.~(\ref{eq: iteration energ no-mem y=1}), by performing the change of variable $w_k=mw_{k-1}+\sqrt{1-m^2}\,u$ and Taylor expanding $G_\lambda^{\rm energ.}$ (around $\sqrt{1-m^2}\,u=0$).
Therefore, at this level of our analysis, it appears that this delocalized manifold disappears when the edge entropy  $s_{{\bf x}_0}$ becomes negative. We will label the threshold $\kappa$ for which this occurs as $\kappa^{\rm no-mem}_{\rm existence}$ (with fixed $\alpha$).

\subsection{Stability of the no-memory $Ansatz$ and generalization of the Franz-Parisi potential}
\label{sec: stab}
Having fixed the memory kernel for the connected minima (with the no-memory $Ansatz$), most of the optimization problem in Eq.~(\ref{eq: opt partition func}) has been eliminated. Therefore, it is not guaranteed that what we have described so far corresponds to a true saddle point of the free energy. We must check for instabilities to determine the extent to which our simplification is acceptable.
In this section, we will first check if the no-memory $Ansatz$ is stable under a global perturbation. In other words, we will compute the quantity
\begin{align}
    \frac{\delta\phi}{\delta {\bf Q}_{{\bf P}_{k},{{\bf P}'_{k'}}}}\, , \quad\forall\, {\bf P}_{k},{\bf P}'_{k'}\,.
\end{align}
If this quantity is null for all sets $\{{\bf P}_{k},{\bf P}'_{k'}\}$, then the no-memory $Ansatz$ is a saddle-point of the free energy.  In fact, with a few computation steps -see the the details computation in App.~\ref{app: annealed stab Ansatz}, we obtain (for all values of $\kappa$ and $\alpha$)
\begin{align}
        \frac{\delta\phi}{\delta {\bf Q}_{{\bf P}_{k},{{\bf P}'_{k'}}}}\!\!>0\, ,\,\, \forall\, {\bf P}_{k},\!{\bf P}'_{k'}\,\,({\rm with}\,\,{\bf P}_{k}\!\neq\!{\bf P}^*_{k}\,\,{\rm and}\,\, k'\!\!<\!k)\,.
\end{align}
In other words, the measure for connected states -see $\mathcal{Z}$ in Eq.~(\ref{eq: opt partition func})- is never dominated by the no-memory geometry. We have now two possibilities with this instability. Either the no-memory manifold simply characterizes a subdominant number of connected minima in this problem, or it represents a non-physical geometry for the solutions manifold. The first case would give algorithmically accessible solutions, while the second case would not prescribe anything regarding minima in the SBP. 

Taking a step back, we can try to answer this question by checking whether a local algorithm could probe the connected no-memory states. We can imagine the best-case scenario: an algorithm is directly initialized inside the no-memory cluster and tries to explore its constitutive states. The trajectory of this algorithm will form a chain of explored configurations ${{\bf x}k}{k\in[![0,t]!]}$; the question is then whether this chain follows a no-memory correlation profile. Contrary to the chains we studied before, the one formed by the algorithm trajectory is built sequentially. This means that ${\bf x}_1$ is obtained after considering the initialization ${\bf x}_0$ as a frozen variable, then ${\bf x}_2$ is obtained after considering both ${\bf x}_1$ and ${\bf x}_0$ as frozen variables, and so on.
Luckily, we already have at our disposal an observable that quantifies such moves: the local entropy $s_{\rm loc}(k)$, with $k\in [![1,k_f]!]$. We recall that the local entropy counts the number of available configurations when performing the move ${\bf x}^{{\bf P}^*_k}\rightarrow {\bf x}^{{\bf P}_k}$, while keeping ${\bf x}^{{\bf P}^*_k}$ fixed, in the connected manifold.

If we perturb this quantity, we can check whether the move from layer $k-1$ to $k$ is indeed dominated by configurations with a no-memory correlation profile, or if we should be entropically attracted by minima with another correlation structure. In other words, we can check if (when sitting on ${\bf x}^{{\bf P}^*_k}$) the next minima ${\bf x}^{{\bf P}_k}$ follow with high probability a no-memory correlation profile or not. If we observe such local instabilities, the chain formed by our algorithm trajectory will also deviate from a no-memory correlation profile: the algorithm will be at some point entropically attracted by other minima. We can consider in this case that our $Ansatz$ is simply non-physical. To test this, we will compute the sign of ($\forall\, k'\in[\![0,k-1]\!]$)
\begin{align}
    \frac{\delta s_{\rm loc}(k)}{\delta {\bf Q}_{{\bf P}_{k},{{\bf P}'_{k'}}}}=\frac{\delta}{{\delta {\bf Q}_{{\bf P}_{k},{{\bf P}'_{k'}}}}}\left[\frac{\left(y_k\partial_{y_k}-y_{k+1}\partial_{y_{k+1}}\right)\phi}{\prod_{j=1}^k y_j}\right]\,,
\end{align}
with ${\bf P}_{k}$ containing ${\bf P}'_{k'}$, i.e. ${\bf x}^{{\bf P}'_{k'}}$ is an ancestor of ${\bf x}^{{\bf P}_{k}}$. If this quantity remains negative for all values of $k$ and $k'$ (with fixed $\alpha$ and $\kappa$), then our paths geometry is stable: we are not entropically attracted by regions where minima start correlating. Conversely, if it becomes positive for a given value of $k$ and $k'$, no-memory paths become locally destabilized.
To remain concise, we will write this quantity as $\delta s_{\rm loc}(k,k')$ in the following:
\begin{align}
    \delta s_{\rm loc}(k,k')=\frac{\delta s_{\rm loc}(k)}{\delta {\bf Q}_{{\bf P}_{k},{{\bf P}'_{k'}}}}\,.
\end{align}

\begin{figure}
    \centering
    \includegraphics[width=0.6\linewidth]{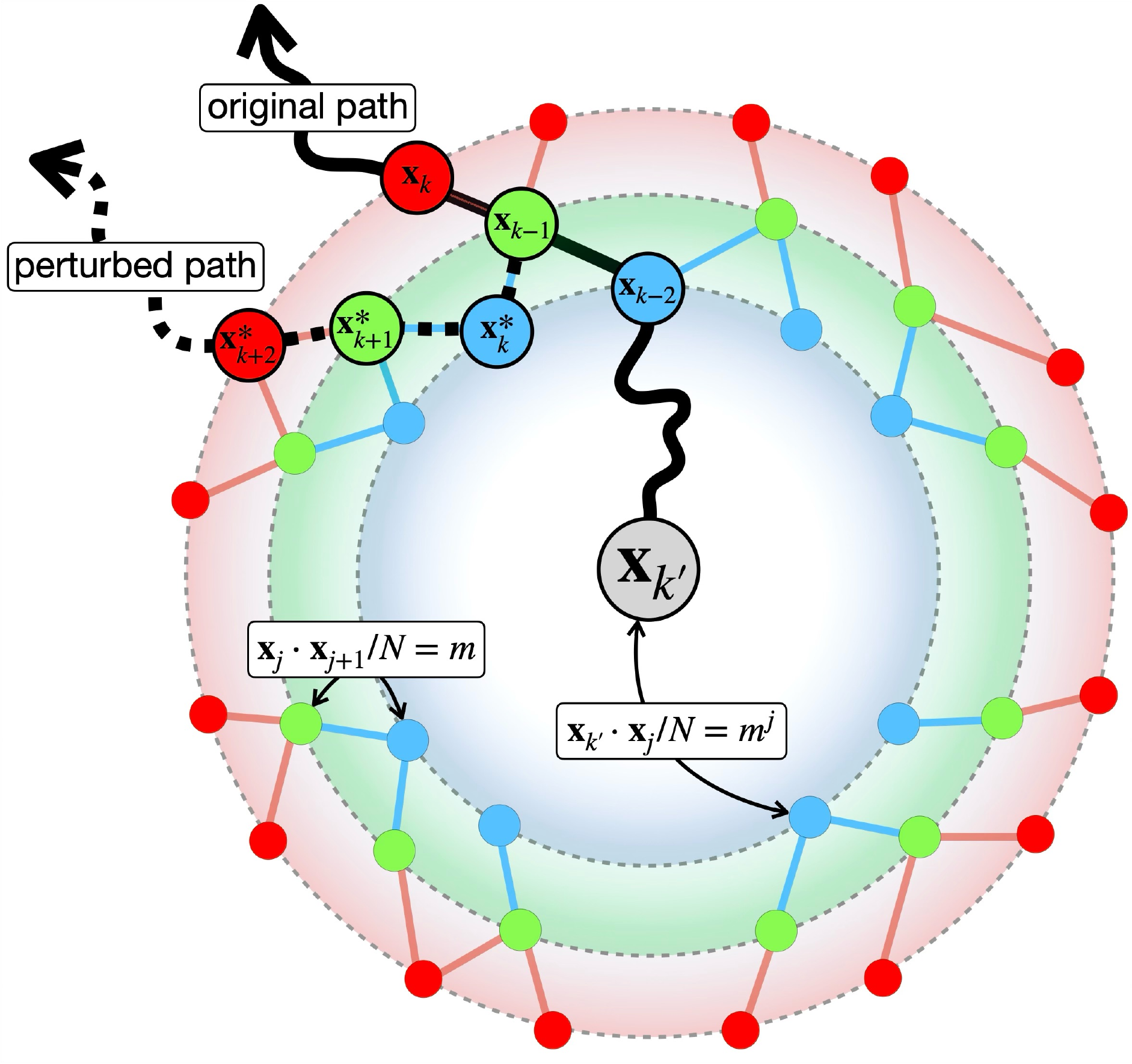}
    \caption{Representation of the perturbation in the no-memory cluster(s). While along a no-memory path the overlap is ${\bf x}^{{\bf P}_{k}}\cdot{\bf x}^{{\bf P}'_{k'}}/N=m^{\vert k-k'\vert}$, we perturb the cluster by recorrelating two configurations ${\bf x}^{{\bf P}_{k}}$ and ${\bf x}^{{\bf P}'_{k'}}$: ${\bf x}^{{\bf P}_{k}}\cdot{\bf x}^{{\bf P}'_{k'}}=m^{\vert k-k'\vert-2}$. In App.\ref{app: simple model FP-gen}, we show that the stability of the local entropy $s_{\rm loc}(\cdot)$ -with this perturbation- can be mapped to a well-known Franz-Parisi potential computation.}
    \label{fig: stability sketch}
\end{figure}

At the perturbative level,  we will have to compute the potential shift when we change the overlap ${\bf Q}_{{\bf P}_{k},{{\bf P}'_{k'}}}$ along a chain of minima as
\begin{align}
    &{\bf Q}_{{\bf P}_{k},{{\bf P}'_{k'}}}=m^{\vert k-k'\vert}
    \rightarrow {\bf Q}_{{\bf P}_{k},{{\bf P}'_{k'}}}=m^{\vert k-k'\vert-2}\, .
\end{align}
As pictured in Fig.~\ref{fig: stability sketch}, this perturbation corresponds to have a slight re-correlation of ${\bf x}^{{\bf P}_{k}}$  with ${\bf x}^{{\bf P}'_{k'}}$. More explicitly, instead of having ${\bf x}^{{\bf P}_{k}}\cdot {\bf x}^{{\bf P}'_{k'}}/N=m^{\vert k-k' \vert }$ , we recall that ${\bf x}^{{\bf P}'_{k'}}$ is an ancestor of ${\bf x}^{{\bf P}_{k}}$, we set ${\bf x}^{{\bf P}_{k}}\cdot {\bf x}^{{\bf P}'_{k'}}/N=m^{\vert k-k' \vert -2}$. As we will mention later -and detail in App.~\ref{app: simple model FP-gen}-, this form for the perturbation facilitates correspondence with other physical objects (for models simpler than SBP).

For the more expert readers, this stability criterion relates to the well-known Franz-Parisi potential \cite{franz1995recipes}. In fact, for problems with a simpler energy landscape (simpler in terms of connectivity property), it can be shown that the clustering transition around a planted configuration ${\bf x}_0$ (captured by a change of curvature in the so-called Franz-Parisi potential) corresponds to a change of sign in ${\delta s_{\rm loc}(k,k')}$. In other words, our stability criterion for connected paths captures already-known clustering transitions (in simple models). As an example of this correspondence, we detail in App.~\ref{app: simple model FP-gen} how the clustering transition of a simple binary model is captured by both the Franz-Parisi potential and our stability criterion. For the case of the SBP, the Franz-Parisi potential and our stability computation make two distinct predictions. We compiled the detailed computation for both quantities (${\delta s_{\rm loc}}$ and the Franz-Parisi potential) in App.~\ref{app: entropy loc} and App.~\ref{app: FP connected minima} respectively. 
 
 In the core of the cluster ($1\ll k\ll k_f$ and $k/k'\sim \mathcal{O}(1)$), the stability criterion becomes translationally invariant.  In other words, the value of the perturbation depends only on $\vert k-k'\vert$ (and not explicitly on $k$ and $k'$). This was actually expected, all connected minima in the core of the manifold are equivalent. As mentioned above, our stability criterion can be compared to more standard statistical mechanics approaches; in particular, to the Franz-Parisi potential. Interestingly, it appears that the Franz-Parisi potential overestimates the clustering transition. We have a range of parameters $\{\alpha,\kappa\}$ for which the core of the connected manifold is locally unstable -$\delta s_{\rm loc}(k,k')<0$-, while the Franz-Parisi computation does not capture any clustering phenomenon for the no-memory minima. This suggests that, for the SBP, our approach can reveal local geometric instabilities that conventional statistical-mechanics methods fail to capture. The reason for this appears to be that the local geometry -probed here with $s_{\rm loc}(k,k')$- is in part governed by the behavior of margins close to the threshold -$\vert w\vert \approx \kappa$-. In usual cases, this contribution is negligible and the stability of $s_{\rm loc}(\cdot)$ can be mapped to a usual Franz-Parisi computation  -see App. \ref{app: simple model FP-gen}-. However, for the SBP, the margins cost function $\log(G^{\rm enrg.}_{\lambda^{\rm top}}[w,m])$ diverges at these edges, making this contribution non-negligible. We will later provide numerical evidence that, when the core of the connected cluster is destabilized, it becomes difficult for an algorithm to reach the typical minima of the no-memory connected cluster (i.e. its edge).

\begin{figure}
    \centering
    \includegraphics[width=0.7\linewidth]{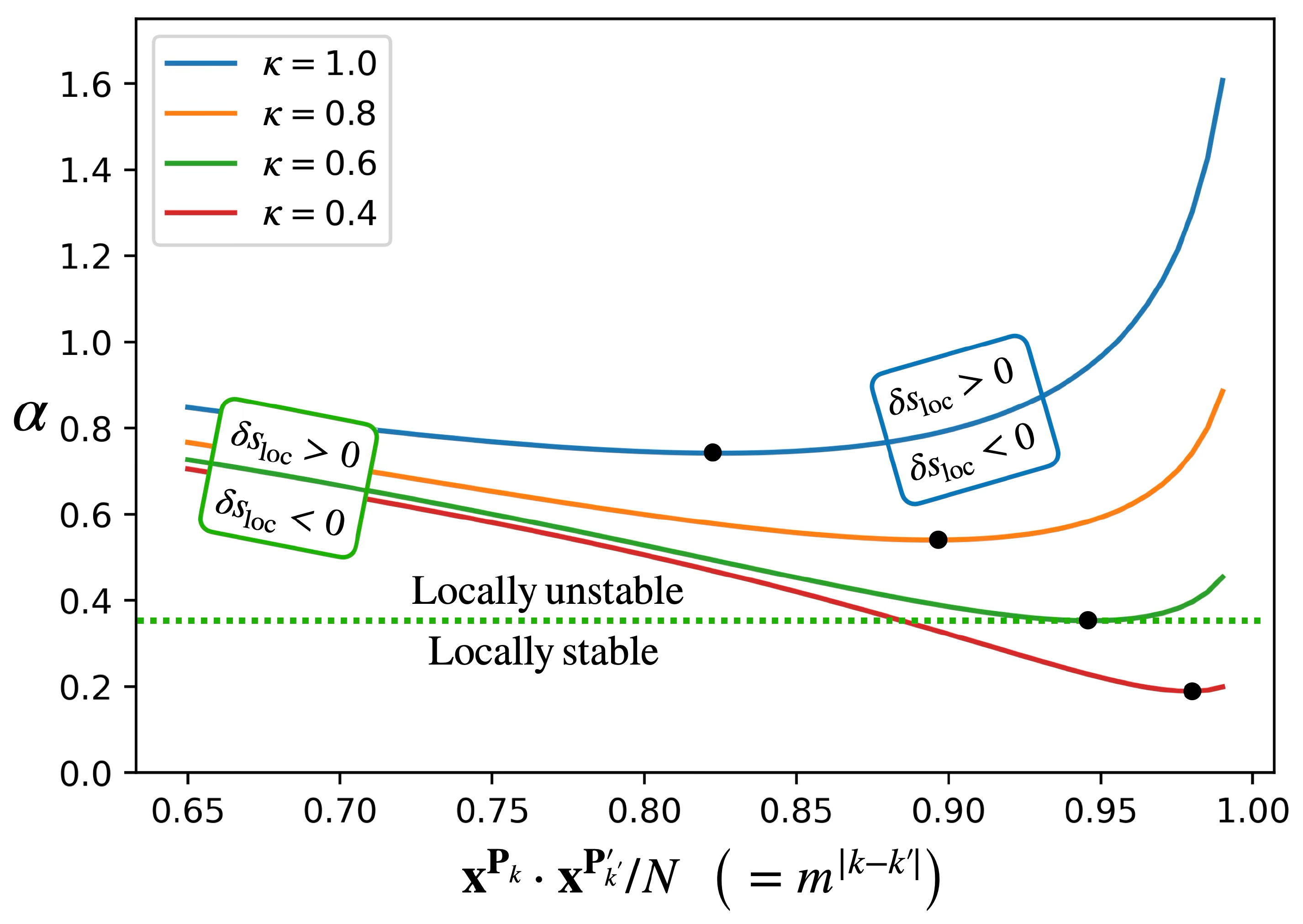}
    \caption{Plot indicating the sign of the perturbation $\delta s_{\rm loc.}$ as a function of $\alpha$ and of the distance between the two recorrelating configurations ${\bf x}^{{\bf P}_k}$ and ${\bf x}^{{\bf P}'_{k'}}$ in the core of the cluster. The full colored lines indicate the change of sign in the local entropy perturbation for four values of $\kappa$ ($\in\{0.4,0.6,0.8,1\}$). For each value of $\kappa$, we highlighted with a black dot the maximum value of $\alpha$ until which no-memory paths are  locally stable everywhere. By this we mean that $\delta s_{\rm loc}$ is negative for all distances. To guide the eye even more, we draw with a dashed green line this critical value of $\alpha$ for $\kappa=0.6$.}
    \label{fig: stability}
\end{figure}

In Fig.~\ref{fig: stability}, we plot the change of sign for $\delta s_{\rm loc}(k,k')$ as a function of the distance ${\bf x}^{{\bf P}_k}\cdot {\bf x}^{{\bf P}'_{k'}}/N\,(=m^{\vert k-k'\vert})$ and of the parameter $\alpha$ -with fixed threshold $\kappa\in \{0.4,0.6,0.8,1\}$-. We highlight with black dots (and a colored dashed line in the case of $\kappa=0.6$) the transition point for which no-memory paths become destabilized. A priori, our results should depend on the value chosen for $m$. Nevertheless, we observe numerically that the limit $\lim_{m\rightarrow 1}\delta s_{\rm loc}(k,k')/(1-m)$ converges, which makes the stability/unstability delimitation lines well-defined for $m\rightarrow1$.  We will label the threshold $\kappa$ for which this destabilization happens as $\kappa^{\rm no-mem}_{\rm loc.\,stab.}$ (with fixed $\alpha$).

\begin{figure}
    \centering
    \includegraphics[width=0.7\linewidth]{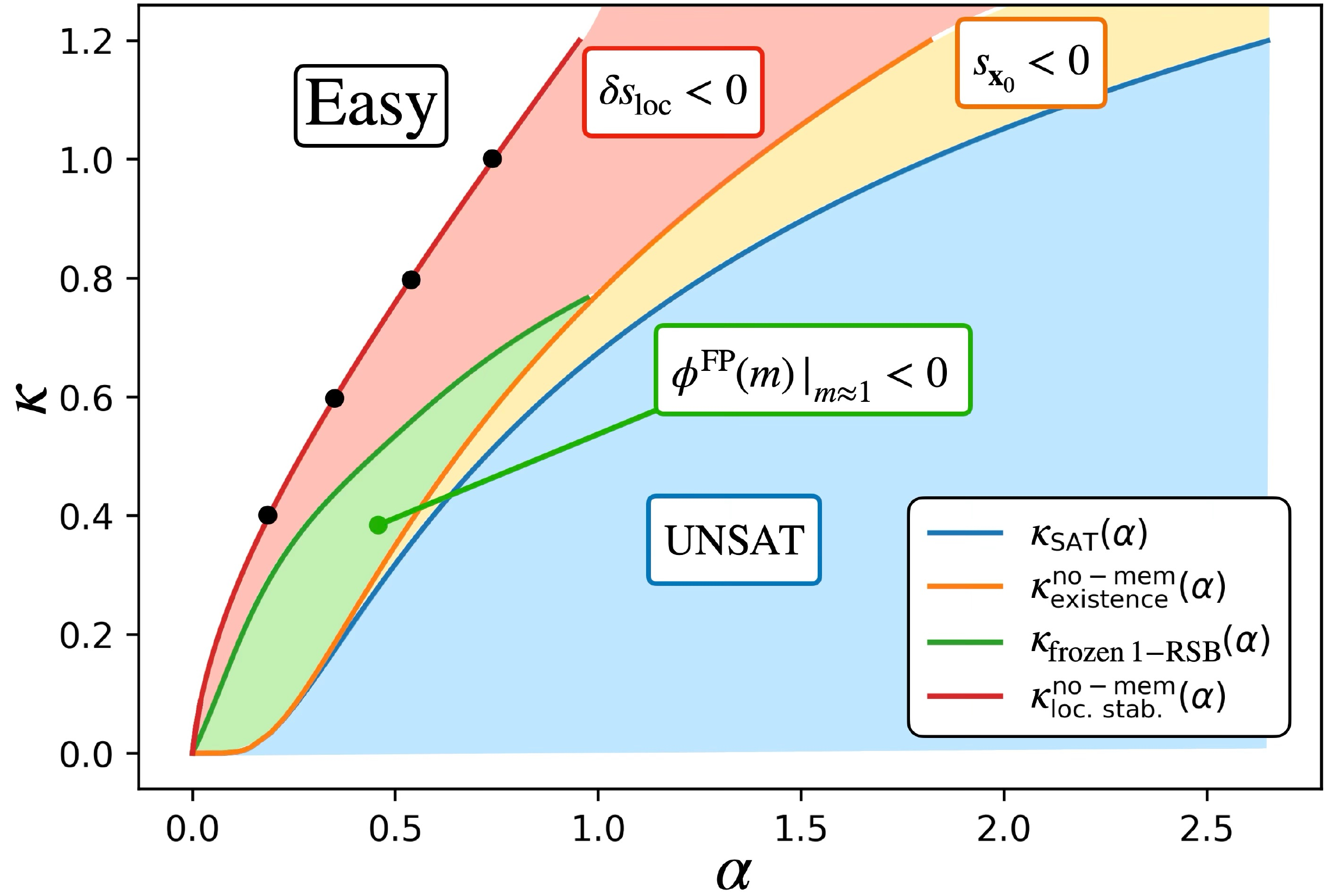}
    \caption{Phase diagram compiling all the different transitions we showed in the SBP solutions manifold. First in blue we have the UNSAT region, in which the SBP admits no solutions -see Eq.~\ref{eq:SAT}-. Then, the orange region corresponds to a regime in which solutions exist. However, the solutions predicted by the no-memory cluster(s) are unphysical, since their entropy $s_{{\bf x}_0}$ is negative. In green, we have the region in which these solutions develop an overlap gap. More particularly, the computation of a Franz-Parisi potential -see App.~\ref{app: FP connected minima}- shows they follow a frozen 1-RSB structure. Finally, in red, we have the regime for which the solutions paths composing the no-memory cluster(s) are unstable. The transition line is obtained by fitting the four critical points highlighted Fig.~\ref{fig: stability} -the black dots, which are also indicated in this figure-.}
    \label{fig: phase diagram}
\end{figure}

\begin{figure*}[t]
  \centering
  \includegraphics[width=\textwidth]{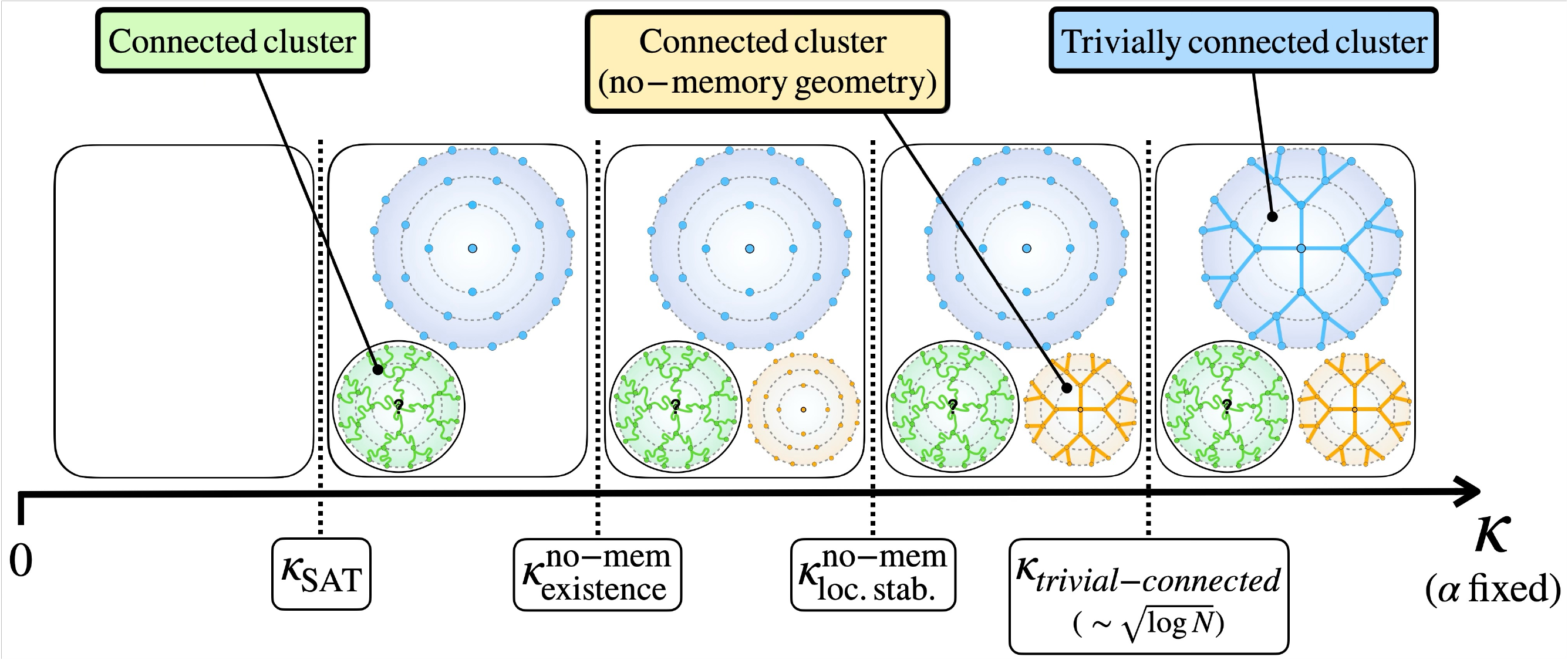}
  \caption{Plot representing the different transitions occurring when tuning $\kappa$ (and keeping $\alpha $ fixed). Going from high to low $\kappa$, we first have the shattering of the trivially connected cluster. Identified in \cite{Barbier2025}, this solutions manifold is composed of no-memory chains that can be explored without applying a local entropy bias. They can in fact be explored with a standard Monte-Carlo. Then, we have the destabilization of the no-memory cluster. This occurs as its core becomes locally unstable. We observe numerically (see Sec.~\ref{sec: numerics}) that this transition also corresponds to the clustering of the minima at its edge. At lower $\kappa$, we also have the transition for which minima with a margins distribution $P^{{\rm edge}}(\cdot)$ disappear ($\kappa=\kappa^{\rm no-mem}_{\rm existence}$). Finally, we have the UNSAT transition for $\kappa=\kappa_{\rm SAT}$. Below this value, the SBP admits no solutions. As the figure indicates, we cannot conclude anything about clusters with geometries other than the no-memory one.} 
  \label{fig: annealing}
\end{figure*}

Finally, in Fig.~\ref{fig: phase diagram} we compile into one phase diagram all the mechanisms affecting the SBP solutions space that we have identified so far. Going from low to high $\kappa$ (with fixed $\alpha$), we first have the UNSAT region in blue. In this regime, the SBP admits no solutions. Then, in the orange section, no solutions exist with the margins distribution $P^{{\rm edge}}(\cdot)$. In the regime depicted by the green demarcation, solutions with the distribution of margins  $P^{{\rm edge}}(\cdot)$ are strictly isolated. By this we mean that these solutions are all separated from each other by a distance scaling with system size. In other words, if we sit on one of these solutions, a finite fraction of bits have to be flipped to join any other equivalent minima. It is also known as a frozen 1-RSB (replica symmetry breaking) phase \cite{krauth89storage}. This result follows from studying the Franz-Parisi potential from App.~\ref{app: FP connected minima}. Finally, we highlight in red the region for which the core of the connected cluster -with a no-memory geometry- is locally unstable. 
In the same vein, we display in Fig.~\ref{fig: annealing} the different transitions that occur in the solutions manifold as we tune $\kappa$. If we consider an annealing procedure ($\kappa$ decreasing from $\infty$ to $0$), we first have a connected manifold -identified in \cite{Barbier2025}- that shatters at high $\kappa$. The presence of this trivially-connected cluster explains why simple algorithms like Monte-Carlo dynamics manage to solve the problem for $\kappa\sim \sqrt{\log N}$. Below this transition, connected regions remain present. In fact, the one with a no-memory geometry becomes unstable for $\kappa=\kappa^{\rm no-mem}_{\rm loc.\, stab.}$. This time, the transition does not depend on the size of the system. After this transition, these atypical solutions remain exponentially numerous until $\kappa=\kappa_{\rm existence}^{\rm no-mem}$. More generally, regardless of any connectivity properties, the solutions manifold remains non-empty until $\kappa=\kappa_{\rm SAT}$.

In the following, we will study how modified Monte-Carlo dynamics can explore the connected manifold we have characterized so far.

\section{Numerical results}
\label{sec: numerics}
 
In this section, we make use of our characterization of the no-memory connected cluster. Rather than emulating more model-agnostic dynamics, such as focused belief propagation \cite{braunstein2006learning} or local-entropy sampling \cite{baldassi2016local}, we design a bias that specifically targets this connected region and implement it within a Monte Carlo algorithm.

The bias we propose is straightforward. Instead of sampling solutions using the original loss $\mathcal{L}_{\rm SBP}(\cdot)$ -see Eq.~(\ref{eq: SBP loss})-, we will target minima in the delocalized cluster(s) using $\mathcal{L}^{\rm eff.}_{\rm SBP}(\cdot)$ from Eq.~(\ref{eq: eff loss deloc}). We know from the local entropy analysis that this loss gives access to the delocalized cluster(s) for at least $\kappa>\kappa^{\rm no-mem.}_{\rm loc.\, stab.}$. Below this critical threshold, we do not yet know whether these minima remain accessible. Unfortunately, in practice, evaluating the loss $\mathcal{L}^{\rm eff.}_{\rm SBP}(\cdot)$ through Eq.~(\ref{eq: iteration energ no-mem y=1}) is numerically expensive. Therefore, we will have to approximate $G_{\lambda^{\rm top}}^{\rm energ.}\left[w,m\right]$ with a simpler function:
\begin{align}
    \lim_{m\rightarrow1} G_{\lambda^{\rm top}}^{\rm energ.}\left[w,m\right]\approx     \tilde{G}_{\lambda^{\rm top}}^{\rm energ.}\left[w\right]
\end{align}
with
\begin{align}
    \tilde{G}_{\lambda^{\rm top}}^{\rm energ.}\left[w\right]=\frac{(\kappa-w)(\kappa+w)}{\kappa^2}\,.
\end{align}
We propose this function after numerically studying the shape $G_{\lambda^{\rm top}}^{\rm energ.}\left[\cdot,\cdot\right]$, for different values of 
$\kappa$, and observing good agreement between the two. This simplification, although abrupt, will allow us to avoid the high cost of recalculating the loss each time we change the value of $\kappa$.

Apart from this modification, we propose to perform a standard annealing in $\kappa$ with a Monte-Carlo dynamics. In more detail, having fixed the size $N$ of the system and the patterns $\{\xi^\mu\}_{\mu \in [\![1,M=\alpha N]\!]}$, we initialize our system in a random configuration ${\bf x}_0$ on the hypercube and set $\kappa=\sqrt{2\alpha\log(2N)}$. With this value for $\kappa$, we ensure that all points of the hypercube are a solution with high probability \cite{book_Potters}. Then, we let the system evolve following the Monte-Carlo iteration\footnote{The code for the Monte-Carlo algorithm is available on https://github.com/dbarbier76/Paths-in-symmetric-Perceptron}:
\begin{align}
{\bf x}^*_t&=\tilde{\bf I}_{i_t}{\bf x}_t\\
    {\bf x}_{t+1}&=\left\{\begin{array}{ccc}
        {\bf x}_t & {\rm if}&  e^{\mathcal{L}_{\rm eff.}^{\rm SBP}({\bf x}_t)-\mathcal{L}_{\rm eff.}^{\rm SBP}({\bf x}^*_t)}<a_t\\
         {\bf x}^*_t& {\rm if}& e^{\mathcal{L}_{\rm eff.}^{\rm SBP}({\bf x}_t)-\mathcal{L}_{\rm eff.}^{\rm SBP}({\bf x}^*_t)}\geq a_t
    \end{array}\right.
\end{align}
with $i_t$ and $a_t$ drawn uniformly in $[\![1,N]\!]$ and $[0,1]$ respectively. $\tilde{\bf I}_{i_t}$ is the $N\times N$ identity matrix with the $i_t^{\rm th}$ coefficient being $-1$ instead of $1$. Whenever we observe that the system is fully decorrelated (${\bf x}_t\cdot {\bf x}_{0}/N<0.05$), we decrease $\kappa$ by a small increment $d\kappa=0.005$. The initialization for the dynamics at this new threshold $\kappa$ is ${\bf x}_{0}={\bf x}_{t}$. Finally, we stop the annealing if we observe no full decorrelation for $t/N<1500$. 
For clarity, we will later call each time the Monte-Carlo dynamics is operated at a fixed $\kappa$ a ``round of annealing''.

\begin{figure}[h]
    \centering
    \includegraphics[width=0.75\linewidth]{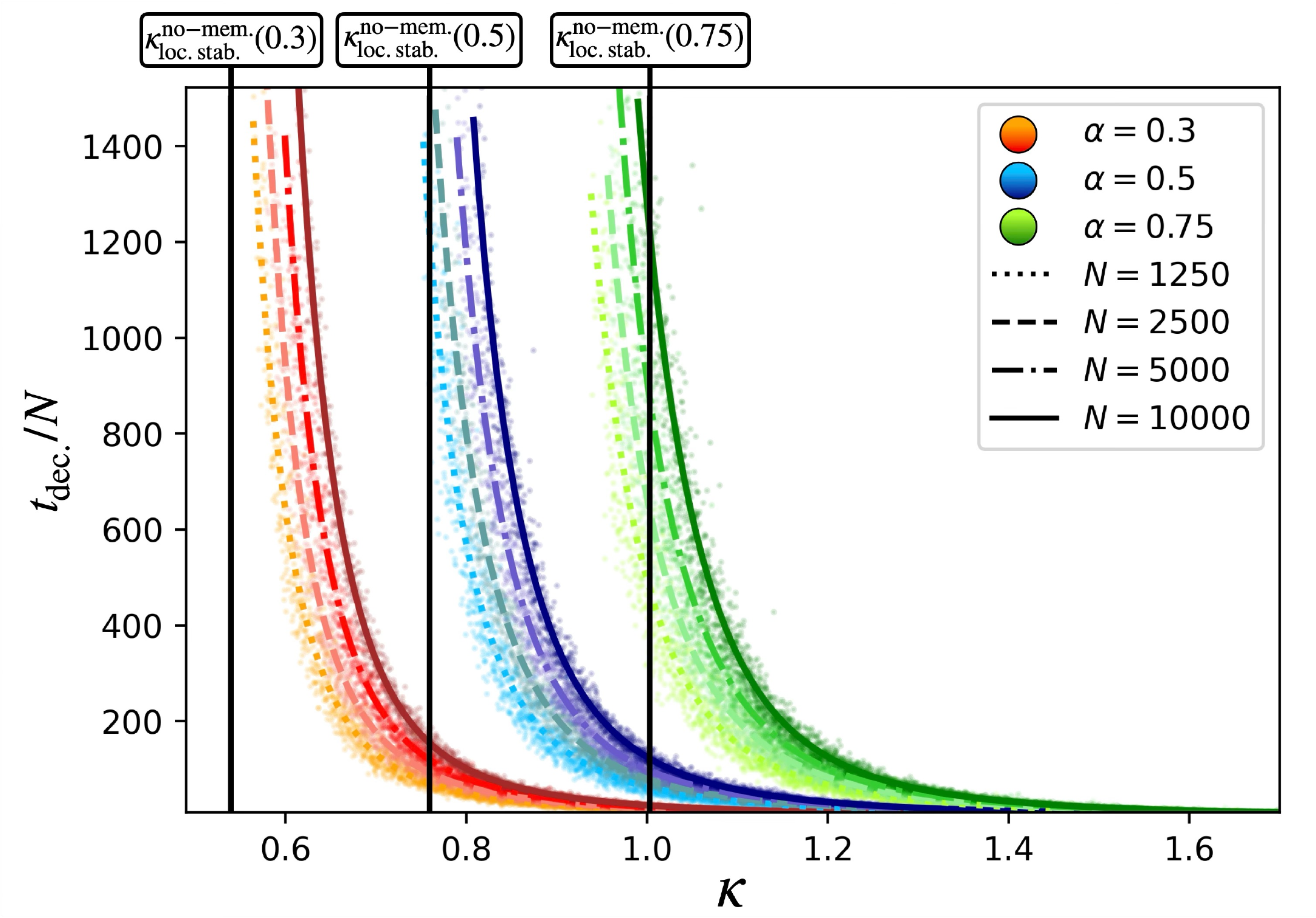}
    \caption{Plot showing the decorrelation time $t_{\rm dec.}$ as a function of $\kappa$. Each annealing setup ($\alpha=\{0.3,0.5,0.75\}$ and $N=\{1250,2500,500,10^4\}$) is simulated five times. Each point in the plot indicates a decorrelation time obtained for a given simulation and a given round of annealing. In the shades of red (respectively blue and green), we have the simulations for $\alpha=0.3$ (respectively $\alpha=0.5$ and $\alpha=0.75$). The different system sizes are highlighted with different shades, the lightest corresponds to $N=1250$, the darkest corresponds to $N=10^4$. We also fitted the points of each setup with a polynomial, we plotted them with corresponding colored line code (with a different pattern for each value of $N$.). Finally, we highlighted with black lines the critical values of $\kappa$ for which we predicted a local instability in the no-memory paths.}
    \label{fig: escape times}
\end{figure}

In Fig.~\ref{fig: escape times}, we plot as a function of $\kappa$ the number of iterations $t_{\rm dec.}$ that were numerically necessary for the system to fully decorrelate. In other words, for each round of annealing, $t_{\rm dec.}$ is defined as ${\bf x}_{t_{\rm dec.}}\cdot {\bf x}_{0}/N=0.05$. The annealing procedure is performed five times for each value of $\alpha(\in\{0.3,0.5,0.75\})$ and $N(\in\{1250,2500,5000,10^4\})$. Polynomial interpolations of the data points are represented as lines (one for each setting $\{\alpha, N\}$). First, the decorrelation time $t_{\rm dec.}/N$ appears to diverge near the local instability transition (i.e., $\kappa<\kappa^{\rm no-mem.}_{\rm loc.\, stab.}$). This indicates that, as the core of the cluster becomes locally unstable, the Monte Carlo dynamics fails to decorrelate efficiently in the edge and remains confined to a small region of the phase space. This is actually an additional insight that was not captured by our theoretical predictions. Indeed, it was not clear from our computation that the core instability would be associated with an algorithmically hard exploration of the cluster edge. It seems that the edge minima that we try to target with the effective loss start to cluster below the critical threshold $\kappa^{\rm no-mem.}_{\rm loc.\, stab.}$. Secondly, we can observe strong finite size effects with this figure. In fact, as we increase the size of the system, the escape time $t_{\rm dec.}/N$ grows significantly. This is because our dynamics not only guides the system through the paths we characterized, but it also generates fluctuations around them. If we are sitting in a minimum ${\bf x}$ with the margins distribution $P^{{\rm edge}}(\cdot)$, we can expect the Monte-Carlo algorithm to accept sometimes a handful of spin flips in a random direction (${\bf x}\rightarrow {\bf x}^*$). However, if such a thing happens, the increase of energy would be
\begin{align}
    \mathcal{L}^{\rm eff.}_{\rm SBP}({\bf x}^*)\!-\!    \mathcal{L}^{\rm eff.}_{\rm SBP}({\bf x})\approx&-\sum_{\mu=1}^M\frac{\xi^\mu\cdot\left({\bf x}^*\!-\!{\bf x}\right)} {\sqrt{N}}\partial_w\log\left(\!G_{\lambda^{\rm top}}^{\rm }\!\left[w\!=\!\frac{\xi^\mu\cdot{\bf x}}{\sqrt{N}},m\rightarrow1\right] \!\right)\\
    &-\sum_{\mu=1}^M\frac{\left[\xi^\mu\cdot\left({\bf x}^*-{\bf x}\right)\right]^2} {{2N}}\partial^2_w\log\left(G_{\lambda^{\rm top}}^{\rm }\left[\frac{\xi^\mu\cdot{\bf x}}{\sqrt{N}},m\rightarrow1\right] \right)\nonumber
\end{align}
where we neglected the contribution from the Heaviside functions as we have $P^{{\rm edge}}( w)\!\underset{\vert w\vert\rightarrow\kappa }{\sim}\!\kappa\!-\!\vert w\vert$. Averaging over the patterns distribution, we have in the thermodynamic limit ($N\rightarrow+\infty$)
\begin{align}
    &\mathcal{L}^{\rm eff.}_{\rm SBP}({\bf x}^*)\!-\!    \mathcal{L}^{\rm eff.}_{\rm SBP}({\bf x})\approx-\frac{\left({\bf x}^*\!-\!{\bf x}\right)^2}{2N}\!\int \!dw\, P^{{\rm edge}}\!(w)\partial^2_w\log\left(\!G_{\lambda^{\rm top}}^{\rm }\!\left[w,m\!\rightarrow\!1\right] \right)\,.
\end{align}
In fact, this quantity diverges as we observe numerically that
\begin{align}
    \partial^2_w\log\left(\!G_{\lambda^{\rm top}}^{\rm }\!\left[w,m\!\rightarrow\!1\right]\right)\underset{\vert w\vert\rightarrow\kappa }{\sim}\frac{1}{(\kappa\!-\!\vert w\vert)^2}\,.
\end{align}
This means that the landscape around connected minima becomes increasingly steep as the system size increases. All the moves corresponding to fluctuations around the paths, and that help decorrelating, get killed by increasing $N$. To demonstrate this further, we studied in App.~\ref{app: modified+ MC} the case where we damped loss around the edges $\vert w\vert\sim \kappa$ -to obtain a convergent value for $\mathcal{L}^{\rm eff.}_{\rm SBP}({\bf x}^*)\!-\!    \mathcal{L}^{\rm eff.}_{\rm SBP}({\bf x})$-. With this bias, we are able to suppress the finite size effect we observe in Fig.~\ref{fig: escape times}.  

\begin{figure}[h]
    \centering
    \includegraphics[width=0.75\linewidth]{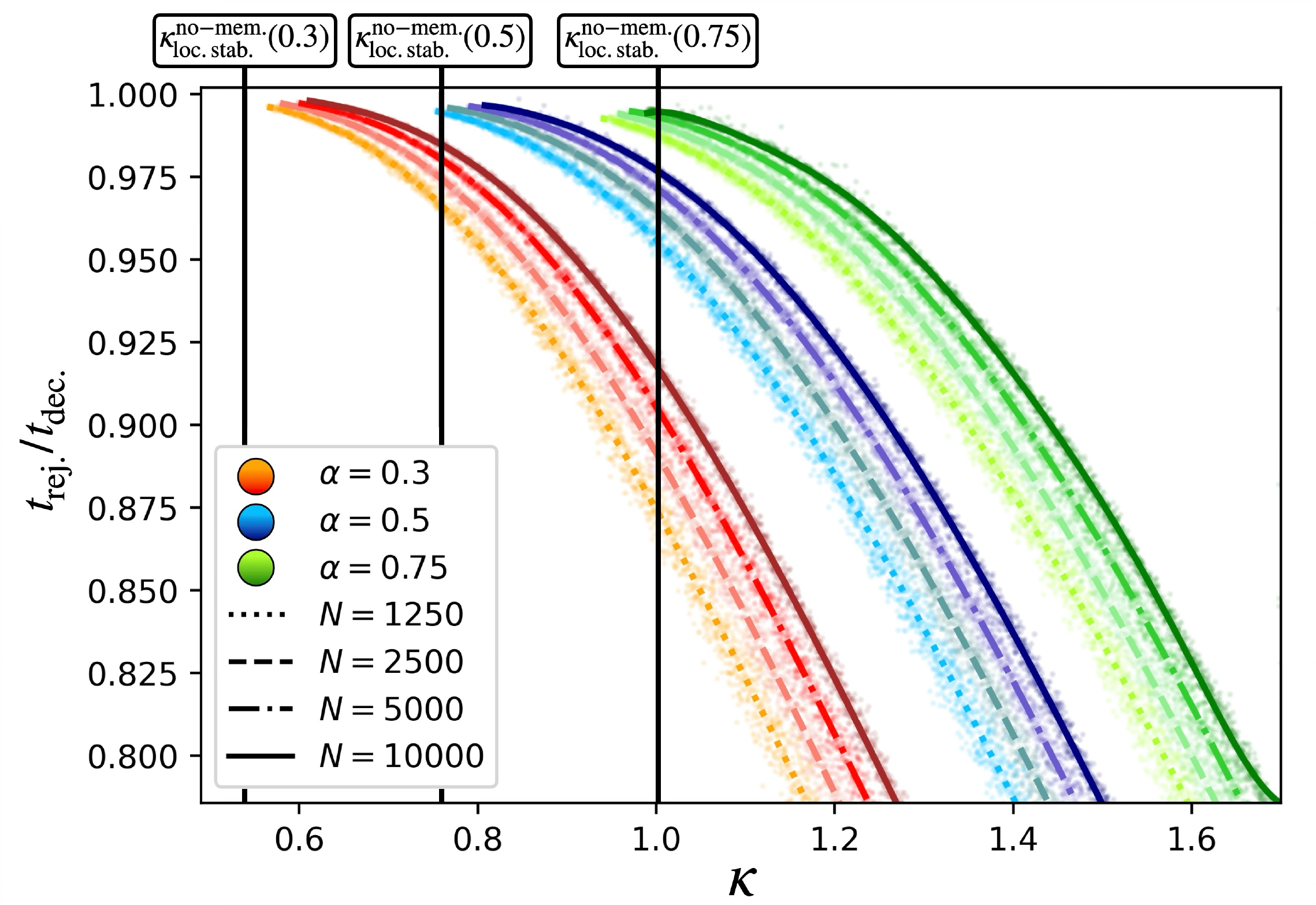}
    \caption{Plot showing the ratio between the total number of rejected spin flips $t_{\rm rej.}$ and the decorrelation time $t_{\rm dec.}$ (for a given round of annealing) as a function of $\kappa$. We simulated five times each annealing setup ($\alpha=\{0.3,0.5,0.75\}$ and $N=\{1250,2500,500,10^4\}$). Each point in the plot indicates a time ratio obtained for a given simulation and a given round of annealing. In red (respectively blue and green), we have the simulations for $\alpha=0.3$ (respectively $\alpha=0.5$ and $\alpha=0.75$). Different shades indicate different system sizes, the lightest corresponds to $N=1250$, the darkest corresponds to $N=10^4$. The data points of each setup have also been fitted with a polynomial, they are indicated with a corresponding colored line (with a different pattern for each value of $N$.). Finally, we highlighted with black lines the critical values of $\kappa$ for which we predicted a local instability in the no-memory cluster(s).}
    \label{fig: ratio rejec}
\end{figure}

To determine the nature of the clustering below $\kappa^{\rm no-mem.}_{\rm loc.\, stab.}$, we plot in Fig.~\ref{fig: ratio rejec} the ratio between the number $t_{\rm rej.}$ of rejected spin flips (${\bf x}_{t+1}={\bf x}_{t}$) and the total number of spin flip trials $t_{\rm dec.}$ for each decorrelation round. We can note that the dynamics slows down because it rejects more and more trials as $\kappa$ decreases. So much so that the procedure ends up rejecting almost all proposed moves when $\kappa$ approaches the local stability transition. Therefore, the clustering phase appears as a frozen 1-RSB glassy phase \cite{krauth89storage}, where the targeted minima are isolated from each other and no moves are accepted to decorrelate from the initialization. 

Let us continue to characterize the dynamics by focusing now on the correlations ${\bf x}_t \cdot{\bf x}_{t'}/N$ during the annealing.
As a reminder, the correlation profile along a path of minima is
\begin{align}
    \frac{{\bf x}_k\cdot{\bf x}_{k'}}{N}&=m^{\vert k-k'\vert }\underset{m\rightarrow 1}{=}e^{-\vert k-k'\vert(1-m) }\,.
\end{align}
As a rough approximation, we can consider that the system decorrelates also with this profile inside the edge of the cluster. Thus, we should expect with a single time rescaling -$t\rightarrow\gamma(\kappa)t$- that dynamics follow the same profile:
\begin{align}
\label{eq: power law decorr}
   \frac{ {\bf x}_t\cdot {\bf x}_{0}}{N}=e^{-\gamma(\kappa) t/N}\,.
\end{align}
The rescaling factor $\gamma(\kappa)$ corresponds to the typical inverse-time required to move inside a path. 

\begin{figure}
    \centering
\includegraphics[width=1\linewidth]{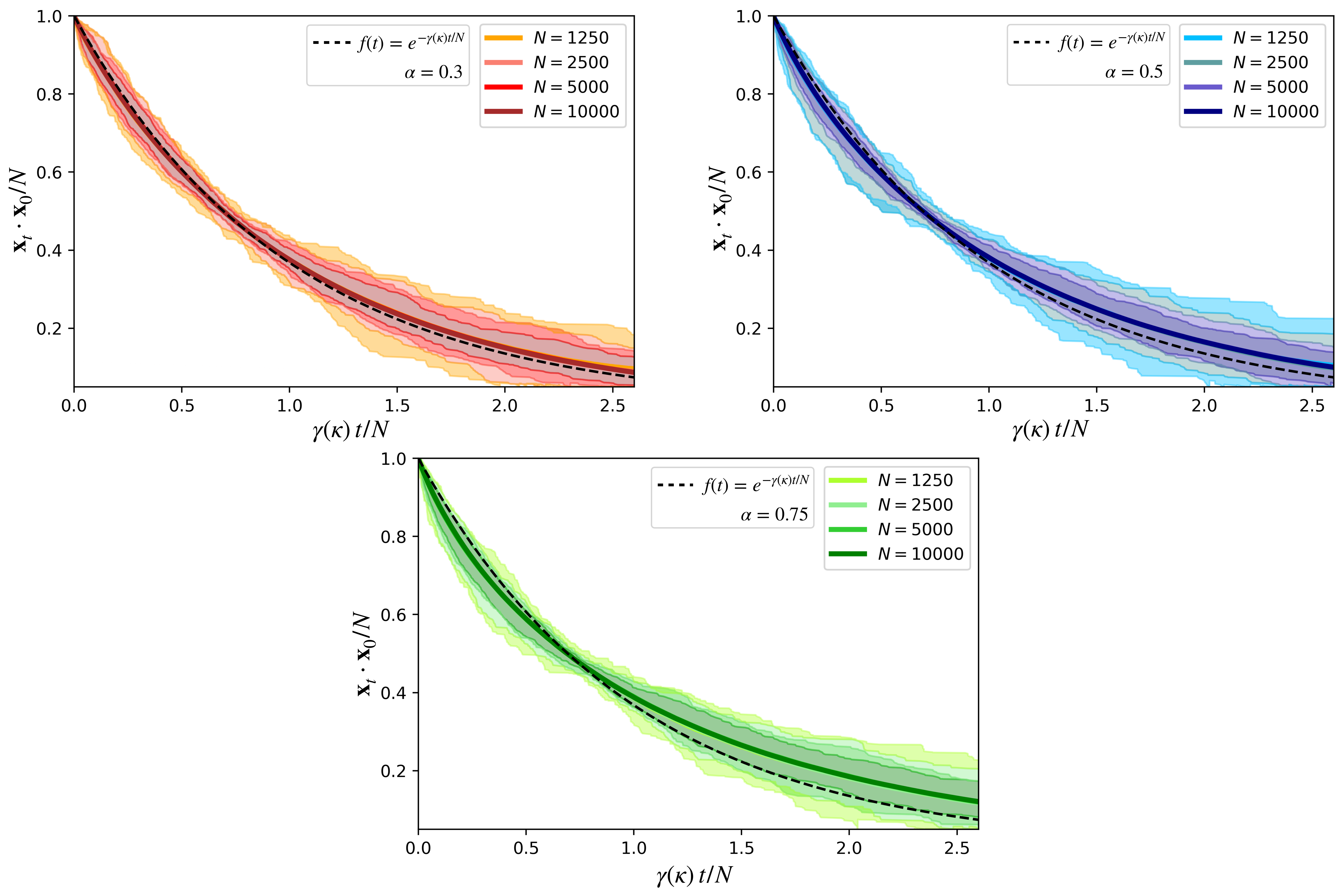}
    \caption{Plots of the correlation functions ${\bf x}_t\cdot {\bf x}_0/N$ (obtained for each round of annealing) as a function of the rescaled time $\gamma(\kappa)t$. Each color corresponds to different value for $\alpha$, red is $\alpha=0.3$, blue is $\alpha=0.5$ and green is $\alpha=0.75$. The solid lines are averages over five entire annealing procedures, for each setup ($\alpha=\{0.3,0.5,0.75\}$ and $N=\{1250,2500,500,10^4\}$). For a given value of $\alpha$ -i.e. a given color-, we can note that  the averaged correlation functions of different system sizes overlap strongly (the solid lines are almost on top of each other). Fixing $N$ and $\alpha$, the colored shade corresponds to the maximum and minimum value obtained for the correlation function after time-rescaling (over the five realization of annealing). In each plots, the dashed line corresponds to a exponential decay, which is the expected decorrelation profile if the dynamics only explores no-memory paths  -see Eq.~(\ref{eq: power law decorr})-.}
    \label{fig: correlation function}
\end{figure}

In Fig.~\ref{fig: correlation function}, we plotted the overlap ${\bf x}_t\cdot {\bf x}_0/N$ -averaged over 5 entire annealing procedure- as a function of the rescaled time $\gamma(k)t/N$. In practice, we set $\gamma(\kappa)$ for each round of annealing by fitting ${\bf x}_t\cdot{\bf x}_0/N$ to the exponential decay from Eq.~(\ref{eq: power law decorr}). The colored shades correspond to the worst fitted points we obtained for each setup $\{\alpha,\kappa\}$. Overall, we observe a good agreement between the simple exponential decay modeling and our simulations. It holds not only for the averaged curve (over entire annealing procedures) but also for the extreme values, as illustrated by the shaded regions. This indicates that the exponential decay -characteristic of a no-memory geometry- remains valid at all times throughout the annealing process, even near the algorithmic transition point. At no point do we observe that our system is attracted towards paths with a different correlation profile. More carefully, we also notice that this modeling deteriorates as $\alpha$ increases. While several explanations are possible, we will later see that this deterioration coincides with the breakdown of our approximation $ \lim_{m\rightarrow1} G_{\lambda^{\rm top}}^{\rm energ.}\left[w,m\right]\approx\tilde{G}_{\lambda^{\rm top}}^{\rm energ.}\left[w\right]$. 

For the interested reader, such decorrelation profiles have also been studied in \cite{Barbier2025} for an unbiased Monte Carlo algorithm initialized in a robust solution of the SBP (like the ones we target with the connected ensemble). In short, it has been observed (Fig. 4 in \cite{Barbier2025}) that an unbiased Monte Carlo algorithm cannot decorrelate from a robust solution in polynomial time. Even worse, its performance deteriorates as the system size increases. This shows how a well-chosen bias can help outperform brute-force algorithms.

\begin{figure}[htbp]
\centering
    \includegraphics[width=1\linewidth]{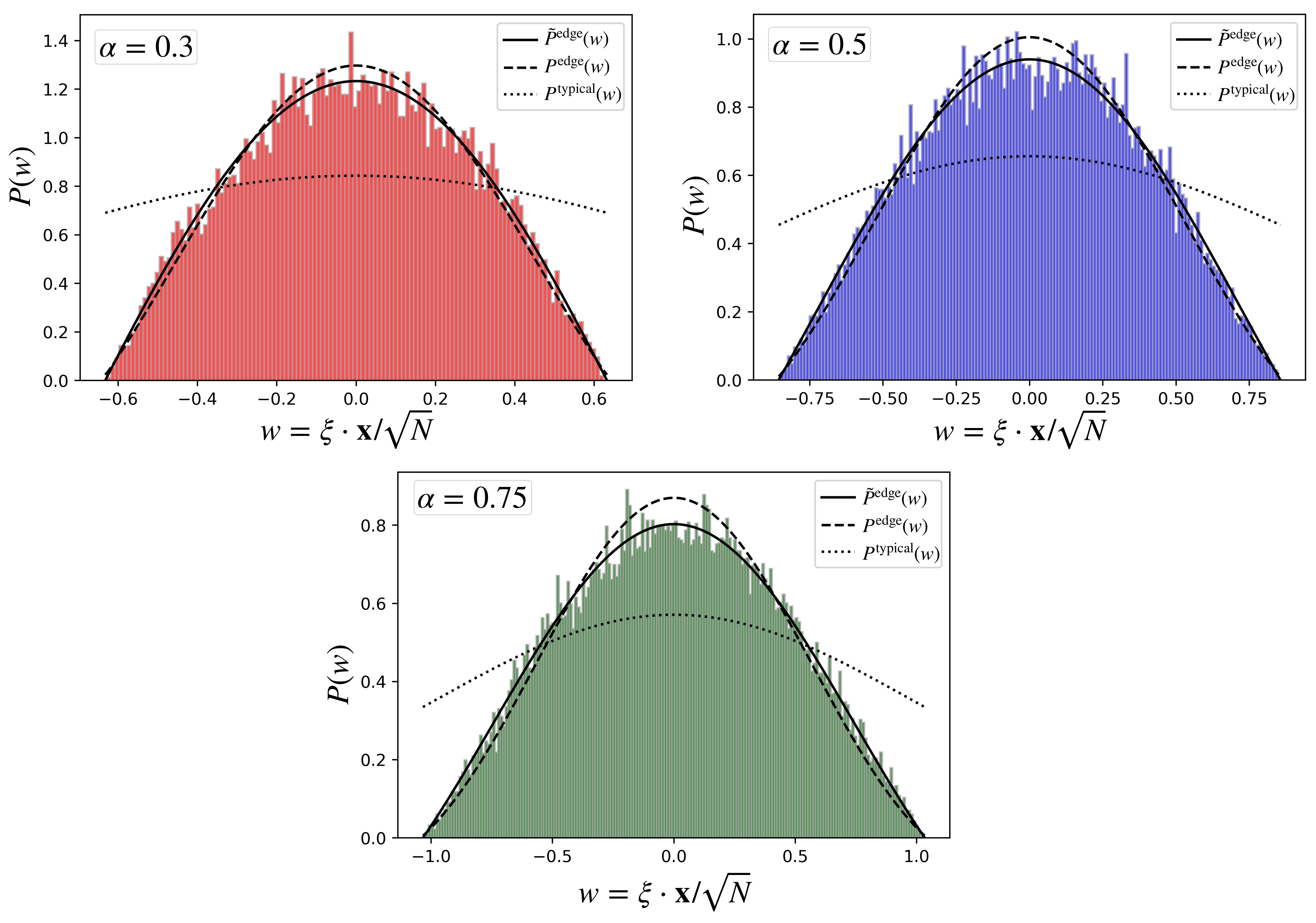}
    \caption{Plots displaying the distribution of margins at the end of annealing procedures, for $\alpha\in\{0.3,0.5,0.75\}$ and $N=10^4$. Each plot corresponds to a different value of $\alpha$, where we averaged over five independent annealing simulations. The full black lines correspond to the expected margins distribution -defined in Eq.~(\ref{eq: modified margin distrib})-. The dashed ones are the theoretical distribution of margins in the no-memory cluster(s) -see Eq.~(\ref{eq: distrib no-mem cluster})-. Finally, the dotted lines correspond to the distribution of margins for SBP typical solutions -see Eq.~(\ref{eq: typical distrib})-.}
    \label{fig: margins distribution}
\end{figure}

Finally, we plot in Fig.~\ref{fig: margins distribution} the margins distribution obtained at the end of five independent annealing procedures for fixed parameters $\alpha\in\{0.3,0.5,0.75\}$ and $N=10^4$. As points of reference, we added the distribution for typical solutions -$P^{\rm typical} (\cdot)$ defined in Eq.~(\ref{eq: typical distrib})-, the one for minima in the no-memory cluster(s) -$P^{{\rm edge}}(\cdot)$- and the one expected with the simplified loss:
\begin{align}
\label{eq: modified margin distrib}
    \tilde{P}^{{\rm edge}}(w)=\frac{e^{-\frac{w^2}{2}}\tilde{G}_{\lambda^{\rm top}}^{\rm energ.}\left[w\right]}{\int_{-\kappa}^\kappa d\underline{w} \,e^{-\frac{\underline{w}^2}{2}}\tilde{G}_{\lambda^{\rm top}}^{\rm energ.}\left[\underline{w}\right]}\,.
\end{align}
Again, we can see that we have a very good agreement between our predictions and the simulations, even if we are extremely close to the instability transition. This shows that the system evolves around the minima we have described, until eventually the core of the connected cluster becomes unstable. The dynamics never drifts into another basin of solutions. We can also observe that our test distribution $\tilde{P}^{{\rm edge}}(\cdot)$ remains overall close to the original one -${P}^{{\rm edge}}(\cdot)$-. Nonetheless, this simplification deteriorates as $\alpha$ increases: the discrepancy between $P^{\rm edge}(\cdot)$ and $\tilde{P}^{\rm edge}(\cdot)$ becomes larger for $\alpha = 0.75$ than for $\alpha = 0.3$. This is consistent with our observations regarding the decorrelation profile of ${\bf x}_t \cdot {\bf x}_0/N$, where the same deterioration is visible. Overall, this approximation appears increasingly ill-suited to correctly target the no-memory cluster as $\alpha$ grows. Still, it remains striking how easily these minima can be reached using only a properly modified Monte Carlo algorithm.

\section{Conclusion}
In conclusion, we have shown how local entropy biases are part of a toolbox that helps target connected regions in a rugged energy landscape. Until now, these biases (appearing in biology \cite{Medhekar2007,Macadangdang2022}, machine learning \cite{baldassi2015subdominant,baldassi2020clustering,baldassi2021unveiling} and computer sciences \cite{gamarnik2022algorithms,gamarnik2021survey}) were only qualitatively understood. In particular, their link with connectivity was never truly unveiled. By generalizing these approaches, we have shown that they are a building block for defining connected clusters within a statistical mechanics framework. Using this new statistical ensemble, we focused our paper on the solutions manifold for the symmetric binary perceptron (SBP). 
With its simplicity, the SBP allows an in-depth study of its landscape, where theoretical predictions can be compared to simple simulations.
Although basic, this model shares important characteristics with more realistic setups. For example, its landscape is dominated by isolated minima and exploring with local algorithms the solutions manifold appears to be a difficult task. 

The statistical mechanics ensemble for connected states allowed us to characterize a delocalized cluster of minima. With its star shape, it resembles the geometry of accessible solutions in the negative margin perceptron \cite{Annesi2023}. We were also able to show its existence for $\kappa>\kappa^{\rm no-mem}_{\rm loc.\,stab.}$, below which the paths it comprises in its core become locally unstable. To bridge the gap between this new approach and more well-known statistical mechanics tools, we also showed that path stability can be mapped exactly to a Franz-Parisi potential for simpler models. This equivalence is lost with the SBP.

This characterization also serves for the design of a modified Monte-Carlo algorithm. It uses a loss that directly targets the solutions located on the edge of the delocalized cluster. Simulations showed that the annealing dynamics starts to fail when the core of the cluster becomes locally unstable. 

As a final note, this first foray into the statistical mechanics of connected configurations leaves the door open to a plethora of subsequent work. Among the more obvious ones, we can mention the extension of our computation to other replica-overlapping structures and to $y_k\neq 1$ ($k\in[\![1,k_f]\!]$). With this, maybe we could target connected regions that go below the critical threshold $\kappa^{\rm no-mem}_{\rm loc.\,stab.}$. Another question concerns the design of algorithms. Can we have a better use of the connected minima characterization than a simple modification of the standard Monte-Carlo dynamics? An idea could be, for example, to test a replicated Monte-Carlo with a chain geometry among copies. Finally, one could be interested in using this new statistical ensemble to probe paths in models whose landscape also evolves over time. This would be important for studying algorithm stability in computer science \cite{gamarnik2021survey} or even for emulating evolution processes which have a time-dependent fitness landscape. On a more global scale, if we can understand how to better navigate these rugged synthetic landscapes, we could perhaps exploit this new knowledge in more applicable areas: phylogenetic tree reconstructions, accelerated evolutionary processes, machine learning, to name a few.

\section*{Acknowledgments}
We would like to thank Riccardo Zecchina, Jérôme Garnier-Brun, Luca Saglietti, Gianmarco Perrupato, Enrico Malatesta and Chris Jones for the many discussions and guidance about this project.

\appendix

\section{Computing the free energy}
\subsection{General computation}
\label{app: General comp}
In this section we detail how to compute the $quenched$ free energy
\begin{align}
    \phi={\rm I\!E}_{\xi}[\log(\mathcal{Z})]\underset{y_0\rightarrow 0}{=}\frac{{\rm I\!E}_{\xi}[\mathcal{Z}^{y_0}]-1}{y_0}\
\end{align}
with
\begin{align}
\hspace{-0.22cm}{\rm I\!E}_\xi[\mathcal{Z}^{y_0}]\!=\!{\rm I\!E}_{\xi}\!\left\{\!\prod_{k=0}^{k_f}\prod_{{\bf P}_{k}}\sum_{{\bf x}^{{\bf P}_{ k}}\in\Sigma^{N,m}_{{\bf x}^{{\bf P}^*_k}}}e^{-\mathcal{L}_{\rm SBP}\left({\bf x}^{{\bf P}_{k}}\right)}\!\right\}\,.
\end{align}
So as to perform the average over the patterns distribution we will introduce the margins as
\begin{align}
    e^{-\mathcal{L}_{\rm SBP}\left({\bf x}^{{\bf P}_k}\right)}&\!=\!\!\prod_{\mu=1}^M\!\Theta\left(\kappa-\left\vert\frac{\xi^\mu\cdot{\bf x}^{{\bf P}_k}}{\sqrt{N}}\right\vert\right)=\!\!\prod_{\mu=1}^M\int_{-\kappa}^{+\kappa} \!\!dw^{{\bf P}_k,\mu}\!\int_{-\infty}^{+\infty}\!\!d\hat{w}^{\,{\bf P}_k,\mu}e^{i\hat{w}^{\,{\bf P}_k,\mu}\left(\!w^{{\bf P}_k,\mu}-\frac{\xi^\mu\cdot{\bf x}^{{\bf P}_k}}{\sqrt{N}}\!\right)}\!.
\end{align}
The successive integration over the patterns $\{\xi^\mu\}_{\mu\in{[\![1,M]\!]}}$ distribution and after over the fields $\{\hat{w}^{{\bf P}_k,\mu}\}_{\mu\in{[\![1,M]\!]}}$ is trivial as it simply involves Gaussian integrals. It yields
\begin{align}
    {\rm I\!E}_\xi[\mathcal{Z}^{y_0}]=\prod_{k=0}^{k_f}\!\prod_{{\bf P}_{k}}\prod_{\mu=1}^M\sum_{{\bf x}^{{\bf P}_{ k}}\in\Sigma^{N,m}_{{\bf x}^{{\bf P}^*_k}}}\int_{-\kappa}^{\kappa}\!\!dw^{{\bf P}_k,\mu}\frac{e^{-\frac{{\bf w}\underline{\bf Q_{\bf x}}^{-1}{\bf w}}{2}}}{\mathcal{N}_{\bf x}}
\end{align}
with
\begin{align}
   \mathcal{N}_{\bf x}=\prod_{k=0}^{k_f}\!\prod_{{\bf P}_{k}}\prod_{\mu=1}^M\int_{-\infty}^{+\infty} \!\!dw^{{\bf P}_k,\mu}{e^{-\frac{{\bf w}\underline{\bf Q_{\bf x}}^{-1}{\bf w}}{2}}}
\end{align}
and
\begin{align}
    \underline{\bf Q_{\bf x}}_{{\bf P}_k,{\bf P}'_{k'}}^{\mu,\mu'}=\delta_{\mu,\mu'}\,\frac{{\bf x}^{{\bf P}_k}\cdot{\bf x}^{{\bf P}'_{k'}}}{N}\, .
\end{align}
Another way to obtain this result is to recognize that (without the loss) the margins are random Gaussian variables, following the means and correlations
\begin{align}
\label{eq: mean margin}
    {\rm I\!E}_{\xi}\left[w^{{\bf P}_k,\mu}\right]&={\rm I\!E}_{\xi}\left[\frac{\xi^\mu\cdot{\bf x}^{{\bf P}_k}}{\sqrt{N}}\right]=0\, ,\\
\label{eq: correlation margin}
    {\rm I\!E}_{\xi}\left[w^{{\bf P}_k,\mu}w^{{\bf P}'_{k'},\mu'}\right]&={\rm I\!E}_{\xi}\left[\frac{(\xi^\mu\cdot{\bf x}^{{\bf P}_k})(\xi^{\mu'}\cdot{\bf x}^{{\bf P}'_{k'}})}{{N}}\right]=\delta_{\mu,\mu'}\,\frac{{\bf x}^{{\bf P}_k}\cdot{\bf x}^{{\bf P}'_{k'}}}{N}\, ,
\end{align}
and to simply introduce this distribution in the partition function. The last step is to decouple the margins contribution from the hypercube summation. For this we introduce magnetic fields as
\begin{align}
    &\prod_{k=0}^{k_f}\prod_{{\bf P}_{k}}\! \sum_{{\bf x}^{{\bf P}_{ k}}\in\Sigma^{N,m}_{{\bf x}^{{\bf P}^*_k}}}f\left(\{{\bf x}^{{\bf P}_k}\cdot{\bf x}^{{\bf P}'_{k'}}\}\right)=\prod_{k,k'=0}^{k_f}\prod_{{\bf P}_{k},{\bf P}'_{k'}}\int \frac{d\hat{\bf Q}_{{\bf P}_{k},{\bf P}'_{k'}}}{2\pi} \!\prod_{k,k'=0}^{k_f}\prod_{\substack{{\bf P}_{k}\\{\bf P}'_{k'}(\neq {\bf P}_k,{\bf P}^*_k)}}\int d{\bf Q}_{{\bf P}_{k},{\bf P}'_{k'}}\nonumber\\
    &\hspace{+1.3cm}\times\! \prod_{k=0}^{k_f}\prod_{{\bf P}_{k}}\sum_{{\bf x}^{{\bf P}_{ k}}\in\Sigma^{N}}\times\!e^{\underset{k,{k'}}{\sum}\,\underset{{\bf P}_k,{\bf P}'_{k'}}{\sum} \!\!i\hat{\bf Q}_{{\bf P}_{k},{\bf P}'_{k'}}\left[{\bf x}^{{\bf P}_k}\cdot{\bf x}^{{\bf P}'_{k'}}-N{\bf Q}_{{\bf P}_{k},{\bf P}'_{k'}}\right]}f\left(\{{\bf Q}_{{\bf P}_{k},{\bf P}'_{k'}}\}\right)\,.
\end{align}
Before writing this decoupling for the SBP model, we can also mention that it can be evaluated with a saddle point approximation when $N\rightarrow+\infty$:
\begin{align}
     &\lim_{N\rightarrow+\infty}\prod_{k=0}^{k_f}\prod_{{\bf P}_{k}}\sum_{{\bf x}^{{\bf P}_{ k}}\in\Sigma^{N,m}_{{\bf x}^{{\bf P}^*_k}}}f\left(\{{\bf x}^{{\bf P}_k}\cdot{\bf x}^{{\bf P}'_{k'}}\}\right)\sim\underset{\underset{\left({\rm s.t.}\,{\bf Q}_{{\bf P}_{k},{\bf P}^*_{k}}=m\,,\, {\bf Q}_{{\bf P}_{k},{\bf P}_{k}}=1\right)}{\hat{\bf Q},{\bf Q}}}{\rm opt}\left\{ \vphantom{\prod_{\sum_{{\bf x}_k^{{\bf j}_{ k}}\!\in\Sigma^{N}}{\bf j}_{k}}}f\left(\{{\bf Q}_{{\bf P}_{k},{\bf P}'_{k'}}\}\right)\right.\\
    &\hspace{4.5cm}\times\prod_{k=0}^{k_f}\prod_{{\bf P}_{k}} \sum_{{\bf x}^{{\bf P}_{ k}}\!\in\Sigma^{N}}\left.\!e^{\underset{k,{k'}}{\sum}\,\underset{{\bf P}_k,{\bf P}'_{k'}}{\sum}\!\!\hat{\bf Q}_{{\bf P}_{k},{\bf P}'_{k'}}\!\left[{\bf x}^{{\bf P}_k}\cdot{\bf x}^{{\bf P}'_{k'}}-N{\bf Q}_{{\bf P}_{k},{\bf P}'_{k'}}\right]}\!\right\}\!.\nonumber
\end{align}
Applying this formula to our setup we obtain
\begin{align}
   \lim_{N\rightarrow+\infty} {\rm I\!E}_\xi[\mathcal{Z}^{y_0}]\sim&\underset{\underset{\left({\rm s.t.}\,{\bf Q}_{{\bf P}_{k},{\bf P}^*_{k}}=m\,,\, {\bf Q}_{{\bf P}_{k},{\bf P}_{k}}=1\right)}{\hat{\bf Q},{\bf Q}}}{\rm opt}\hspace{-0.5cm}\left\{\tilde{\mathcal Z} \right\}
\end{align}
with
\begin{align}
    \tilde{\mathcal Z}=&\prod_{k=0}^{k_f}\prod_{{\bf P}_{k}}\prod_{\mu=1}^M\int_{-\kappa}^{\kappa}{dw^{{\bf P}_{ k},\mu}}\frac{e^{-\frac{{\bf w}\underline{\bf Q}^{-1}{\bf w}}{2}}}{\mathcal{N}}\prod_{k=0}^{k_f}\prod_{{\bf P}_{k}}\sum_{{\bf x}^{{\bf P}_{ k}}\in\Sigma^{N}} e^{\underset{k,{k'}}{\sum}\,\underset{{\bf P}_k,{\bf P}'_{k'}}{\sum}\!\!\hat{\bf Q}_{{\bf P}_{k},{\bf P}'_{k'}}\!\left[{\bf x}^{{\bf P}_k}\cdot{\bf x}^{{\bf P}'_{k'}}-N{\bf Q}_{{\bf P}_{k},{\bf P}'_{k'}}\right]} ,
\end{align}
the normalization $\mathcal{N}$ being
\begin{align}
\mathcal{N}=\prod_{k=0}^{k_f}\prod_{{\bf P}_{k}}\prod_{\mu=1}^M\int_{-\infty}^{+\infty}d{w^{{\bf P}_{ k},\mu}} e^{-\frac{{\bf w}\underline{\bf Q}^{-1}{\bf w}}{2}}
\end{align}
and $\underline{\bf Q}^{-1}$ being the inverse of the overlap matrix
\begin{align}
     \underline{\bf Q}_{{\bf P}_k,{\bf P}'_{k'}}^{\mu,\mu'}=\delta_{\mu,\mu'}\,  {\bf Q}_{{\bf P}_k,{\bf P}'_{k'}}\, .
\end{align}
Finally, we can recognize that all directions $i\in[\![1,N]\!]$ of the hypercube $\Sigma^N$ are decoupled -same for the margins directions $\mu\in[\![1,M]\!]$-, which yields
\begin{align}
    \tilde{\mathcal Z}=&\left\{\prod_{k=0}^{k_f}\prod_{{\bf P}_{k}}\int_{-\kappa}^{\kappa}d{w^{{\bf P}_{ k}}}\frac{e^{-\frac{{\bf w}{\bf Q}^{-1}{\bf w}}{2}}}{\mathcal{N}}\right\}^M\!\!\times\!\left\{\!\prod_{k=0}^{k_f}\prod_{{\bf P}_{k}}\sum_{{x}^{{\bf P}_{ k}}=\pm 1}e^{\underset{k,{k'}}{\sum}\,\underset{{\bf P}_k,{\bf P}'_{k'}}{\sum}\!\!\hat{\bf Q}_{{\bf P}_{k},{\bf P}'_{k'}}\!\left[{x}^{{\bf P}_k}{ x}^{{\bf P}'_{k'}}-{\bf Q}_{{\bf P}_{k},{\bf P}'_{k'}}\right]}\!\right\}^{\!N}\!\!\! ,
\end{align}
with the new normalization
\begin{align}
\mathcal{N}=\prod_{k=0}^{k_f}\prod_{{\bf P}_{k}}\int_{-\infty}^{+\infty}d{w^{{\bf P}_{ k}}}e^{-\frac{{\bf w}{\bf Q}^{-1}{\bf w}}{2}}
\end{align}
and
\begin{align}
    {\bf w}=\left[w^{{\bf P}_0},w^{{\bf P}'_0},\dots,w^{{\bf P}_k},w^{{\bf P}'_k}\right]
\end{align}
This completes the computation of the free energy in the general setting. In the following section, we will present the simplifications resulting from the no-memory $Ansatz$.

\subsection{Simplification with the no-memory \emph{Ansatz}}
\label{app: no-mem computational steps}
The no-memory $Ansatz$ consists in supposing that a configuration ${\bf x}^{{\bf P}_k}$ only couples with its direct ancestor. This means that we have
\begin{align}
    {\bf Q}^{-1}_{{\bf P}_k,{{\bf P}'_{k'}}}&\neq 0 \quad \mbox{iff}\quad  {\bf P}'_{k'}\in\{{\bf P}_{k},{\bf P}^*_{k}\}\, \,(k'\leq k)\, ,\\
        \hat{\bf Q}_{{\bf P}_k,{{\bf P}'_{k'}}}&\neq 0 \quad \mbox{iff}\quad  {\bf P}'_{k'}\in\{{\bf P}_{k},{\bf P}^*_{k}\}\, \,(k'\leq k)\, .
\end{align}
with the condition that
\begin{align}
    {\bf Q}_{{\bf P}_k,{{\bf P}_{k}}}&={\rm I\!E}_{\xi,{\bf x}^{{\bf P}_k}}\left[{\bf x}^{{\bf P}_k}\cdot {\bf x}^{{\bf P}_{k}}/N\right]=1\, ,\\
    {\bf Q}_{{\bf P}_k,{{\bf P}^*_{k}}}&={\rm I\!E}_{\xi,{\bf x}^{{\bf P}_k},{\bf x}^{{\bf P}^*_k}}\left[{\bf x}^{{\bf P}_k}\cdot {\bf x}^{{\bf P}^*_{k}}/N\right]=m\,,
\end{align}
with the expectation
\begin{align}
    {\rm I\!E}_{\xi,{\bf x}^{{\bf P}_k},{\bf x}^{{\bf P}'_{k'}}}\left[f\left(\{\xi^\mu\},{\bf x}^{{\bf P}_k},{\bf x}^{{\bf P}'_{k'}}\right)\right]={\rm I\!E}_{\xi}\left[\frac{\prod_{j=0}^{k_f}\prod_{{\bf P}_{j}}\! \sum_{{\bf x}^{{\bf P}_{ j}}\in\Sigma^{N,m}_{{\bf x}^{{\bf P}^*_j}}}e^{-\mathcal{L}_{\rm SBP}\left({\bf x}^{{\bf P}_{ j}}\right)}f\left(\{\xi^\mu\},{\bf x}^{{\bf P}_k},{\bf x}^{{\bf P}'_{k'}}\right)}{\prod_{j'=0}^{k_f}\prod_{{\bf P}_{j'}}\! \sum_{\underline{\bf x}^{{\bf P}_{j'}}\in\Sigma^{N,m}_{\underline{\bf x}^{{\bf P}^*_{j'}}}}e^{-\mathcal{L}_{\rm SBP}\left(\underline{\bf x}^{{\bf P}_{ j}}\right)}}\right]\,.
\end{align}
In fact, it is the minimal number of interactions that can ensure the connectivity constraint ${\bf x}^{{\bf P }_k}\cdot{\bf x}^{{\bf P }^*_k}/N=m$. As detailed in \cite{Barbier2025}, this $Ansatz$ yields for the margins contribution
\begin{align}
\label{eq: iteration energ app}
    &\prod_{k=0}^{k_f}\prod_{{\bf P}_{k}}\int_{-\kappa}^{\kappa}d{w^{{\bf P}_{ k}}}\frac{e^{-\frac{{\bf w}{\bf Q}^{-1}{\bf w}}{2}}}{\mathcal{N}}=\left\{\int_{-\kappa}^{\kappa}dw_0\frac{e^{-\frac{\left({w_0}\right)^2}{2}}}{\sqrt{2\pi}}G_0^{\rm energ.}\left[w_0,m,\{y_l\}_{l\in[\![1,k_f]\!]}\right]\right\}^{y_0}
\end{align}
with
\begin{align}
    G_k^{\rm energ.}\!\left[w_k,m,\{y_l\}_{l\in[\![k+1,k_f]\!]}\right]\!=\left\{\int_{-\kappa}^\kappa \!\frac{dw_{k+1}e^{-\frac{(w_{k+1}-mw_k)^2}{2(1-m^2)}}}{\sqrt{2 \pi (1-m^2)}}G_{k+1}^{\rm energ.}\left[w_{k+1},m,\{y_l\}_{l\in[\![k+2,k_f]\!]}\right]\right\}^{y_{k+1}}
\end{align}
The technique to obtain this result is to recognize that the margins (without the loss) are correlated Gaussian random variables following the correlations from Eq.~(\ref{eq: mean margin}, \ref{eq: correlation margin}). They can be rewritten as a linear combination of Gaussian i.i.d. variables:
\begin{align}
    w^{{\bf P}_0}&\sim{\mathcal N}(0,1)\,,\\
    w^{{\bf P }_k}&=mw^{{\bf P }^*_k}+\sqrt{1-m^2}u^{{\bf P }_k}\,\,\, {\rm with}\,\,\, u^{{\bf P }_k}\sim{\mathcal N}(0,1).
\end{align}
This allows us to write
\begin{align}
&\prod_{k=0}^{k_f}\prod_{{\bf P}_{k}}\int_{-\kappa}^{\kappa}d{w^{{\bf P}_{ k}}}\frac{e^{-\frac{{\bf w}{\bf Q}^{-1}{\bf w}}{2}}}{\mathcal{N}}=\prod_{{\bf P}_0}\int_{-\kappa}^{\kappa}dw^{{\bf P}_0}\frac{e^{-\frac{\left({w^{{\bf P}_0}}\right)^2}{2}}}{\sqrt{2\pi}}\prod_{k=1}^{k_f}\!\prod_{{\bf P}_{k}}\int\mathcal{D}{u^{{\bf P}_{ k}}}\,\Theta\left(\kappa-\left\vert w^{{\bf P}_k}\right\vert\right)\,.
\end{align}
Due to the no-memory $Ansatz$, the partition function corresponds to a summation on a Bethe tree. Starting with the last level $k=k_f$, all contributions from the paths ${\bf P}_{k_f}$ can be factorized. For each direct ancestor ${\bf P}^*_{k_f}$ there are exactly $y_{k_f}$ non-interacting descendants. This means that we have 
\begin{align}
\prod_{k=0}^{k_f}\prod_{{\bf P}_{k}}\int_{-\kappa}^{\kappa}d{w^{{\bf P}_{ k}}}\frac{e^{-\frac{{\bf w}{\bf Q}^{-1}{\bf w}}{2}}}{\mathcal{N}}=&\prod_{{\bf P}_0}\int_{-\kappa}^{\kappa}dw^{{\bf P}_0}\frac{e^{-\frac{\left({w^{{\bf P}_0}}\right)^2}{2}}}{\sqrt{2\pi}}\prod_{k=1}^{k_f}\!\prod_{{\bf P}_{k}}\int\mathcal{D}{u^{{\bf P}_{ k}}}\,\Theta\left(\kappa-\left\vert w^{{\bf P}_k}\right\vert\right)\\
&\times\!\!\prod_{{\bf P}_{k_f-1}}\!\!\int\!\!\nonumber\mathcal{D}u^{{\bf P}_{k_f-1}}\Theta\left(\kappa\!-\!\left\vert w^{{\bf P}_{k-1}}\right\vert\right)G^{\rm energ.}_{k_f-1}\!\!\left[w^{{\bf P}_{k-1}}\!,m,y_{k_f}\right]
\end{align}
with
\begin{align}
    G^{\rm energ.}_{k_f-1}\!\left[w^{{\bf P}_{k-1}}\!,m,y_{k_f}\right]&\!\!=\!\left[\int \!\mathcal{D}u_{k_f}\Theta\!\left(\kappa\!-\!\vert w_{k_f} \vert\right)\!\right]^{\!y_{k_f}}\hspace{-0.2cm} ,\\
    w_{k_f}&\!\!=mw^{{\bf P}_{k-1}}\!+\!\sqrt{1\!-\!m^2}\,u_{k_f}\,.
\end{align}
These computation steps can be iterated from the level $k=k_f-1$ to $k=1$. This gives
\begin{align}
&\prod_{k=0}^{k_f}\prod_{{\bf P}_{k}}\int_{-\kappa}^{\kappa}d{w^{{\bf P}_{ k}}}\frac{e^{-\frac{{\bf w}{\bf Q}^{-1}{\bf w}}{2}}}{\mathcal{N}}=\left\{\int_{-\kappa}^{\kappa}dw_0\frac{e^{-\frac{\left(w_0\right)^2}{2}}}{\sqrt{2\pi}}
G^{\rm energ.}_{0}\left[w_0,m,\{y_{l}\}_{l\in[\![1,\dots,k_f]\!]}\right]\right\}^{y_0}
\end{align}
with
\begin{align}
    G^{\rm energ.}_{k}\!\left[w_k,m,\{y_{l}\}_{l\in[\![k+1,\dots,k_f]\!]}\right]\!=\!\Bigg\{\!\int \!\mathcal{D}u_{k+1}\Theta\left(\kappa\!-\!\vert w_{k+1} \vert\right) G^{\rm energ.}_{k+1}\!\left[w_{k+1},m,\{y_{l}\}_{l\in[\![k+2,\dots,k_f]\!]}\right]\!\Bigg\}^{\!y_{k_f}}\hspace{-0.2cm} \, ,
\end{align} 
and
\begin{align}
    w_{k}&=mw_{k-1}+\sqrt{1-m^2}\,u_{k}\,.
\end{align}

Finally, by reversing the change of variable $u_k=(w_k-mw_{k-1})/\sqrt{1-m^2}$, we obtain Eq.~(\ref{eq: iteration energ app}).

The hypercube contribution verifies the same Bethe tree geometry. In fact, we can start  by integrating the configurations at level $k_f$ and obtain
\begin{align}
   &\prod_{k=0}^{k_f}\prod_{{\bf P}_{k}} \sum_{{x}^{{\bf P}_{ k}}=\pm 1}e^{\underset{k,{k'}}{\sum}\,\underset{{\bf P}_k,{\bf P}'_{k'}}{\sum}\!\!\hat{\bf Q}_{{\bf P}_{k},{\bf P}'_{k'}}{x}^{{\bf P}_k}{ x}^{{\bf P}'_{k'}}}=\prod_{k=0}^{k_f-2}\!\prod_{{\bf P}_{k}}\sum_{{x}^{{\bf P}_{ k}}=\pm 1}e^{\overset{k_f-2}{\underset{k=1}{\sum}}\,\underset{{\bf P}_k}{\sum}\,\hat{\bf Q}_{{\bf P}_{k},{\bf P}^*_{k}}{x}^{{\bf P}_k}{ x}^{{\bf P}^*_{k}}}\\
   &\hspace{1cm}\times\!\prod_{{\bf P}_{k_f-1}}\sum_{{x}^{{\bf P}_{ k_f-1}}=\pm 1}e^{\underset{{\bf P}_{k_f-1}}{\sum}\hat{\bf Q}_{{\bf P}_{k_f-1},{\bf P}^*_{k_f-1}}{x}^{{\bf P}_{k_f-1}}{ x}^{{\bf P}^*_{k_f-1}}} G^{\rm entr.}_{k_f-1}\left[x^{{\bf P}_{k_f-1}},\hat{\bf Q}_{{\bf P}_{k_f},{\bf P}^*_{k_f}},y_{k_f}\right]\nonumber
\end{align}
with
\begin{align}
    &G^{\rm entr.}_{k_f-1}\left[x^{{\bf P}_{k_f-1}},\hat{\bf Q}_{{\bf P}_{k_f},{\bf P}^*_{k_f}},y_{k_f}\right]\!=\!\left\{\sum_{x_{k_f}=\pm 1}e^{\hat{\bf Q}_{{\bf P}_{k_f},{\bf P}^*_{k_f}}x_{k_f}x^{{\bf P}_{k_f-1}}}\right\}^{y_{k_f}}\!\!\!=\left\{2{\rm cosh}\left[\hat{\bf Q}_{{\bf P}_{k_f},{\bf P}^*_{k_f}}\right]\right\}^{y_{k_f}}\, .
\end{align}
We specify that, due to the symmetry between the paths ${\bf P}_{k}$, the magnetic fields now depend only on the layer over which they apply: $\hat{\bf Q}_{{\bf P}_{k},{\bf P}^*_{k}}\rightarrow \hat{\bf Q}_{{k},{k-1}}$ .
The rest of the layers can be integrated iteratively from the level $k=k_f-1$ to $k=1$ following the same computation steps, and in the end it yields
\begin{align}
       &\prod_{k=0}^{k_f}\prod_{{\bf P}_{k}}\sum_{{x}^{{\bf P}_{ k}}=\pm 1}e^{\underset{k,{k'}}{\sum}\,\underset{{\bf P}_k,{\bf P}'_{k'}}{\sum}\!\!\hat{\bf Q}_{{\bf P}_{k},{\bf P}'_{k'}}{x}^{{\bf P}_k}{ x}^{{\bf P}'_{k'}}}=2^{y_0}\prod_{k=1}^{k_f}\left[2{\rm cosh}\left(\hat{\bf Q}_{{k},{k-1}}\right)\right]^{\overset{\hspace{-0.26cm}k}{\underset{l=0}{\prod}y_l}}\,.
\end{align}

Putting together the contributions from the margins and from the hypercube we get
\begin{align}
   \lim_{N\rightarrow+\infty} {\rm I\!E}_\xi[\mathcal{Z}^{y_0}]\sim&\underset{\hat{\bf Q}_{{k},{k-1}}}{\rm opt}\left\{\tilde{\mathcal Z} \right\}
\end{align}
with
\begin{align}
    &\tilde{\mathcal Z}\!=\!2^{y_0}e^{-\overset{k_f}{\underset{k=1}{\sum}} \left[\left(\overset{\hspace{-0.26cm}k}{\underset{l=0}{\prod}y_l}\right)\hat{\bf Q}_{{k},{k-1}}m\right]}\prod_{k=1}^{k_f}\left[2{\rm cosh}\left(\hat{\bf Q}_{{k},{k-1}}\right)\right]^{\overset{\hspace{-0.26cm}k}{\underset{l=0}{\prod}y_l}}\\
    &\hspace{0.5cm}\times\left\{\int_{-\kappa}^{\kappa}dw_0\frac{e^{-\frac{\left(w_0\right)^2}{2}}}{\sqrt{2\pi}}
G^{\rm energ.}_{0}\left[w_0,m,\{y_{l}\}_{l\in[\![1,\dots,k_f]\!]}\right]\right\}^{y_0}\nonumber
\end{align}
Finally, the optimization over the magnetic fields is a standard result from statistical physics:
\begin{align}
    \hat{\bf Q}_{{k},{k-1}}={\rm arctanh}(m)\, ,\quad \forall k\in[\![1,k_f]\!]\,.
\end{align}
With this we obtain the very last simplification
\begin{align}
    \hspace{-0.2cm}-m\hat{\bf Q}_{{k},{k-1}}\!+\!\log\!\left[2\cosh\left(\hat{\bf Q}_{{k},{k-1}}\right)\right]\!=\!\tilde{\phi}^{\rm ent.}(m)
\end{align}
with $\tilde{\phi}^{\rm ent.}(\cdot)$ defined in Eq.~(\ref{eq: ent binary}).

\section{Computing the local entropy $s_{\rm loc}$ and the edge entropy $s_{{\bf x}_0}$}
\label{app: entropy loc}
To compute the local entropy, we recall that we use the definition
\begin{align}
    s_{\rm loc}(k)= \frac{\left(y_k\partial_{y_k}-y_{k+1}\partial_{y_{k+1}}\right)\phi}{\prod_{j=1}^k y_j}
\end{align}
If we first focus on the margins contribution of  $\phi$, we can see that (we drop the dependencies in $G^{\rm energ.}_k$ for compact notations)
\begin{align}
  \partial_{y_{k}} \log\left\{\int_{-\kappa}^\kappa dw_0\,\frac{e^{\frac{-w_0^2}{2}}}{\sqrt{2\pi}}\,G_0^{\rm energ.}\right\}&=\frac{\int_{-\kappa}^\kappa dw_0\,\frac{e^{\frac{-w_0^2}{2}}}{\sqrt{2\pi}}\,\partial_{y_k}G_0^{\rm energ.}}{\int_{-\kappa}^\kappa d\underline{w}_0\,\frac{e^{\frac{-\underline{w}_0^2}{2}}}{\sqrt{2\pi}}\,G_0^{\rm energ.}}\\
  &=y_1\frac{\int_{-\kappa}^\kappa dw_0\,\frac{e^{\frac{-w_0^2}{2}}}{\sqrt{2\pi}}\,G_0^{\rm energ.}}{\int_{-\kappa}^\kappa d\underline{w}_0\,\frac{e^{\frac{-\underline{w}_0^2}{2}}}{\sqrt{2\pi}}\,G_0^{\rm energ.}}\!\times\!\frac{\int_{-\kappa}^\kappa dw_1\, \frac{e^{\frac{-(w_1-mw_0)^2}{2(1-m^2)}}}{\sqrt{2\pi(1-m^2)}}\partial_{y_k}G_1^{\rm energ.}}{\int_{-\kappa}^\kappa d\underline{w}_1\, \frac{e^{\frac{-(\underline{w}_1-mw_0)^2}{2(1-m^2)}}}{\sqrt{2\pi(1-m^2)}}G_1^{\rm energ.}}\nonumber\\
  &=y_{k'}\int dw_0\,P^{\rm edge}(w_0)\prod_{k'=1}^{k-1}\left[y_k' \int dw_{k'}\,\tilde{T}_{k'}(w_{k'},w_{k'-1})\right]\nonumber\\
  &\hspace{1.5cm}\times\log \left[\int_{-\kappa}^\kappa dw_k\,\frac{e^{\frac{-(w_k-mw_{k-1})^2}{2(1-m^2)}}}{\sqrt{2\pi(1-m^2)}} G_{k}^{\rm energ.}\right]\nonumber
\end{align}
with
\begin{align}
    P^{\rm edge}(w_0)=\frac{e^{\frac{-w_0^2}{2}}\Theta(\kappa-\vert w_0\vert)\,G_0^{\rm energ.}}{\int_{-\kappa}^\kappa d\underline{w}_0\,{e^{\frac{-\underline{w}_0^2}{2}}}\,G_0^{\rm energ.}}
\end{align}
and 
\begin{align}
\hspace{-0.3cm}\tilde{T}_j(w_{j},w_{j-1})=\frac{ {e^{\frac{-(w_{j}-mw_{j-1})^2}{2(1-m^2)}}\Theta(\kappa-\vert w_{j}\vert)}\,G_j^{\rm energ.}}{\int_{-\kappa}^\kappa d\underline{w}_j\, {e^{\frac{-(\underline{w}_{j}-mw_{j-1})^2}{2(1-m^2)}}}\,G_j^{\rm energ.}} .
\end{align}
We can recognize here the no-memory chain process introduced in \cite{Barbier2025}. The diffusion is governed by the sequence of propagators $\{\tilde{T}_j\}_{j=[\![1,k-1]\!]}$, while $P^{\rm edge}(w_0)$ is the distribution of interactions for the initial point in the chain. Therefore, if we propagate the distribution $P^{\rm edge}(w_0)$ from $w_0$ to $w_{k-1}$, we obtain
\begin{align}
    \partial_{y_{k}} \log\left\{\int_{-\kappa}^\kappa dw_0\,\frac{e^{\frac{-w_0^2}{2}}}{\sqrt{2\pi}}\,G_0^{\rm energ.}\right\}&=\prod_{k'=1}^{k-1}y_{k'}\int dw_{k-1}P^{k-1}(w_{k-1})\\
  &\hspace{1cm}\times\log \left\{\int_{-\kappa}^\kappa dw_k\,\frac{e^{\frac{-(w_k-mw_{k-1})^2}{2(1-m^2)}}}{\sqrt{2\pi(1-m^2)}} G_{k}^{\rm energ.}\right\}\nonumber\, .
\end{align}
With this the local entropy becomes 
\begin{align}
    s_{\rm loc}(k)=& \frac{\left(y_{k}\partial_{y_{k}}-{y_{k+1}\partial_{y_{k+1}}}\right)\phi}{\prod_{j=1}^k y_j}\\
    =& N\tilde{\phi}^{\rm ent.}(m)+M\int dw_{k-1}P^{k-1}(w_{k-1})\log \left\{\int_{-\kappa}^\kappa dw_k\,\frac{e^{\frac{-(w_k-mw_{k-1})^2}{2(1-m^2)}}}{\sqrt{2\pi(1-m^2)}} G_{k}^{\rm energ.}\right\}\nonumber\\
    &-y_{k+1}M\int dw_{k}P^{k}(w_{k})\log \left\{\int_{-\kappa}^\kappa dw_{k+1}\,\frac{e^{\frac{-(w_{k+1}-mw_{k})^2}{2(1-m^2)}}}{\sqrt{2\pi(1-m^2)}} G_{k+1}^{\rm energ.}\right\}\nonumber
\end{align}
with the propagated distribution
\begin{align}
    P^{k}(w_{k})=\int dw_0\,P^{\rm edge}(w_0)\prod_{k'=1}^k\left[\int dw_{k'}\,\tilde{T}_{k'}(w_{k'},w_{k'-1})\right].
\end{align}
In the special case of large cliques (${k}_f\rightarrow+\infty$) and all Lagrange multipliers equal to one, we mentioned in Sec.~\ref{sec: deloc cluster} that the obtain the steady state
\begin{align}
     G_{k}^{\rm energ.}\propto G_{\lambda^{\rm top}}^{\rm energ.}\,.
\end{align}
and
\begin{align}
    \frac{G_{k}^{\rm energ.}}{G_{k+1}^{\rm energ.}}=\lambda
\end{align}
Thus, all propagators $\{\tilde{T}_j\}_{j=[\![1,k-1]\!]}$ are equal. In the core ($1\ll k\ll k_f$), all minima are equivalent and $P^{k}(w_k)=P^{{\rm core}}(w_k)$ is exactly the steady state distribution of this diffusion process. From Eq.~(\ref{eq: iteration energ no-mem y=1}), it follows that we have in this particular situation
\begin{align}
    s_{\rm loc}(k)=& \frac{\left(y_{k}\partial_{y_{k}}-{y_{k+1}\partial_{y_{k+1}}}\right)\phi}{\prod_{j=1}^k y_j}\\
    =& N\tilde{\phi}^{\rm ent.}(m)+M\int dw_{k-1}P^{\rm core}(w_{k-1})\log \left\{\lambda\int_{-\kappa}^\kappa dw_k\,\frac{e^{\frac{-(w_k-mw_{k-1})^2}{2(1-m^2)}}}{\sqrt{2\pi(1-m^2)}} G_{k+1}^{\rm energ.}\right\}\nonumber\\
    &-M\int dw_{k}P^{\rm core}(w_{k})\log \left\{\int_{-\kappa}^\kappa dw_{k+1}\,\frac{e^{\frac{-(w_{k+1}-mw_{k})^2}{2(1-m^2)}}}{\sqrt{2\pi(1-m^2)}} G_{k+1}^{\rm energ.}\right\}\nonumber\\
    =&N\left[\tilde{\phi}^{\rm ent.}(m)+\alpha\log(\lambda)\right]\nonumber
\end{align}
where $\lambda$ is defined in Eq.~(\ref{eq: iteration energ no-mem y=1}).

Finally, to compute the edge entropy, we use the definition
\begin{align}
 s_{{\bf x}_0}^{\rm no-memory}=\left(1-{y_{1}\partial_{y_{1}}}\right)\phi
    \nonumber\, .
\end{align}
This computation is simpler than the local entropy evaluation, it yields
\begin{align}
     s_{{\bf x}_0}=&N\log(2)+M\log\left\{\int_{-\kappa}^\kappa dw_0\,\frac{e^{\frac{-w_0^2}{2}}}{\sqrt{2\pi}}\,G_{0}^{\rm energ.}\right\}\\
     &-y_1M\!\int\! dw_{0}P^{\rm edge}(w_0)\log \left\{\!\int_{-\kappa}^\kappa \!dw_{1}\!\frac{e^{\frac{-(w_{1}-mw_{0})^2}{2(1-m^2)}}G_{1}^{\rm energ.}}{\sqrt{2\pi(1-m^2)}} \!\right\}\nonumber\\
     \hspace{+1.17cm}\underset{m\rightarrow 1}{=}&N\log(2)+M\log\left\{\int_{-\kappa}^\kappa dw_0\,\frac{e^{\frac{-w_0^2}{2}}}{\sqrt{2\pi}}\,G_{0}^{\rm energ.}\right\}\nonumber\\
     &-y_1M\int dw_{0}P^{\rm edge}(w_0)\log \left\{ G_{1}^{\rm energ.}\right\}\nonumber\, .
\end{align}
Again, if we take the special case of large cliques with all $y$'s equal to one we obtain
\begin{align}
     s_{{\bf x}_0}&\hspace{-0.05cm}\underset{m\rightarrow 1}{=}N\log(2)\!+\!M\!\log\left\{\int_{-\kappa}^\kappa \!dw_0\,\frac{e^{\frac{-w_0^2}{2}}}{\sqrt{2\pi}}\,G_{\lambda^{\rm top}}^{\rm energ.}\right\}-M\int dw_{0}P^{{\rm edge}}(w_{0})\log \left\{ G_{\lambda^{\rm top}}^{\rm energ.}\right\}\, .
\end{align}

\section{Computation steps for the stability criteria}
\label{app: stab Ansatz}

\subsection{Global stability}
\label{app: annealed stab Ansatz}
In this section, we develop the computation steps to obtain the global stability criterion introduced in Sec.~\ref{sec: stab} as
\begin{align}
    \frac{\delta\phi}{\delta {\bf Q}_{{\bf P}_{k},{{\bf P}'_{k'}}}}\, , \quad\forall\, {\bf P}_{k},{\bf P}'_{k'}\,.
\end{align}
For this, we will focus on the case ${\bf P}'_{k'}\neq {\bf P}^*_{k}$ ($k'<k$). Following Eq.~(\ref{eq: free energ true}), we have
\begin{align}
 &\left.   \frac{\delta\phi}{\delta {\bf Q}_{{\bf P}_{k},{{\bf P}'_{k'}}}}\right\vert_{\underset{{\bf P}'_{k'}\neq {\bf P}^*_{k}}{\rm no-mem.}}\!\!=-\frac{M}{2}\!\sum_{k_1,k_2}\,\sum_{{\bf P}_{k_1},{{\bf P}_{k_2}}}\!\!\!\frac{\delta{\bf Q}^{-1}_{{\bf P}_{k_1},{{\bf P}_{k_2}}}}{\delta{\bf Q}_{{\bf P}_{k},{{\bf P}'_{k'}}}}\left[\left\langle w^{{\bf P}_{k_1}}w^{{\bf P}_{k_2}}\right\rangle_\kappa-\left\langle w^{{\bf P}_{k_1}}w^{{\bf P}_{k_2}}\right\rangle_{\rm free}\right]
\end{align}
with the measures
\begin{align}
    \langle A\rangle_k=\frac{\prod_{k=0}^{k_f}\prod_{{\bf P}_k} \int_{-\kappa}^\kappa\! d{{ w}^{{\bf P}_k}}A e^{\frac{-\underset{{k,\,{k'}}}{\sum}\,\underset{{{\bf P}_k,\,{\bf P}'_{k'}}}{\sum}{\bf Q}^{-1}_{{\bf P}_k,{{\bf P}'_{k'}}}w^{{\bf P}_k}w^{{\bf P}'_{k'}}}{2}}}{\prod_{k=0}^{k_f}\prod_{{\bf P}_k} \int_{-\kappa}^\kappa\! d{{ w}^{{\bf P}_k}}e^{\frac{-\underset{{k,\,{k'}}}{\sum}\,\underset{{{\bf P}_k,\,{\bf P}'_{k'}}}{\sum}{\bf Q}^{-1}_{{\bf P}_k,{{\bf P}'_{k'}}}w^{{\bf P}_k}w^{{\bf P}'_{k'}}}{2}}}
\end{align}
and
\begin{align}
    \langle A\rangle_{\rm free}=\frac{\prod_{k=0}^{k_f}\prod_{{\bf P}_k} \int\! d{{ w}^{{\bf P}_k}}A e^{\frac{-\underset{{k,\,{k'}}}{\sum}\,\underset{{{\bf P}_k,\,{\bf P}'_{k'}}}{\sum}{\bf Q}^{-1}_{{\bf P}_k,{{\bf P}'_{k'}}}w^{{\bf P}_k}w^{{\bf P}'_{k'}}}{2}}}{\prod_{k=0}^{k_f}\prod_{{\bf P}_k} \int\! d{{ w}^{{\bf P}_k}}e^{\frac{-\underset{{k,\,{k'}}}{\sum}\,\underset{{{\bf P}_k,\,{\bf P}'_{k'}}}{\sum}{\bf Q}^{-1}_{{\bf P}_k,{{\bf P}'_{k'}}}w^{{\bf P}_k}w^{{\bf P}'_{k'}}}{2}}}\,.
\end{align}
With the no-memory $Ansatz$ and $y_k=1$ for $k\in[\![1,k_f]\!]$, this further simplifies as
\begin{align}
     \left.   \frac{\delta\phi}{\delta {\bf Q}_{{\bf P}_{k},{{\bf P}'_{k'}}}}\right\vert_{\underset{{\bf P}'_{k'}\neq {\bf P}^*_{k}}{\rm no-mem.}}\!\!=-\frac{MQ^{-2}}{2}\Bigg\{&\left[\left\langle w^{{\bf P}^*_{k}}w^{{\bf P}_{k}}\right\rangle_\kappa-\left\langle w^{{\bf P}{'^*}_{k'}}w^{{\bf P}'_{k'}}\right\rangle_{\rm free}\right]\\
    &+\left[\left\langle w^{{\bf P}^*_{k}}w^{{\bf P}_{k}}\right\rangle_\kappa-\left\langle w^{{\bf P}_{k'+1}}w^{{\bf P}'_{k'}}\right\rangle_{\rm free}\right]\Big\vert_{{\bf P}'_{k'}={\bf P}^*_{k'+1}}\nonumber\\
     &+\left[\left\langle w^{{\bf P}_{k+1}}w^{{\bf P}_{k}}\right\rangle_\kappa-\left\langle w^{{\bf P}^*_{k'}}w^{{\bf P}'_{k'}}\right\rangle_{\rm free}\right]\Big\vert_{{\bf P}_{k}={\bf P}^*_{k+1}}\nonumber\\
     &+\left[\left\langle w^{{\bf P}_{k+1}}w^{{\bf P}_{k}}\right\rangle_\kappa-\left\langle w^{{\bf P}'_{k'+1}}w^{{\bf P}'_{k'}}\right\rangle_{\rm free}\right]\Big\vert_{\underset{{\bf P}'_{k'}={\bf P}^*_{k'+1}}{{\bf P}_{k}={\bf P}^*_{k+1}}}\Bigg\}\nonumber
\end{align}
where we used the formula
\begin{align}
    \frac{\delta{\bf Q}^{-1}_{{\bf P}_{k_1},{{\bf P}_{k_2}}}}{\delta{\bf Q}_{{\bf P}_{k},{{\bf P}'_{k'}}}}=-{\bf Q}^{-1}_{{\bf P}_{k_1},{{\bf P}_{k}}}{\bf Q}^{-1}_{{\bf P}'_{k'},{{\bf P}_{k_2}}}
\end{align}
and (for all ${\bf P}_{k}$)
\begin{align}
    {\bf Q}^{-1}_{{\bf P}^*_{k},{{\bf P}_{k}}}&=Q^{-1}\\
    &=\nonumber-\frac{2m}{1-m^2}\,.
\end{align}
Remarking that we have trivially $\left\langle w^{{\bf P}_{k_1}}w^{{\bf P}_{k_2}}\right\rangle_\kappa-\left\langle w^{{\bf P}_{k_1}}w^{{\bf P}_{k_2}}\right\rangle_{\rm free}<0$ (for all ${\bf P}_{k_1}$ and ${\bf P}_{k_2}$), the derivative of the free energy with respect to the overlap matrix is never null around the no-memory $Ansatz$. Thus, this form of ${\bf Q}$ is never a saddle-point.

\subsection{Local stability}
\label{app: quenched stab Ansatz}
In this section we characterize the local stability of the no-memory cluster. For this, we want to compute the stability of the local entropy $s_{\rm loc}(k)$ when we allow ${\bf x}_k$ to re-correlate with ${\bf x}_{k'}$. More practically, we want to determine numerically (for $y_k=1$ for $k\in[\![1,k_f]\!]$)
\begin{align}
    s_{\rm loc}^{\rm pert.}(k)&=(y_k\partial_{y_k}-y_{k+1}\partial_{y_{k+1}})\phi^{\rm pert.}\\
    &=N\sum_{x_0=\pm 1}\frac{1}{2}\prod_{j=1}^{k-1}\left[\sum_{x_j=\pm 1}\frac{e^{\underline{h}x_jx_{j-1}}}{2\mbox{cosh}(\underline{h})}\right]\log\left[\sum_{x_k=\pm 1}e^{h(k,k-1)[x_kx_{k-1}-m]+h(k,k')[x_k x_{k'}-m(k,k')]}\right]\nonumber\\
    &\hspace{-1.3cm}+M\int dw_0 P^{\rm edge}(w_0)\prod_{j=1}^{k-1}\left[\int dw_j \,\tilde{T}_j(w_j,w_{j-1})\right]\!\log\!\left[\!\int_{-\kappa}^\kappa \!\!dw_{k}\frac{e^{\frac{-\left[w_{k}-mw_{k-1}-\mathcal{C}\left(w_{k'}-m^{k-k'-1}w_{k-1}\right)\right]^2}{2\tilde{\mathcal{C}}}}}{\sqrt{2\pi\tilde{\mathcal{C}}}}G_{k}^{\rm energ.}\right]\nonumber
\end{align}
with $\underline{h}$ the magnetic field given by the no-memory $Ansatz$, i.e. $\underline{h}=\mbox{arctanh}(m)$. 
In addition, we used the fact that the decomposition for $w_k$ in the perturbed case is the linear combination
\begin{align}
    w_k=&mw_{k-1}+\frac{m(k,k')-m^{k-k'}}{1-m^{2(k-k'-1)}}\left(w_{k'}-m^{k-k'-1}w_{k-1}\right)+\sqrt{1-m^2-\frac{[m(k,k')-m^{k-k'}]^2}{1-m^{2(k-k'-1)}}}u_k\nonumber\\
    =&mw_{k-1}+\mathcal{C}\left(w_{k'}-m^{k-k'-1}w_{k-1}\right)+\sqrt{\tilde{\mathcal{C}}}\,u_k
\end{align}
with $u_k\sim\mathcal{N}(0,1)$.
This choice ensures ${\rm I\!E}_\xi[w_k w_k]=1$, ${\rm I\!E}_\xi[w_k w_{k-1}]=m$ and ${\rm I\!E}_\xi[w_k w_{k'}]=m(k,k')$.
We directly see here a Markov process with a feedback interaction between $w_k$ and $w_{k'}$.
Again, we recognize a Markov-chain process from $x_0/w_0$ to $x_{k'-1}/w_{k'-1}$. In the case where $1\ll k\ll k_f$ and $k/k'\sim\mathcal{O}(1)$, this process collapses into the core of the cluster -where the margin distribution is $P^{\rm core}(\cdot)$ everywhere-. Thus, we can further simplify the previous expression in this case as 
\begin{align}
     s_{\rm loc}^{\rm pert.}(k)&=N\sum_{x_{k'}=\pm 1}\frac{1}{2}\sum_{x_{k-1}=\pm 1}\frac{e^{\tilde{h}x_{k-1}x_{k'}}}{2\mbox{cosh}(\tilde{h})}\log\left[\sum_{x_k=\pm 1}e^{h(k,k-1)[x_kx_{k-1}-m]+h(k,k')[x_k x_{k'}-m(k,k')]}\right]\nonumber\\
     &\hspace{0.25cm}+M\int dw_{k'} P^{{\rm core}}(w_{k'})\prod_{j=k'+1}^{k-1}\left[\int dw_j \,\tilde{T}_j(w_j,w_{j-1})\right]\\
     &\hspace{1cm}\times\log\!\left[\!\int_{-\kappa}^\kappa \!\!dw_{k}\frac{e^{\frac{-\left[w_{k}-mw_{k-1}-\mathcal{C}\left(w_{k'}-m^{k-k'-1}w_{k-1}\right)\right]^2}{2\tilde{\mathcal{C}}}}}{\sqrt{2\pi\tilde{\mathcal{C}}}}G_{\lambda^{\rm top}}^{\rm energ.}\right]\nonumber\\
     &=N\sum_{x_{k'}=\pm 1}\frac{1}{2}\sum_{x_{k-1}=\pm 1}\frac{e^{\tilde{h}x_{k-1}x_{k'}}}{2\mbox{cosh}(\tilde{h})}\log\left[\sum_{x_k=\pm 1}e^{h(k,k-1)[x_kx_{k-1}-m]+h(k,k')[x_k x_{k'}-m(k,k')]}\right]\nonumber\\
     &\hspace{0.35cm}+M\,\mbox{Tr}\left[\tilde{T}^{\vert k-k'\vert-2}\,T^{\rm edge}_{\rm loc.}\right]\nonumber
\end{align}
with $\tilde{h}=\mbox{arctanh}(m^{\vert k-k'\vert -1})$ and 
\begin{align}
       T^{\rm edge}_{\rm loc.}[w_{k-1},w_{k'}]=&P^{{\rm core}}(w_{k'})\log\!\left[\!\int_{-\kappa}^\kappa \!dw_{k}\frac{e^{\frac{-\left[w_{k}-mw_{k-1}-\mathcal{C}\left(w_{k'}-m^{k-k'-1}w_{k-1}\right)\right]^2}{2\tilde{\mathcal{C}}}}}{\sqrt{2\pi\tilde{\mathcal{C}}}}\,G_{\lambda^{\rm top}}^{\rm energ.}\!\right] .
\end{align}

As with the global stability, the perturbation is translationally invariant, in other words it depends only on the distance $\vert k-k'\vert$. Seating on the configuration ${\bf x}_{k-1}$, the perturbed local entropy represents the number of solutions locally available for re-correlating  with ${\bf x}_{k'}$. If this number becomes greater than the unperturbed local entropy, algorithms start to be entropically attracted by configurations they explored at previous times. To write it more clearly this destabilization appears for
\begin{align}
    s_{\rm loc}^{\rm pert.}(k)-s_{\rm loc}^{\rm no-mem.}(k)>0\,.
\end{align}

\section{Franz-Parisi computations}
\subsection{The case of connected minima (with no-memory geometry)}
\label{app: FP connected minima}
In this section, we will calculate the typical number of solutions $\bf x$ located at a Hamming distance $N(1-m)/2$ from a given reference vector ${\bf x}_0$. In particular, we will consider that both system -$\bf x$ and ${\bf x}_0$- have the same distribution of margins, and more particularly:
\begin{align}
\label{eq: marg FP}
    P_{{\bf x}_0}(w)=P_{{\bf x}}(w)=P^{{\rm edge}}(w)\, .
\end{align}
This computation will help us highlighting when no-memory connected minima are isolated, using more traditional statistical mechanics tools. More practically, we will determine if a so-called frozen 1-RSB structure emerges with these solutions -also known as an overlap-gap property (OGP) \cite{gamarnik2021survey}-.
We will see that this approach underestimates this ''isolated'' regime (in terms of the range of parameters $\alpha$ and $\kappa$) compared to the local stability criterion from Sec.~\ref{sec: stab}. 

This typical number of solutions is given by the Franz-Parisi potential \cite{franz1995recipes}
\begin{align}
\label{eq: FP connected manifold}
    \phi^{\rm FP}\!(m)\!=\!{\rm I\!E}_\xi\!\!\left[\!\sum_{{\bf x}_0\in\Sigma^N}\!\!\!\frac{e^{-\mathcal{L}^{\rm eff.}_{\rm SBP}\left({\bf x}_0\right)}}{\mathcal{Z}}\log\!\left(\sum_{{\bf x}\in\Sigma_{{\bf x}_0}^{N,m}}\hspace{-0.3cm}e^{-\underline{\mathcal{L}}({\bf x})}\!\right)\!\!\right]
\end{align}
where $\mathcal{L}^{\rm eff.}_{\rm SBP}$ is defined in Eq.~(\ref{eq: eff loss deloc}), $\mathcal{Z}$ is the normalization
\begin{align}
    \mathcal{Z}=\sum_{{\bf x}_0\in\Sigma^N}{e^{-\mathcal{L}^{\rm eff.}_{\rm SBP}\left({\bf x}_0\right)}}
\end{align}
and $\underline{\mathcal{L}}(\cdot)$ is tuned to ensure the correct margins distributions for $\bf x$, see Eq.~(\ref{eq: marg FP}). Using again replica to evaluate this potential, we consider an $annealed$ symmetry, i.e.
\begin{align}
    {\rm I\!E}_{\xi,{\bf x}_0^a,{\bf x}_0^b}\left[\frac{{\bf x}_0^a\cdot {\bf x}_0^b}{N}\right]&={\rm I\!E}_{\xi}\left[\sum_{{\bf x}_0^a,{\bf x}_0^b\in\Sigma^N}\!\!\!\frac{e^{-\mathcal{L}^{\rm eff.}_{\rm SBP}\left({\bf x}_0^a\right)-\mathcal{L}^{\rm eff.}_{\rm SBP}\left({\bf x}_0^b\right)}{\bf x}_0^a\cdot{\bf x}_0^b}{N\mathcal{Z}^2}\right]=\delta_{a,b}\,,\\
    {\rm I\!E}_{\xi,{\bf x}^a,{\bf x}^b}\left[\frac{{\bf x}^a\cdot {\bf x}^b}{N}\right]&={\rm I\!E}_{\xi}\left[\sum_{{\bf x}_0\in\Sigma^N}\!\!\!\frac{e^{-\mathcal{L}^{\rm eff.}_{\rm SBP}\left({\bf x}_0\right)}}{\mathcal{Z}}\times\frac{\sum_{{\bf x^a},{\bf x^b}\in\Sigma_{{\bf x}_0}^{N,m}}e^{-\underline{\mathcal{L}}({\bf x}^a)-\underline{\mathcal{L}}({\bf x}^b)}{\bf x}^a\cdot{\bf x}^b}{N\left(\sum_{\underline{\bf x}\in\Sigma_{{\bf x}_0}^{N,m}}e^{-\underline{\mathcal{L}}(\underline{\bf x})}\right)^2}\right]\\
    &=m^2+(1-m^2)\delta_{a,b}\nonumber
\end{align}
and
\begin{align}
    {\rm I\!E}_{\xi,{\bf x}^a,{\bf x}_0^b}\left[\frac{{\bf x}^a\cdot {\bf x}_0^b}{N}\right]={\rm I\!E}_{\xi}\left[\sum_{{\bf x}_0^b\in\Sigma^N}\!\!\!\frac{e^{-\mathcal{L}^{\rm eff.}_{\rm SBP}\left({\bf x}_0^b\right)}}{\mathcal{Z}}\times\frac{\sum_{{\bf x^b}\in\Sigma_{{\bf x}_0}^{N,m}}e^{-\underline{\mathcal{L}}({\bf x}^a)}{\bf x}^a\cdot{\bf x}_0^b}{N\sum_{\underline{\bf x}\in\Sigma_{{\bf x}_0}^{N,m}}e^{-\underline{\mathcal{L}}(\underline{\bf x})}}\right]=m\delta_{a,b}
\end{align}
where indices $a$ and $b$ designate two replica of the system. The computation with this $Ansatz$ is standard and has already been detailed in many works \cite{aubin2019storage,barbier2023atypical}. After some computational steps, it yields
\begin{align}
    \phi^{\rm FP}(m)=&N\tilde{\phi}^{\rm ent.}(m)+\!M\!\int dw_0P^{{\rm edge}}(w_0)\log\!\left(\!\int \!dw\frac{e^{-\frac{(w-w_0)^2}{2(1-m^2)}-\underline{\mathcal{L}}(w)}}{\sqrt{2(1-m^2)}} \!\right)
\end{align}
and 
\begin{align}
    P_{\bf x}(w)&=\int dw_0 P^{{\rm edge}}(w_0)\frac{e^{-\frac{(w-w_0)^2}{2(1-m^2)}-\underline{\mathcal{L}}(w)}}{\mathcal{Z}_w(w_0)}\, ,\\ \mathcal{Z}_w(w_0)&=\int dw' e^{-\frac{(w'-w_0)^2}{2(1-m^2)}-\underline{\mathcal{L}}(w')}\,.
\end{align}
In fact, it is because of this dependence of $P_{\bf x}(\cdot)$ on $\underline{\mathcal{L}}(\cdot)$ that we can set $P_{{\bf x}}(\cdot)=P_{{\bf x}_0}(\cdot)=P^{{\rm edge}}(\cdot)$ by correctly tuning the loss.  

To detect a frozen 1-RSB phase, our procedure is as follows. Browsing $m$ from $1$ to $0$, we tune $\underline{\mathcal{L}}(\cdot)$ to obtain $P_{{\bf x}}(\cdot)=P^{{\rm edge}}(\cdot)$ for all values of $m$. Then, we evaluate $\phi^{\rm FP}(m)$ with the correct loss and determine when the system develops an overlap-gap -$\phi^{\rm FP}(m)<0$-. Because of our setting, this gap appears for $m$ infinitely close to one. This is a conjunction of the $annealed$ hypothesis (which imposes a self-overlap $q=m^2$ between replica) and the Nishimori condition $ P_{{\bf x}_0}(\cdot)=P_{{\bf x}}(\cdot)$ (which implies $\phi^{\rm FP}(\cdot)$ to be maximized for $q=m$). Putting the two constraints together, we have that $\phi^{\rm FP}(\cdot)$ can only develop a maxima for $m=0$ or $m=1$. As we have trivially $\phi^{\rm FP}(m=1)=0$ (i.e. if $ {\rm I\!E_\xi}\left[{{\bf x}^a\cdot {\bf x}_0^a}/{N}\right]=1$ the only available solution is ${\bf x}_0$ itself), having a maxima in $m=1$ gives $\phi^{\rm FP}(m\approx1)<0$.

\begin{figure}[h!]
    \centering
    \includegraphics[width=0.8\linewidth]{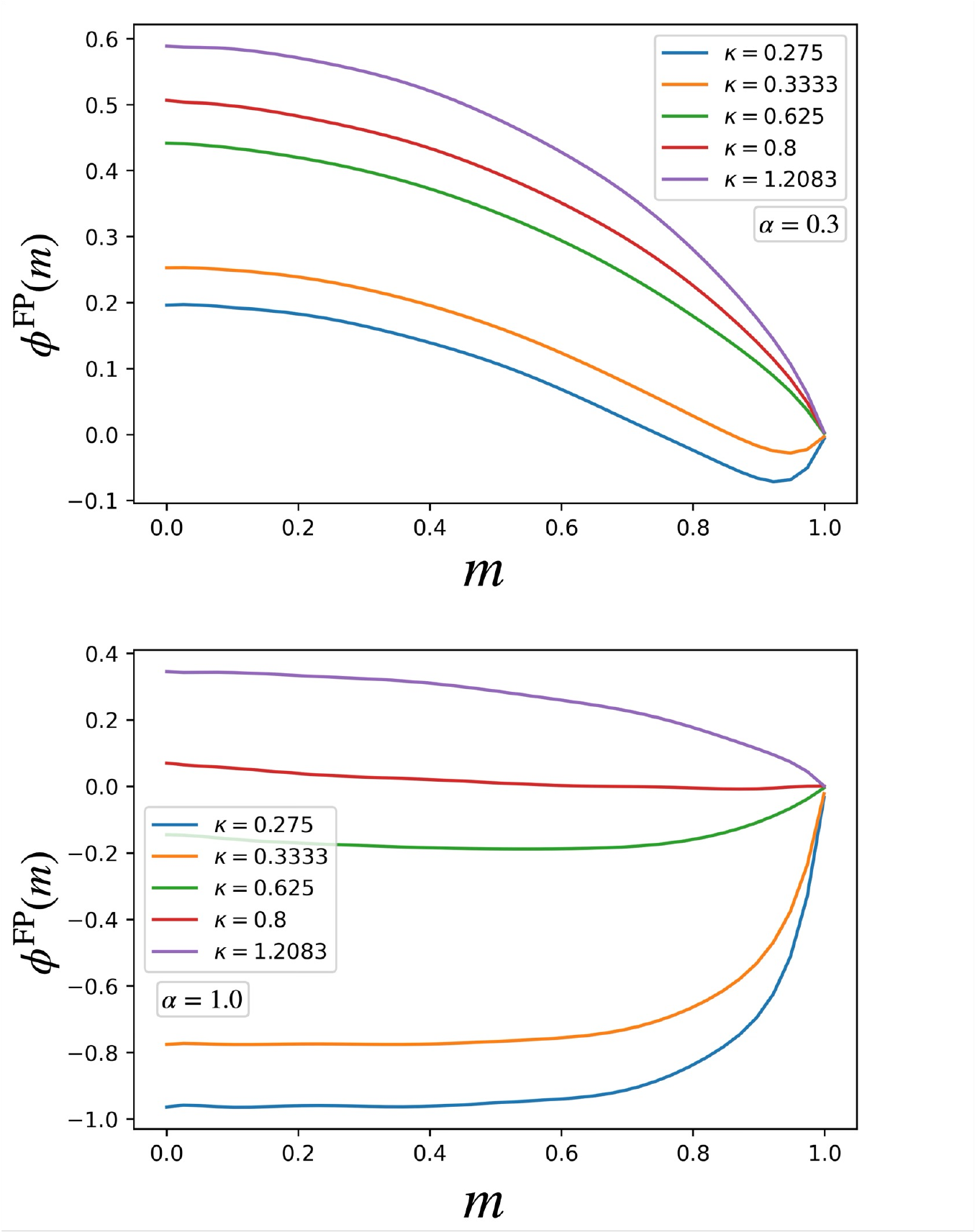}
    \caption{Plots displaying the Franz-Parisi potential defined in Eq.~(\ref{eq: FP connected manifold}) as a function of the overlap $m$ with the planted configuration. The top panel corresponds to $\alpha=0.3$, while the bottom one is $\alpha=1.0$. As explained, the conjunction of the $annealed$ hypothesis and the Nishimori condition gives that the potential is maximized for either $m=1$ or $m=0$. Therefore, when tuning $\kappa$, we observe a transition between a true $annealed$ phase (in which minima are not overlapping and not isolated) and a frozen 1-RSB phase (in which minima are also not overlapping but are now isolated).}
    \label{fig: FP connected minima}
\end{figure}

In Fig.\ref{fig: FP connected minima}, we plot the Franz-Parisi potential as a function of the overlap $m$, for two values of $\alpha$. As predicted, we see that it can only develop a maximum in $m=0$ and $m=1$ (depending on the value of $\kappa$ we have set). When a maximum exists for $m=1$, solutions are isolated -as we get with it $\phi^{\rm FP}(m\approx1)<0$-. However, the local stability for connected path predicts that these minima are isolated for a broader range of parameters $\alpha$ and $\kappa$ than predicted here, see Fig.~\ref{fig: phase diagram}.

\subsection{The case of a simple model}
\label{app: simple model FP-gen}
\begin{figure}[h!]
    \centering
    \includegraphics[width=0.7\linewidth]{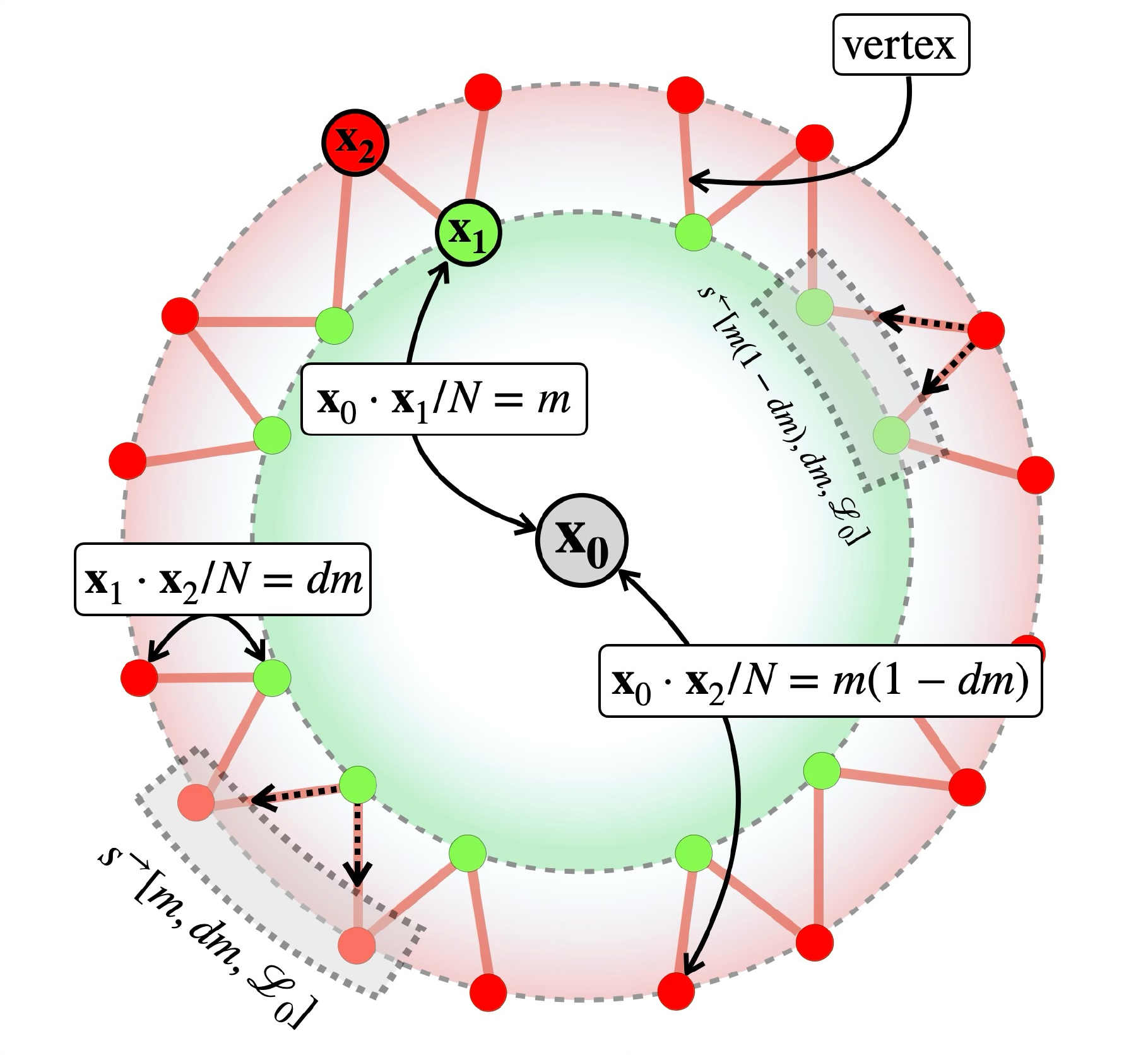}
    \caption{Schematic representation for the landscape of equilibrated systems (${\bf x}_1$ and ${\bf x}_2$) around a fixed configuration ${\bf x}_0$. The red lines represent what we call vertices, i.e. a path connecting a typical configuration ${\bf x}_1$ to a configuration ${\bf x}_2$.}
    \label{fig: simple model}
\end{figure}

In this section we consider a simple model on the hypercube (${\bf x}\in \Sigma^N$) with the potential
\begin{align}
    \mathcal{L}({\bf x})=\sum_{\mu=1}^M\left(\frac{\xi^\mu\cdot{\bf x}}{\sqrt{N}} \right)^2
\end{align}
and i.i.d. Gaussian random patterns, $\xi^\mu\sim \mathcal{N}(0,1)$.
In the following, we set a planted configuration ${\bf x}_0\in \Sigma^N$ with a given distribution of margins $P^0(w_0=\xi^\mu\cdot{\bf x}_0/\sqrt{N})$ and an energy $\mathcal{L}_0=\mathcal{L}({\bf x}_0)$. We want to compute the number of vertices (i.e short paths) linking a typical configuration ${\bf x}_1$ -given that ${\rm I\!E}_\xi[{\bf x}_1\cdot {\bf x}_0/N]=m$ and $\mathcal{L}({\bf x}_1)=\mathcal{L}_0$- to typical states ${\bf x}_2$ -with ${\rm I\!E}_\xi[{\bf x}_2\cdot {\bf x}_0/N]=m(1-dm)$ and $\mathcal{L}({\bf x}_2)=\mathcal{L}_0$-. In short, we want to compute the number of paths linking the typical configurations sitting on two close ``overlap''-slices ($m$ and $m(1-dm)$), given that the energy remains constant. To picture this a better, we schematized our construction in Fig.~\ref{fig: simple model}.

The number of vertices can be computed in two ways, either we count the number of configurations ${\bf x}_1$ and count how many solutions ${\bf x}_2$ they have around them, or we do the inverse -we count the number of configurations ${\bf x}_2$ and count how many solutions ${\bf x}_1$ they have around them-. In other words, we have the two equalities
\begin{align}
    N_{\rm vertex}(m,{\bf x}_0)=&\sum_{{{\bf x}_1\in\Sigma^{N,m}_{{\bf x}_0}}}e^{-\beta[\mathcal{L}({\bf x}_1)-\mathcal{L}_0]}\sum_{\underset{{\rm s.t.}\,{\bf x}_2\cdot {\bf x}_1/N=dm}{{\bf x}_2\in\Sigma^{N,m(1-dm)}_{{\bf x}_0}}}e^{-\beta'[\mathcal{L}({\bf x}_2)-\mathcal{L}_0]}
\end{align}
and
\begin{align}
    N_{\rm vertex}(m,{\bf x}_0)=&\sum_{{{\bf x}_2\in\Sigma^{N,m(1-dm)}_{{\bf x}_0}}}e^{-\beta[\mathcal{L}({\bf x}_2)-\mathcal{L}_0]}\sum_{\underset{{\rm s.t.}\,{\bf x}_2\cdot {\bf x}_1/N=dm}{{\bf x}_1\in\Sigma^{N,m}_{{\bf x}_0}}}e^{-\beta'[\mathcal{L}({\bf x}_1)-\mathcal{L}_0]}\,.
\end{align}
where $\beta$ and $\beta'$ are inverse temperatures fixing the average energy of each system to be equal to $\mathcal{L}_0$.
The previous equations directly imply that
\begin{align}
    &{\rm I\!E}_\xi\left\{\log\left[\sum_{{{\bf x}_1\in\Sigma^{N,m}_{{\bf x}_0}}}e^{-\beta[\mathcal{L}({\bf x}_1)-\mathcal{L}_0]}\right]\right\}\\&+{\rm I\!E}_\xi\left\{\log\left[\sum_{{{\bf x}_1\in\Sigma^{N,m}_{{\bf x}_0}}}\frac{e^{-\beta[\mathcal{L}({\bf x}_1)-\mathcal{L}_0]}}{\sum_{{\underline{{\bf x}}_1\in\Sigma^{N,m}_{{\bf x}_0}}}e^{-\beta[\mathcal{L}(\underline{{\bf x}}_1)-\mathcal{L}_0]}}\sum_{\underset{{\rm s.t.}{\bf x}_2\cdot {\bf x}_1/N=dm}{{\bf x}_2\in\Sigma^{N,m(1-dm)}_{{\bf x}_0}}}e^{-\beta'[\mathcal{L}({\bf x}_2)-\mathcal{L}_0]}\right]\right\}\nonumber\\
    =&{\rm I\!E}_\xi\left\{\log\left[\sum_{{{\bf x}_2\in\Sigma^{N,m(1-dm)}_{{\bf x}_0}}}e^{-\beta[\mathcal{L}({\bf x}_2)-\mathcal{L}_0]}\right]\right\}\nonumber\\
    &+\!{\rm I\!E}_\xi\!\left\{\!\log\!\!\left[\sum_{{{\bf x}_2\in\Sigma^{N,m(1-dm)}_{{\bf x}_0}}}\!\frac{e^{-\beta[\mathcal{L}({\bf x}_2)-\mathcal{L}_0]}}{\sum_{{\underline{\bf x}_2\in\Sigma^{N,m(1-dm)}_{{\bf x}_0}}}\!e^{-\beta[\mathcal{L}(\underline{\bf x}_2)-\mathcal{L}_0]}}\sum_{\underset{{\rm s.t.}{\bf x}_2\cdot {\bf x}_1/N=dm}{{\bf x}_1\in\Sigma^{N,m}_{{\bf x}_0}}}e^{-\beta'[\mathcal{L}({\bf x}_1)-\mathcal{L}_0]}\right]\right\}\nonumber
\end{align}
or written differently
\begin{align}
\label{eq: vertex equality}
    &\phi^{\rm FP}\left[m,\mathcal{L}_0\right]+s^\rightarrow\left[m,dm,\mathcal{L}_0\right]=\phi^{\rm FP}\left[m(1-dm),\mathcal{L}_0\right]+s^\leftarrow\left[m(1-dm),dm,\mathcal{L}_0\right]
\end{align}
with the definitions for the different entropies
\begin{align}
&\hspace{-0.2cm}\phi^{\rm FP}[m,\mathcal{L}_0]={\rm I\!E}_\xi\left\{\log\left[\sum_{\underset{{\bf x}\cdot {\bf x}_0=Nm}{{\bf x}\in\Sigma^N}}e^{-\beta[\mathcal{L}({\bf x})-\mathcal{L}_0]}\right]\right\}\,,\\
&s^\rightarrow\left[m,dm,\mathcal{L}_0\right]=\\
&\hspace{0.8cm}{\rm I\!E}_\xi\left\{\log\left[\sum_{{{\bf x}_1\in\Sigma^{N,m}_{{\bf x}_0}}}\frac{e^{-\beta[\mathcal{L}({\bf x}_1)-\mathcal{L}_0]}}{\sum_{{\underline{{\bf x}}_1\in\Sigma^{N,m}_{{\bf x}_0}}}e^{-\beta[\mathcal{L}(\underline{{\bf x}}_1)-\mathcal{L}_0]}}\sum_{\underset{{\rm s.t.}{\bf x}_2\cdot {\bf x}_1/N=dm}{{\bf x}_2\in\Sigma^{N,m(1-dm)}_{{\bf x}_0}}}e^{-\beta'[\mathcal{L}({\bf x}_2)-\mathcal{L}_0]}\right]\right\}\nonumber\,,\nonumber\\
&s^\leftarrow\left[m(1-dm),dm,\mathcal{L}_0\right]=\\
&\hspace{0.8cm}{\rm I\!E}_\xi\!\left\{\!\log\!\!\left[\sum_{{{\bf x}_2\in\Sigma^{N,m(1-dm)}_{{\bf x}_0}}}\!\frac{e^{-\beta[\mathcal{L}({\bf x}_2)-\mathcal{L}_0]}}{\sum_{{\underline{\bf x}_2\in\Sigma^{N,m(1-dm)}_{{\bf x}_0}}}\!e^{-\beta[\mathcal{L}(\underline{\bf x}_2)-\mathcal{L}_0]}}\sum_{\underset{{\rm s.t.}{\bf x}_2\cdot {\bf x}_1/N=dm}{{\bf x}_1\in\Sigma^{N,m}_{{\bf x}_0}}}e^{-\beta'[\mathcal{L}({\bf x}_1)-\mathcal{L}_0]}\right]\right\}\,.\nonumber
\end{align}
The entropy $s^\rightarrow[\cdot]$ represents the number of configurations ${\bf x}_2$ we can enumerate when sitting on a typical configuration ${\bf x}_1$. Conversely, the entropy $s^\leftarrow[\cdot]$ counts the number of available configurations ${\bf x}_1$ when sitting on a typical configuration ${\bf x}_2$. Therefore, we notice that the difference between these two quantities corresponds to the local entropy stability $\delta s_{\rm loc}$ introduced in Sec.~\ref{sec: stab}. Finally, the entropy $\phi^{\rm FP}[\cdot]$ is simply the number of solutions $\bf x$ available at a Hamming distance $N(1-m)/2$ from the planted configuration ${\bf x}_0$ -it corresponds to a Franz-Parisi potential-. 

Again, evaluating these entropies with an $annealed$ approximation for replica is standard \cite{aubin2019storage,barbier2023atypical}. With this $Ansatz$ we have
\begin{align}
\label{eq: FP pot simple model}
   \frac{\phi^{\rm FP}[m,\mathcal{L}_0]}N&=\underset{h,\beta}{\rm opt}\left\{\!-hm\!+\!\log\!\left[\sum_{x=\pm1}\!e^{hx}\right]\!\!+\!\alpha\beta\mathcal{L}_0+\alpha\int \!dw_0 P^{\rm edge}(w_0)\log\!\left[\int \!\mathcal{D}u\,e^{-\beta\left(mw_0+\sqrt{1-m^2}u\right)^2}\right]\!\right\}\nonumber\\
    &=\tilde{\phi}^{\rm ent.}(m)+\underset{\beta}{\rm opt}\left\{-\frac{\alpha}{2}\log\left(\Delta\right)-\frac{\alpha\beta m^2 \langle w_0^2\rangle}{\Delta}+\alpha\beta\mathcal{L}_0\right\}\, ,
\end{align}
\begin{align}
    \frac{s^\rightarrow\left[m,\mathcal{L}_0\right]}N&=\underset{h',h,\beta,\beta'}{\rm opt}\left\{-h'(1-dm)+\sum_{x}\frac{e^{hx}}{2{\rm cosh}(h)} \log\left[\sum_{x'=\pm 1}e^{h'x'x}\right] \right.\\
    &\hspace{+2cm}+\!\alpha\beta'\mathcal{L}_0+\!\alpha\!\!\int\! dw_0 P^0[w_0]\frac{\int \mathcal{D}u\,e^{-\beta\left(mw_0+\sqrt{1-m^2}u\right)^2}}{\int \mathcal{D}\underline{u}\,e^{-\beta\left(mw_0+\sqrt{1-m^2}\underline{u}\right)^2}}\nonumber\\
    &\left.\hspace{+3.8cm}\times\!\log\!\left[\!\int\! \!\mathcal{D}u'e^{-\beta'\!\left[m(1\!-\!dm)w_0+(1\!-\!dm)\sqrt{1-m^2}u+\sqrt{1-(1\!-\!dm)^2}u'\right]^2}\right]\!\right\}\nonumber\\
    &=\!\tilde{\phi}^{\rm ent.}(dm)\!+\!\underset{\beta,\beta'}{\rm opt}\left\{\!-\frac{\alpha}{2}\log\left[1\!+\!2\beta(1\!-\!{m^*}^2)\right]-\frac{\alpha\beta'{m^*}^2[\Delta(1-m^2)-m^2\langle w_0^2\rangle]}{\Delta^2[1+2\beta'(1-{m^*}^2)]}+\alpha\beta'\mathcal{L}_0\right\}\, ,\nonumber
\end{align}
and
\begin{align}
        \frac{s^\leftarrow\left[m,dm,\mathcal{L}_0\right]}N&=\underset{h'_1,h'_2,h,\beta,\beta'}{\rm opt}\left\{-h_1'(1-dm)-\frac{h_2'm}{1-dm}+\sum_{x}\frac{e^{hx}}{2{\rm cosh}(h)}\log\left[\sum_{x'=\pm 1}e^{h_1'x'x+h_2'x'}\right]\right.\nonumber\\
        &\hspace{2.2cm}+\alpha\beta'\mathcal{L}_0 +\alpha\int dw_0 P^0[w_0]\frac{\int \mathcal{D}u\,e^{-\beta\left(mw_0+\sqrt{1-m^2}u\right)^2}}{\int \mathcal{D}\underline{u}\,e^{-\beta\left(mw_0+\sqrt{1-m^2}\underline{u}\right)^2}}\\
    &\left.\hspace{+4cm}\times\log\left[\int \mathcal{D}u'\,e^{-\beta'\left[\frac{m}{1-dm}w_0+\mathcal{N}_1 u+\sqrt{\mathcal{N}_2}u'\right]^2+\beta\mathcal{L}_0}\right]\right\}\nonumber\\
    &\hspace{-0.2cm}\underset{dm\rightarrow 0}{=}\tilde{\phi}^{\rm ent.}(dm)+\left(m-\frac{m}{m^*}\right){\rm arctanh}(m)+\underset{\beta,\beta'}{\rm opt}\left\{-\frac{\alpha}{2}\log\left[1+2\beta'\mathcal{N}_2\right]\right.\nonumber\\
    &\hspace{-1.4cm}-\frac{\alpha\beta'\left[\Delta^2\left(\frac{m}{m^*}\right)^2\langle w_0^2\rangle-4\beta\Delta\frac{m^2}{m^*}\sqrt{1-m^2}\mathcal{N}_1\langle w_0^2\rangle\right]}{(1+2\beta'\mathcal{N}_2)[1+2\beta(1-m^2)]^2}\left.-\frac{\alpha\beta'{\mathcal{N}_1}^2(\Delta+4\beta^2(1-m^2)m^2\langle w_0^2\rangle)}{(1+2\beta'\mathcal{N}_2)[1+2\beta(1-m^2)]^2}\right\}\nonumber
\end{align}
with
\begin{align}
\langle w_0^2\rangle&=\int dw_0 P^0[w_0] w_0^2\, , \\
\Delta&=1+2\beta(1-m^2)\, , \\
m^*&=1-dm\, ,\\
\mathcal{N}_1&=\frac{1-dm+m^2/(1-dm)}{\sqrt{1-m^2}}\, , \\
\mathcal{N}_2&=1-m^2/(1-dm)^2-{\mathcal{N}_1}^2\, .
\end{align}

Note that we used again the decompositions into Gaussian processes as -$u\, , u'\sim\mathcal{N}(0,1)$-
\begin{align}
    w\Big\vert_{{\rm I\!E}_\xi[{ w}_0{ w}']={m}}=mw_0+\sqrt{1-m^2}u
\end{align}
and -${{\rm I\!E}_\xi[{w}{ w}']=d{m}}$, ${{\rm I\!E}_\xi[{w}_0{ w}']=\tilde{m}}$-
\begin{align}
    &w'=dm\,w+\frac{\tilde{m}-mdm}{1-m^2}(w_0-mw)+\sqrt{1-m^2-\frac{(\tilde{m}-mdm)^2}{1-m^2}}\, u'\\
    &\hspace{-1.2cm}\implies\!\!
    \left\{\!\!\begin{array}{l}
         w'\Big\vert_{{\rm I\!E}_\xi[{ w}_0{ w}']=m(1-dm)}=m(1-dm)w_0+(1-dm)\sqrt{1-m^2}u\\
         \hspace{3.51cm}+\sqrt{1-(1-dm)^2}u'\\
         w'\Big\vert_{{\rm I\!E}_\xi[{ w}_0{ w}']=\frac{m}{1-dm}}=\frac{m}{1-dm}w_0+\mathcal{N}_1 u+\sqrt{\mathcal{N}_2}u' 
    \end{array}\right. \!\!.\nonumber
\end{align}

\begin{figure}[h!]
\centering
    \includegraphics[width=0.75\linewidth]{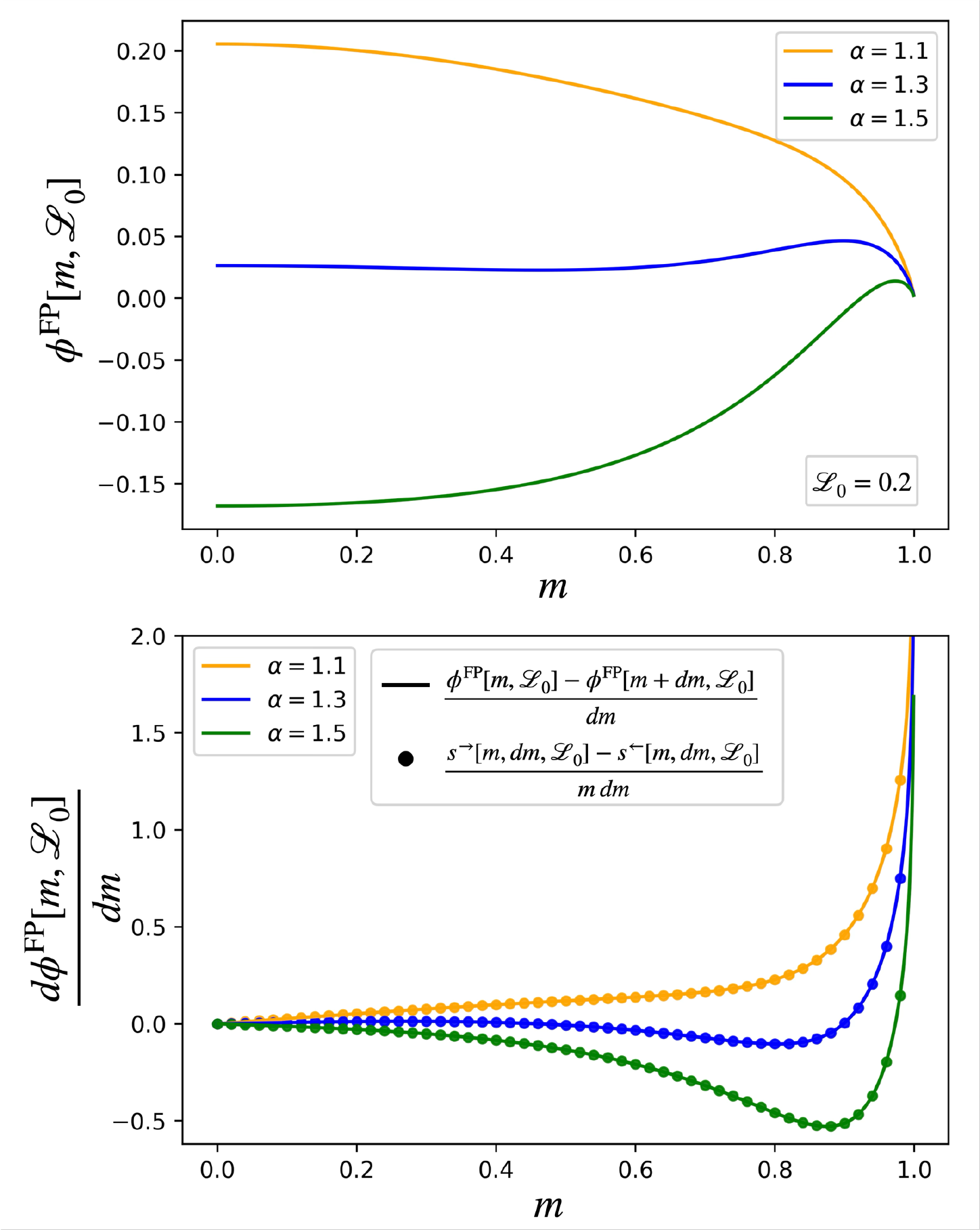}
    \caption{Plots displaying the behavior of the Franz-Parisi potential defined in Eq.~(\ref{eq: FP pot simple model}) as a function of the overlap $m$ with the planted configuration ${\bf x}_0$. The top panel simply shows to the potential dependence with $m$, while the bottom one shows its derivative $d\phi^{\rm FP}[m,\mathcal{L}_0]/dm$. To show the equivalence with the local entropies, we added in this last plot the value of $(s^\rightarrow\left[m,dm,\mathcal{L}_0\right]\!-\!s^\leftarrow\left[m,dm,\mathcal{L}_0\right])/(m\,dm)$ as colored dots (each color corresponding to a value of $\alpha$). }
    \label{fig: model simple}
\end{figure}

Taking the limit $dm\rightarrow0$ in Eq.~(\ref{eq: vertex equality}), we should obtain 
\begin{align}
    \frac{d\phi^{\rm FP}}{dm}\!\left[m,\mathcal{L}_0\right]&\!=\!\frac{s^\rightarrow\left[m,dm,\mathcal{L}_0\right]\!-\!s^\leftarrow\left[m(1\!-\!dm),dm,\mathcal{L}_0\right]}{m\,dm}\\
    &\approx\!\frac{s^\rightarrow\left[m,dm,\mathcal{L}_0\right]\!-\!s^\leftarrow\left[m,dm,\mathcal{L}_0\right]}{m\,dm} .\nonumber
\end{align}
In Fig.~\ref{fig: model simple}, we present cases with different values of $\alpha$ and fixed loss ($\mathcal{L}_0=0.2$). We can observe that the above equality is verified. This shows (for simple models) that the Franz-Parisi potential and the local stability criterion -$\delta s_{\rm loc}\propto s^\rightarrow\left[m,dm,\mathcal{L}_0\right]\!-\!s^\leftarrow\left[m,dm,\mathcal{L}_0\right]$ in this case- are in fact equivalent observables.

\section{Another modification for Monte-Carlo algorithms}
\label{app: modified+ MC}
\begin{figure}[htbp]
    \centering
    \includegraphics[width=0.75\linewidth]{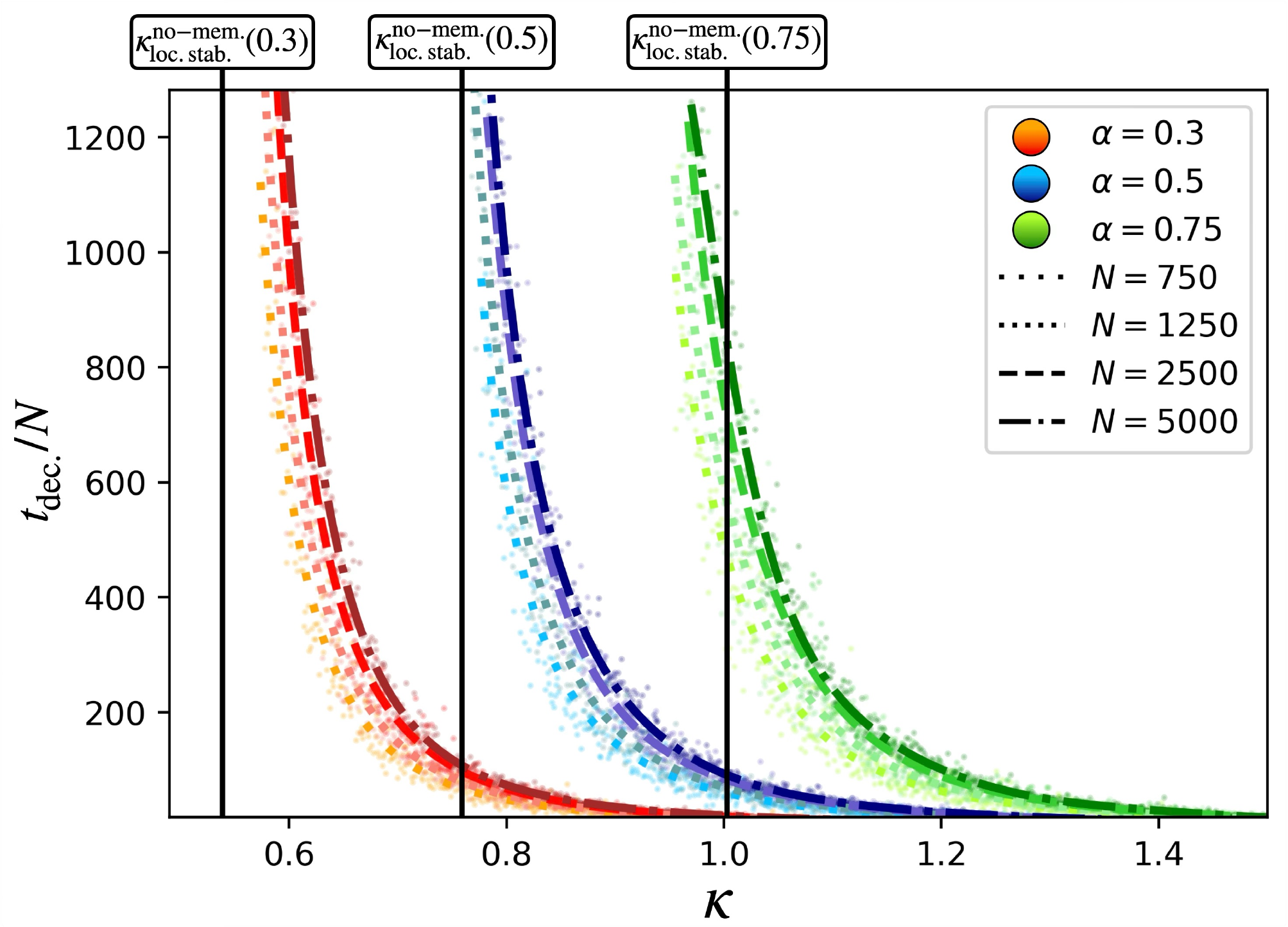}
    \caption{Plot showing the decorrelation time $t_{\rm dec.}$ as a function of $\kappa$ for a single annealing procedure. We tested the annealing setups ($\alpha=\{0.3,0.5,0.75\}$ and $N=\{750,1250,2500,5000\}$) with the new cost $\hat{G}^{\rm energ}_{\lambda^{\rm top}}[\cdot]$ -see Eq.~(\ref{eq: effective cost app})-. Each point in the plot indicates a decorrelation time obtained for a given round of annealing. In the shades of red (respectively blue and green), we have the simulations for $\alpha=0.3$ (respectively $\alpha=0.5$ and $\alpha=0.75$). The different system sizes are highlighted with different shades of the same color, the lightest corresponds to $N=750$, the darkest corresponds to $N=5000$. We also fitted the points of each setup with a polynomial, they are plotted as colored line (with a different pattern for each value of $N$). The critical values of $\kappa$ for which we predicted a local instability -in the no-memory paths- are highlighted with vertical black lines.}
    \label{fig: decorrelation time app}
\end{figure}

We showed in Sec.~\ref{sec: numerics} that solutions within the no-memory cluster have an infinitesimally steep landscape in their immediate surroundings. By this we mean that if we first sit in one of these minima and flip a spin at random, the energy of the system will increase infinitely. This is due to the behavior of $G^{\rm energ}_{\lambda^{\rm top}}$ close to its edge, i.e. $G^{\rm energ}_{\lambda^{\rm top}}[w,m]\underset{\vert w\vert\rightarrow\kappa}{\sim}\kappa-\vert w\vert$. As a remedy, we propose a new test function $\hat{G}^{\rm energ}_{\lambda^{\rm top}}[w]$ ensuring that the energetic shift
\begin{align}
    &\mathcal{L}^{\rm eff.}_{\rm SBP}({\bf x}^*)\!-\!    \mathcal{L}^{\rm eff.}_{\rm SBP}({\bf x})\approx\frac{-\left({\bf x}^*-{\bf x}\right)^2}{2N}\!\int \!dw\, \hat{P}^{\lambda^{\rm top}}\!(w)\partial^2_w\log\left(\!\hat{G}_{\lambda^{\rm top}}^{\rm }\!\left[w\right] \right)
\end{align}
converges, with
\begin{align}
        \hat{P}^{\lambda^{\rm top}}(w)=\frac{e^{-\frac{w^2}{2}}\hat{G}_{\lambda^{\rm top}}^{\rm energ.}[w]}{\int_{-\kappa}^\kappa d\underline{w} \,e^{-\frac{\underline{w}^2}{2}}\hat{G}_{\lambda^{\rm top}}^{\rm energ.}[\underline{w}]}\,.
\end{align}
For this, we take
\begin{align}
\label{eq: effective cost app}
   \hat{G}^{\rm energ}_{\lambda^{\rm top}}[w]=&\frac{(\kappa+w)(\kappa-w)}{\kappa^2}\times\frac{H(\kappa,w,0.1)-H(\kappa,\kappa,0.1)}{H(\kappa,0,0.1)-H(\kappa,\kappa,0.1)}
\end{align}
with
\begin{align}
    H(x,y,z)=\frac{1}{2}{\rm erf}\left(\frac{x-y}{\sqrt{2z}}\right)+\frac{1}{2}{\rm erf}\left(\frac{x+y}{\sqrt{2z}}\right)
\end{align}
and we have with it  $\hat{G}^{\rm energ}_{\lambda^{\rm top}}[w]\underset{\vert w\vert\rightarrow\kappa}{\sim}(\kappa-\vert w\vert)^2$. In Fig.~\ref{fig: decorrelation time app} we plot the decorrelation time $t_{\rm dec.}$ as a function of $\kappa$ (for a single annealing procedure and different setups $\{\alpha,N\}$). As a reminder, the decorrelation time corresponds to the number of iterations required at each round of annealing to decorrelate from the initialization -${\bf x}_{\rm t_{dec.}}\cdot{\bf x}_0/N=0.05$-. Compared to Fig.~\ref{fig: escape times}, we observe that finite size effects are reduced. This shows that the gradual shift in $t_{\rm dec.}$ with the first loss was indeed caused by fluctuations around the no-memory paths -and more particularly caused by spin flips in random directions-.

\newpage
\bibliographystyle{unsrt}
\bibliography{apssamp}

@article{barbier2023atypical,
      title={On the Atypical Solutions of the Symmetric Binary Perceptron}, 
      author={Damien Barbier and Ahmed El Alaoui and Florent Krzakala and Lenka Zdeborová},
    publisher = {IOP Publishing},
    volume = {57},
    number = {19},
    pages = {195202},
    year = {2024},
    journal = {Journal of Physics A: Mathematical and Theoretical},
    doi={10.1088/1751-8121/ad3a4a}
}

@article{franz1995recipes,
  title={Recipes for metastable states in spin glasses},
  author={Franz, Silvio and Parisi, Giorgio},
  journal={Journal de Physique I},
  volume={5},
  number={11},
  pages={1401--1415},
  year={1995},
  publisher={EDP Sciences}
}

@article{krzakala2007gibbs,
  title={Gibbs states and the set of solutions of random constraint satisfaction problems},
  author={Krzaka{\l}a, Florent and Montanari, Andrea and Ricci-Tersenghi, Federico and Semerjian, Guilhem and Zdeborov{\'a}, Lenka},
  journal={Proceedings of the National Academy of Sciences},
  volume={104},
  number={25},
  pages={10318--10323},
  year={2007},
  publisher={National Acad Sciences}
}

@article{zdeborova2008locked,
  title={Locked constraint satisfaction problems},
  author={Zdeborov{\'a}, Lenka and M{\'e}zard, Marc},
  journal={Physical review letters},
  volume={101},
  number={7},
  pages={078702},
  year={2008},
  doi = {10.1103/PhysRevLett.101.078702},
  publisher={APS}
}

@article{martin2004frozen,
  title={Frozen glass phase in the multi-index matching problem},
  author={Martin, OC and M{\'e}zard, M},
  journal={Physical review letters},
  volume={93},
  number={21},
  pages={217205},
  year={2004},
  publisher={APS},
  doi={10.1103/PhysRevLett.93.217205}
}

@book{mezard1987spin,
  title={Spin glass theory and beyond: An Introduction to the Replica Method and Its Applications},
  author={M{\'e}zard, Marc and Parisi, Giorgio and Virasoro, Miguel Angel},
  volume={9},
  year={1987},
  publisher={World Scientific Publishing Company},
  doi = {10.1142/0271}
}

@article{bansal2020line,
  title={On-line balancing of random inputs},
  author={Bansal, Nikhil and Spencer, Joel H},
  journal={Random Structures \& Algorithms},
  volume={57},
  number={4},
  pages={879--891},
  year={2020},
  publisher={Wiley Online Library},
  doi={10.1002/rsa.20955}
}

@article{baldassi2016local,
	author = {Baldassi, Carlo and Ingrosso, Alessandro and Lucibello, Carlo and Saglietti, Luca and Zecchina, Riccardo},
	journal = {Journal of Statistical Mechanics: Theory and Experiment},
	number = {2},
	pages = {023301},
	publisher = {IOP Publishing},
	title = {Local entropy as a measure for sampling solutions in constraint satisfaction problems},
	volume = {2016},
    doi={10.1088/1742-5468/2016/02/023301},
	year = {2016}}

@article{baldassi2015subdominant,
	author = {Baldassi, Carlo and Ingrosso, Alessandro and Lucibello, Carlo and Saglietti, Luca and Zecchina, Riccardo},
	journal = {Physical review letters},
	number = {12},
	pages = {128101},
	publisher = {APS},
	title = {Subdominant dense clusters allow for simple learning and high computational performance in neural networks with discrete synapses},
	volume = {115},
    doi = {10.1103/PhysRevLett.115.128101},
	year = {2015}}

@article{huang2014origin,
	author = {Huang, Haiping and Kabashima, Yoshiyuki},
	journal = {Physical Review E},
	number = {5},
	pages = {052813},
	publisher = {APS},
	title = {Origin of the computational hardness for learning with binary synapses},
	volume = {90},
    doi = {10.1103/PhysRevE.90.052813},
	year = {2014}}

@article{huang2013entropy,
	author = {Huang, Haiping and Wong, KY Michael and Kabashima, Yoshiyuki},
	journal = {Journal of Physics A: Mathematical and Theoretical},
	number = {37},
	pages = {375002},
	publisher = {IOP Publishing},
	title = {Entropy landscape of solutions in the binary perceptron problem},
	volume = {46},
    doi = {10.1088/1751-8113/46/37/375002},
	year = {2013}}

@article{krauth89storage,
	author = {Krauth, Werner and M{\'e}zard, Marc},
	doi = {10.1051/jphys:0198900500200305700},
	hal_id = {jpa-00211124},
	hal_version = {v1},
	journal = {{Journal de Physique}},
	number = {20},
	pages = {3057-3066},
	title = {{Storage capacity of memory networks with binary couplings}},
	url = {https://hal.archives-ouvertes.fr/jpa-00211124},
	volume = {50},
	year = {1989}}

@article{braunstein2006learning,
	author = {Braunstein, Alfredo and Zecchina, Riccardo},
	journal = {Physical review letters},
	number = {3},
	pages = {030201},
	publisher = {APS},
	title = {Learning by message passing in networks of discrete synapses},
	volume = {96},
    doi = {10.1103/PhysRevLett.96.030201},
	year = {2006}}

@article{baldassi2020clustering,
	author = {Baldassi, Carlo and Della Vecchia, Riccardo and Lucibello, Carlo and Zecchina, Riccardo},
	journal = {Journal of Statistical Mechanics: Theory and Experiment},
	number = {7},
	pages = {073303},
	publisher = {IOP Publishing},
	title = {Clustering of solutions in the symmetric binary perceptron},
	volume = {2020},
    doi={10.1088/1742-5468/ab99be},
	year = {2020}}

@article{baldassi2021unveiling,
	author = {Baldassi, Carlo and Lauditi, Clarissa and Malatesta, Enrico M and Perugini, Gabriele and Zecchina, Riccardo},
	journal = {Physical Review Letters},
	number = {27},
	pages = {278301},
	publisher = {APS},
	title = {Unveiling the structure of wide flat minima in neural networks},
	volume = {127},
	year = {2021}}

@article{baldassi2016unreasonable,
  title={Unreasonable effectiveness of learning neural networks: From accessible states and robust ensembles to basic algorithmic schemes},
  author={Baldassi, Carlo and Borgs, Christian and Chayes, Jennifer T and Ingrosso, Alessandro and Lucibello, Carlo and Saglietti, Luca and Zecchina, Riccardo},
  journal={Proceedings of the National Academy of Sciences},
  volume={113},
  number={48},
  pages={E7655--E7662},
  year={2016},
  doi={10.1073/pnas.1608103113},
  publisher={National Acad Sciences}
}

@article{aubin2019storage,
	author = {Aubin, Benjamin and Perkins, Will and Zdeborova, Lenka},
	journal = {Journal of Physics A: Mathematical and Theoretical},
	number = {29},
	pages = {294003},
	publisher = {IOP Publishing},
	title = {Storage capacity in symmetric binary perceptrons},
	volume = {52},
    doi = {10.1088/1751-8121/ab227a},
	year = {2019}}

@article{gamarnik2022algorithms,
	author = {Gamarnik, David and K{\i}z{\i}lda{\u{g}}, Eren C and Perkins, Will and Xu, Changji},
	journal = {arXiv preprint arXiv:2203.15667},
	title = {Algorithms and Barriers in the Symmetric Binary Perceptron Model},
    doi={10.48550/arXiv.2203.15667},
	year = {2022}}

@article{gardner1988optimal,
	author = {Gardner, Elizabeth and Derrida, Bernard},
	journal = {Journal of Physics A: Mathematical and general},
	number = {1},
	pages = {271},
	publisher = {IOP Publishing},
	title = {Optimal storage properties of neural network models},
	volume = {21},
	year = {1988}}

@article{gamarnik2021survey,
	author = {Gamarnik, David},
	journal = {Proceedings of the National Academy of Sciences},
	number = {41},
	publisher = {National Academy of Sciences},
	title = {The overlap gap property: A topological barrier to optimizing over random structures},
	volume = {118},
	year = {2021}}

@inproceedings{gamarnik2014limits,
	author = {Gamarnik, David and Sudan, Madhu},
	booktitle = {Proceedings of the 5th conference on Innovations in theoretical computer science},
	organization = {ACM},
	pages = {369--376},
	title = {Limits of local algorithms over sparse random graphs},
	year = {2014}}

@article{thouless1977solution,
	author = {Thouless, David J. and Anderson, Philip W. and Palmer, Richard G.},
	journal = {Philosophical Magazine},
	number = {3},
	pages = {593--601},
	publisher = {Taylor \& Francis Group},
	title = {Solution of'Solvable model of a spin glass'},
	volume = {35},
	year = {1977}}

@article{Semerjian,
    author = {Budzynski, Louise and Semerjian, Guilhem},
    year = {2020},
    month = {12},
    pages = {},
    title = {The Asymptotics of the Clustering Transition for Random Constraint Satisfaction Problems},
    volume = {181},
    doi={10.1007/s10955-020-02635-8},
    journal = {Journal of Statistical Physics}}

@article{parisi1979infinite,
	author = {Parisi, Giorgio},
	journal = {Physical Review Letters},
	number = {23},
	pages = {1754},
	publisher = {APS},
	title = {Infinite number of order parameters for spin-glasses},
	volume = {43},
	year = {1979}}

@unpublished{KrzakalaEtAl,
	author = {F.~Krzakala and M.~M\'ezard and F.~Sausset and Y.~Sun and L.~Zdeborova},
	note = {{\sf arXiv:1109.4424}},
	title = {{Statistical physics-based reconstruction in compressed sensing}},
	year = {2011}}

@article{addario2018algorithmic,
	author = {Addario-Berry, Louigi and Maillard, Pascal},
	journal = {{\sf arXiv:1810.05129}},
	title = {The algorithmic hardness threshold for continuous random energy models},
	year = {2018}}

@article{DMM09,
	author = {David~L.~Donoho and Arian~Maleki and Andrea~Montanari},
	journal = {Proceedings of the National Academy of Sciences},
	pages = {18914-18919},
	title = {{Message Passing Algorithms for Compressed Sensing}},
	volume = {106},
	year = {2009}}

@article{mezard2002analytic,
	author = {M{\'e}zard, Marc and Parisi, Giorgio and Zecchina, Riccardo},
	journal = {Science},
	number = {5582},
	pages = {812--815},
	publisher = {American Association for the Advancement of Science},
	title = {Analytic and algorithmic solution of random satisfiability problems},
	volume = {297},
	year = {2002}}

@book{book_Potters,
author = {Potters, Marc and Bouchaud, Jean-Philippe},
year = {2020},
month = {11},
pages = {},
title = {A First Course in Random Matrix Theory: for Physicists, Engineers and Data Scientists},
doi={10.1017/9781108768900}
}

@article{Franz_2013,
doi = {10.1088/1742-5468/2013/02/P02003},
year = {2013},
month = {feb},
publisher = {IOP Publishing and SISSA},
volume = {2013},
number = {02},
pages = {P02003},
author = {Silvio Franz and Giorgio Parisi},
title = {Quasi-equilibrium in glassy dynamics: an algebraic view},
journal = {Journal of Statistical Mechanics: Theory and Experiment},
}

@article{Franz_2015,
doi = {10.1088/1742-5468/2015/10/P10010},
url = {https://dx.doi.org/10.1088/1742-5468/2015/10/P10010},
year = {2015},
month = {oct},
publisher = {IOP Publishing and SISSA},
volume = {2015},
number = {10},
pages = {P10010},
author = {Silvio Franz and Giorgio Parisi and Federico Ricci-Tersenghi and Pierfrancesco Urbani},
title = {Quasi equilibrium construction for the long time limit of glassy dynamics},
journal = {Journal of Statistical Mechanics: Theory and Experiment},
}

@article{PhysRevLett.71.173,
  title = {Analytical solution of the off-equilibrium dynamics of a long-range spin-glass model},
  author = {Cugliandolo, L. F. and Kurchan, J.},
  journal = {Phys. Rev. Lett.},
  volume = {71},
  issue = {1},
  pages = {173--176},
  numpages = {0},
  year = {1993},
  month = {Jul},
  publisher = {American Physical Society},
  doi = {10.1103/PhysRevLett.71.173},
  url = {https://link.aps.org/doi/10.1103/PhysRevLett.71.173}
}

@article{Ros2021,
   author = {Valentina Ros and Giulio Biroli and Chiara Cammarota},
   doi = {10.21468/SciPostPhys.10.1.002},
   issn = {2542-4653},
   issue = {1},
   journal = {SciPost Physics},
   month = {1},
   pages = {002},
   title = {Dynamical instantons and activated processes in mean-field glass models},
   volume = {10},
   year = {2021}
}

@article{Annesi2023,
   author = {Brandon Livio Annesi and Clarissa Lauditi and Carlo Lucibello and Enrico M. Malatesta and Gabriele Perugini and Fabrizio Pittorino and Luca Saglietti},
   doi = {10.1103/PhysRevLett.131.227301},
   issn = {0031-9007},
   issue = {22},
   journal = {Physical Review Letters},
   month = {11},
   pages = {227301},
   title = {Star-Shaped Space of Solutions of the Spherical Negative Perceptron},
   volume = {131},
   year = {2023}
}

@article{Baldassi2020,
   author = {Carlo Baldassi and Enrico M Malatesta and Matteo Negri and Riccardo Zecchina},
   doi = {10.1088/1742-5468/abcd31},
   issn = {1742-5468},
   issue = {12},
   journal = {Journal of Statistical Mechanics: Theory and Experiment},
   month = {12},
   pages = {124012},
   title = {Wide flat minima and optimal generalization in classifying high-dimensional Gaussian mixtures},
   volume = {2020},
   year = {2020}
}

@article{Baldassi2020_,
   author = {Carlo Baldassi and Riccardo Della Vecchia and Carlo Lucibello and Riccardo Zecchina},
   doi = {10.1088/1742-5468/ab99be},
   issn = {1742-5468},
   issue = {7},
   journal = {Journal of Statistical Mechanics: Theory and Experiment},
   month = {7},
   pages = {073303},
   title = {Clustering of solutions in the symmetric binary perceptron},
   volume = {2020},
   year = {2020}
}

@article{Wu2016,
   author = {Nicholas C Wu and Lei Dai and C Anders Olson and James O Lloyd-Smith and Ren Sun},
   doi = {10.7554/eLife.16965},
   editor = {Richard A Neher},
   issn = {2050-084X},
   journal = {eLife},
   month = {7},
   pages = {e16965},
   publisher = {eLife Sciences Publications, Ltd},
   title = {Adaptation in protein fitness landscapes is facilitated by indirect paths},
   volume = {5},
   url = {https://doi.org/10.7554/eLife.16965},
   year = {2016}
}

@article{Papkou2023,
   author = {Andrei Papkou and Lucia Garcia-Pastor and José Antonio Escudero and Andreas Wagner},
   doi = {10.1101/2023.02.27.530293},
   journal = {bioRxiv},
   publisher = {Cold Spring Harbor Laboratory},
   title = {A rugged yet easily navigable fitness landscape of antibiotic resistance},
   url = {https://www.biorxiv.org/content/early/2023/02/28/2023.02.27.530293},
   year = {2023}
}

@article{Greenbury2022,
   author = {S F Greenbury and A A Louis and S E Ahnert},
   issue = {11},
   journal = {Nature Ecology and Evolution},
   pages = {1742-1752},
   publisher = {Springer Nature},
   title = {The structure of genotype-phenotype maps makes fitness landscapes navigable},
   volume = {6},
   year = {2022}
}

@article{Macadangdang2022,
   author = {Benjamin R. Macadangdang and Sara K. Makanani and Jeff F. Miller},
   doi = {10.1146/annurev-micro-030322-040423},
   issn = {0066-4227},
   issue = {1},
   journal = {Annual Review of Microbiology},
   month = {9},
   pages = {389-411},
   title = {Accelerated Evolution by Diversity-Generating Retroelements},
   volume = {76},
   year = {2022}
}

@article{Mauri2023,
   author = {Eugenio Mauri and Simona Cocco and Rémi Monasson},
   doi = {10.1103/PhysRevE.108.024141},
   issn = {2470-0045},
   issue = {2},
   journal = {Physical Review E},
   month = {8},
   pages = {024141},
   title = {Transition paths in Potts-like energy landscapes: General properties and application to protein sequence models},
   volume = {108},
   year = {2023}
}

@article{Medhekar2007,
   author = {Bob Medhekar and Jeff F Miller},
   doi = {10.1016/j.mib.2007.06.004},
   issn = {13695274},
   issue = {4},
   journal = {Current Opinion in Microbiology},
   month = {8},
   pages = {388-395},
   title = {Diversity-generating retroelements},
   volume = {10},
   year = {2007}
}

@article{Fear1998,
   author = {Karen K. Fear and Trevor Price},
   doi = {10.2307/3546365},
   issn = {00301299},
   issue = {3},
   journal = {Oikos},
   month = {9},
   pages = {440},
   title = {The Adaptive Surface in Ecology},
   volume = {82},
   year = {1998}
}

@article{Hatton2024,
   author = {Ian A. Hatton and Onofrio Mazzarisi and Ada Altieri and Matteo Smerlak},
   doi = {10.1126/science.adg8488},
   issn = {0036-8075},
   issue = {6688},
   journal = {Science},
   month = {3},
   title = {Diversity begets stability: Sublinear growth and competitive coexistence across ecosystems},
   volume = {383},
   year = {2024}
}

@article{Stadler2002,
   author = {Bärbel M. R. Stadler and Peter F. Stadler},
   doi = {10.1021/ci0100898},
   issn = {0095-2338},
   issue = {3},
   journal = {Journal of Chemical Information and Computer Sciences},
   month = {5},
   pages = {577-585},
   title = {Generalized Topological Spaces in Evolutionary Theory and Combinatorial Chemistry},
   volume = {42},
   year = {2002}
}

@article{Gamarnik2022,
   author = {David Gamarnik and Cristopher Moore and Lenka Zdeborová},
   doi = {10.1088/1742-5468/ac9cc8},
   issn = {1742-5468},
   issue = {11},
   journal = {Journal of Statistical Mechanics: Theory and Experiment},
   month = {11},
   pages = {114015},
   title = {Disordered systems insights on computational hardness},
   volume = {2022},
   year = {2022}
}

@article{wr32,
author = {Wright, S.},
title = {The Roles of Mutation, Inbreeding, crossbreeding and Selection in Evolution},
journal={Proceedings of the XI International Congress of Genetics},
volume={8},
pages = {209-222},
year = {1932}
}

@article{Hopfield1982,
   author = {J J Hopfield},
   doi = {10.1073/pnas.79.8.2554},
   issn = {0027-8424},
   issue = {8},
   journal = {Proceedings of the National Academy of Sciences},
   month = {4},
   pages = {2554-2558},
   title = {Neural networks and physical systems with emergent collective computational abilities.},
   volume = {79},
   year = {1982}
}

@article{Achlioptas2011,
   author = {Dimitris Achlioptas and Amin Coja‐Oghlan and Federico Ricci‐Tersenghi},
   doi = {10.1002/rsa.20323},
   issn = {1042-9832},
   issue = {3},
   journal = {Random Structures \& Algorithms},
   month = {5},
   pages = {251-268},
   title = {On the solution‐space geometry of random constraint satisfaction problems},
   volume = {38},
   year = {2011}
}

@article{Rannala1996,
   author = {Bruce Rannala and Ziheng Yang},
   doi = {10.1007/BF02338839},
   issn = {0022-2844},
   issue = {3},
   journal = {Journal of Molecular Evolution},
   month = {9},
   pages = {304-311},
   title = {Probability distribution of molecular evolutionary trees: A new method of phylogenetic inference},
   volume = {43},
   year = {1996}
}

@article{Huelsenbeck2001,
   author = {John P. Huelsenbeck and Fredrik Ronquist and Rasmus Nielsen and Jonathan P. Bollback},
   doi = {10.1126/science.1065889},
   issn = {0036-8075},
   issue = {5550},
   journal = {Science},
   month = {12},
   pages = {2310-2314},
   title = {Bayesian Inference of Phylogeny and Its Impact on Evolutionary Biology},
   volume = {294},
   year = {2001}
}

@inproceedings{Huang2024,
   author = {Brice Huang},
   doi = {10.1109/FOCS61266.2024.00074},
   isbn = {979-8-3315-1674-1},
   booktitle = {2024 IEEE 65th Annual Symposium on Foundations of Computer Science (FOCS)},
   month = {10},
   pages = {1126-1136},
   publisher = {IEEE},
   title = {Capacity Threshold for the Ising Perceptron},
   year = {2024}
}

@article{Barbier2025,
   author = {Damien Barbier},
   doi = {10.21468/SciPostPhys.18.3.115},
   issn = {2542-4653},
   issue = {3},
   journal = {SciPost Physics},
   month = {3},
   pages = {115},
   title = {How to escape atypical regions in the symmetric binary perceptron: A journey through connected-solutions states},
   volume = {18},
   year = {2025}
}

@article{Barrat1997,
   author = {Alain Barrat and Silvio Franz and Giorgio Parisi},
   doi = {10.1088/0305-4470/30/16/006},
   issn = {0305-4470},
   issue = {16},
   journal = {Journal of Physics A: Mathematical and General},
   month = {8},
   pages = {5593-5612},
   title = {Temperature evolution and bifurcations of metastable states in mean-field spin glasses, with connections with structural glasses},
   volume = {30},
   year = {1997}
}

@inproceedings{Barbier2023,
   author = {D. Barbier and C. Lucibello and L. Saglietti and F. Krzakala and L. Zdeborová},
   doi = {10.1109/ITW55543.2023.10161684},
   isbn = {979-8-3503-0149-6},
   booktitle = {2023 IEEE Information Theory Workshop (ITW)},
   month = {4},
   pages = {323-328},
   publisher = {IEEE},
   title = {Compressed sensing with l0-norm: statistical physics analysis \& algorithms for signal recovery},
   year = {2023}
}

@article{SaraoMannelli2020,
   author = {Stefano Sarao Mannelli and Giulio Biroli and Chiara Cammarota and Florent Krzakala and Pierfrancesco Urbani and Lenka Zdeborová},
   doi = {10.1103/PhysRevX.10.011057},
   issn = {2160-3308},
   issue = {1},
   journal = {Physical Review X},
   month = {3},
   pages = {011057},
   title = {Marvels and Pitfalls of the Langevin Algorithm in Noisy High-Dimensional Inference},
   volume = {10},
   year = {2020}
}

@article{Baldassi2020___,
   author = {Carlo Baldassi and Fabrizio Pittorino and Riccardo Zecchina},
   doi = {10.1073/pnas.1908636117},
   issn = {0027-8424},
   issue = {1},
   journal = {Proceedings of the National Academy of Sciences},
   month = {1},
   pages = {161-170},
   title = {Shaping the learning landscape in neural networks around wide flat minima},
   volume = {117},
   year = {2020}
}

@article{Baldassi2018,
   author = {Carlo Baldassi and Riccardo Zecchina},
   doi = {10.1073/pnas.1711456115},
   issn = {0027-8424},
   issue = {7},
   journal = {Proceedings of the National Academy of Sciences},
   month = {2},
   pages = {1457-1462},
   title = {Efficiency of quantum vs. classical annealing in nonconvex learning problems},
   volume = {115},
   year = {2018}
}

@article{Laurenceau2025,
   author = {Raphael Laurenceau and Paul Rochette and Elena Lopez-Rodriguez and Catherine Fan and Amandine Maire and Paul Vittot and Karol Melissa Cerdas-Mejias and Auguste Bouvier and Thea Chrysostomou and David Bikard},
   doi = {10.1101/2025.03.24.644984},
   journal = {bioRxiv},
   publisher = {Cold Spring Harbor Laboratory},
   title = {Harnessing Diversity Generating Retroelements for in vivo targeted hyper-mutagenesis},
   year = {2025}
}

@article{Kosterlitz1976,
   author = {J. M. Kosterlitz and D. J. Thouless and Raymund C. Jones},
   doi = {10.1103/PhysRevLett.36.1217},
   issn = {0031-9007},
   issue = {20},
   journal = {Physical Review Letters},
   month = {5},
   pages = {1217-1220},
   title = {Spherical Model of a Spin-Glass},
   volume = {36},
   year = {1976}
}
\end{document}